\documentclass[acmsmall]{acmart}

\usepackage{pdflscape}
\AtBeginDocument{%
  \providecommand\BibTeX{{%
    \normalfont B\kern-0.5em{\scshape i\kern-0.25em b}\kern-0.8em\TeX}}}

\usepackage{amsmath,amsfonts}
\usepackage{algorithmic}
\usepackage{graphicx}
\usepackage{textcomp}
\usepackage{xcolor}
\usepackage{soul}


\def\BibTeX{{\rm B\kern-.05em{\sc i\kern-.025em b}\kern-.08em
    T\kern-.1667em\lower.7ex\hbox{E}\kern-.125emX}}
 \usepackage{rotating}   
  \usepackage{multirow}  
  \usepackage{rotating}
\usepackage[english]{babel}
\usepackage[utf8]{inputenc} 
\usepackage[T1]{fontenc}
\usepackage{tabularx}
\usepackage{pdfpages}
\usepackage{balance}
\usepackage{url}
\usepackage{adjustbox}
\usepackage{amsthm}
\newcolumntype{R}[2]{%
    >{\adjustbox{angle=#1,lap=\width-(#2)}\bgroup}%
    l%
    <{\egroup}%
}

\usepackage{lscape}

\usepackage{pdflscape}

\usepackage[most]{tcolorbox}

\usepackage{listings}
\usepackage{color}
\usepackage[labelfont=bf]{caption} 
\usepackage{ragged2e}
\usepackage{lscape} 
\usepackage{subcaption}
\usepackage{pgfplots} \pgfplotsset{compat=1.14}

\newcommand{\eg}{{\textit{e.g., }}}
\newcommand{\ie}{{\textit{i.e., }}}
\newcommand{\al}{\textit{et al. }}

\usepackage{amsmath}
\usepackage{tcolorbox}
\definecolor{lightGrey}{rgb}{0.9, 0.9, 0.9}

\usepackage{xcolor}

\definecolor{airforceblue}{rgb}{0.36, 0.54, 0.66}
\definecolor{dkgreen}{rgb}{0,0.6,0}
\definecolor{gray}{rgb}{0.5,0.5,0.5}
\definecolor{mauve}{rgb}{0.58,0,0.82}
\definecolor{antiquefuchsia}{rgb}{0.57, 0.36, 0.51}
\definecolor{applegreen}{rgb}{0.55, 0.71, 0.0}
\definecolor{asparagus}{rgb}{0.53, 0.66, 0.42}
\usepackage[graphicx]{realboxes}
\usepackage{color, colortbl} 
 \definecolor{Gray}{gray}{0.9}
  \usepackage{listings,xcolor,tikz}

\usepackage{xspace}
\usepackage{epstopdf} 
\usepackage{amsmath}
\usepackage{tikz}
\usepackage{multirow} 

\begin{document}

\title{Are Multi-language Design Smells Fault-prone? An Empirical Study}


\author{Mouna Abidi}
\affiliation{%
  \institution{DGIGL, Polytechnique Montreal}
  \city{Montreal, QC}
  \country{Canada}}
\email{mouna.abidi@polymtl.ca}

\author{Md Saidur Rahman}
\affiliation{%
  \institution{DGIGL, Polytechnique Montreal}
  \city{Montreal, QC}
  \country{Canada}}
\email{saidur.rahman@polymtl.ca}

\author{Moses Openja}
\affiliation{%
  \institution{DGIGL, Polytechnique Montreal}
  \city{Montreal, QC}
  \country{Canada}}
    \email{moses.openja@polymtl.ca}
    
\email{moses.openja@polymtl.ca}

\author{Foutse Khomh}
\affiliation{%
  \institution{DGIGL, Polytechnique Montreal}
  \city{Montreal, QC}
  \country{Canada}}
\email{foutse.khomh@polymtl.ca}

\renewcommand{\shortauthors}{Abidi et al.}



\newcommand{\RQOne}{\newtext{Do Multi-language design smells occur frequently in open source projects?}}
\newcommand{\RQTwo}{Are some specific Multi-language design smells more frequent than others in open source projects?} 

\newcommand{\RQThree}{Are files with Multi-language design smells more fault-prone than files without?} 

\newcommand{\RQFour}{Are some specific Multi-language design smells more fault-prone than others?} 

\newcommand{\RQFive}{What are the activities that are more likely to introduce bugs in smelly files?} 

\newcommand{\Foutse}[1]{\textcolor{red}{{\it [Foutse: #1]}}}
\newcommand{\Saidur}[1]{\textcolor{green}{{\it [SR: #1]}}}
\newcommand{\Mouna}[1]{\textcolor{green}{{\it [MA: #1]}}}
\newcommand{\Moses}[1]{\textcolor{orange}{{\it [MO: #1]}}}

\newcommand{\TODO}[1]{\textcolor{cyan}{{\it [TODO: #1]}}}
\newcommand{\cc}[1]{\cellcolor{gray!15}{{#1}}}
\newcommand{\newtext}[1]{{{#1}}}
\newcommand{\rocksdb}{\textit{Rocksdb}}
\newcommand{\vlc}{\textit{VLC-android}}
\newcommand{\realm}{\textit{Realm}}
\newcommand{\conscrypt}{\textit{Conscrypt}}
\newcommand{\pljava}{\textit{Pljava}}
\newcommand{\javacpp}{\textit{Javacpp}}
\newcommand{\zstd}{\textit{Zstd-jni}}
\newcommand{\jpype}{\textit{Jpype}}
\newcommand{\javasmt}{\textit{Java-smt}}

\newcommand{\excessivecom}{\textit{Excessive Inter-language Communication}}
\newcommand{\scatter}{\textit{Too Much Scattering}}
\newcommand{\cluster}{\textit{Too Much Clustering}}
\newcommand{\declaration}{\textit{Unused Method Declaration}}
\newcommand{\implementation}{\textit{Unused Method Implementation}}
\newcommand{\unusedParameters}{\textit{Unused Parameters}}
\newcommand{\SafeReturnValue}{\textit{Assuming Safe Return Value}}
\newcommand{\excessiveObject}{\textit{Excessive Objects}}
\newcommand{\exception}{\textit{Not Handling Exceptions}}
\newcommand{\cachingObjects}{\textit{Not Caching Objects}}
\newcommand{\notSecuringLib}{\textit{Not Securing Libraries}}
\newcommand{\hardCodingLib}{\textit{Hard Coding Libraries}}
\newcommand{\relativePath}{\textit{Not Using Relative Path}}
\newcommand{\memoryMagement}{\textit{Memory Management Mismatch}}
\newcommand{\localReference}{\textit{Local References Abuse}}

\begin{abstract}
Nowadays, modern applications are developed using components written in different programming languages and technologies. \newtext{The cost benefits of reuse and the advantages of each programming language are two main incentives behind the proliferation of such systems. 
However, as the number of languages increases, so does the challenges related to the development and maintenance of these systems.} \newtext{In such situations, developers may introduce design smells 
(\emph{i.e.,} anti-patterns and code smells) which are symptoms of poor design and implementation choices.}
Design smells are defined as poor design and coding choices that can negatively impact the quality of a software program despite satisfying functional requirements. 
Studies on mono-language systems suggest that the presence of design smells may indicate a higher risk of future bugs and affects code comprehension, thus making systems harder to maintain. However, the impact of multi-language design smells on software quality such as fault-proneness is yet to be investigated.

In this paper, we present an approach to detect multi-language design smells in the context of JNI systems. We then investigate the prevalence of those design smells and their impacts on fault-proneness. Specifically, we detect 15 design smells in 98 releases of nine open source JNI projects.
Our results show that the design smells are prevalent in the selected projects and persist throughout the releases of the systems. 
We observe that in the analyzed systems, 33.95\% of the files involving communications between Java and C/C++ contain occurrences of multi-language design smells. Some kinds of smells are more prevalent than others, \eg \unusedParameters, \scatter, \declaration. 
 Our results suggest that files with multi-language design smells can often be more associated with bugs than files without these smells, and that specific smells are more correlated to fault-proneness than others. From analyzing fault-inducing commit messages, we also extracted activities that are more likely to introduce bugs in smelly files. We believe that our findings are important for practitioners as it can help them prioritize design smells during the maintenance of multi-language systems. 
\end{abstract}


\maketitle

\section{Introduction}
\label{sec:Intro}

\newtext{
 Modern applications are moving from the use of a single programming language to build a single application towards the use of more than one programming language  \cite{linos2003tool,kontogiannis2006comprehension,kochhar2016large}. Capers Jones reported in his book published in 1998, that at least one third of the software application at that time were written using two programming languages. He estimated that 10\% of the applications were written with three or more programming languages \cite{jones1998estimating}. Kontogiannis argued that these percentages are becoming higher with the technological advances \cite{kontogiannis2006comprehension}. Developers often leverage the strengths and take benefits of several programming languages to cope with the pressure of the market.}

A common approach \newtext{to develop multi-language system} is to write the source code in multiple languages to capture additional functionality and efficiency not available in a single language. 
For example, a mobile development team might combine Java, C/C++, JavaScript, SQL, and HTML5 to develop a fully-functional application. 
The core logic of the application might be written in Java, with some routines written in C/C++, and using some scripting languages or other domain specific languages to develop the user interface \cite{matthews2009operational}. \newtext{The cost benefits of reuse and the advantages of each programming language are increasingly powerful reasons behind the proliferation of multi-language systems.}

However, despite the numerous advantages of multi-language systems, they are not without some challenges. 
During 2013, famous web sites, \eg Business Insider, Huffington Post, and Salon were inaccessible, redirecting visitors to a Facebook error page. 
This was due to a bug related to the integration of components written in different programming languages. The bug was in JavaScript widgets embedded in Facebook and their interactions with Facebook's servers.\footnote{\url{https://www.wired.com/2013/02/facebook-widget-snafu/}} 
Another example related to multi-language design smells is a bug reported early in 2018, which was due to the misuse of the guideline specification when using the Java Native Interface (JNI), to combine Java with C/C++ in \texttt{libguests}.\footnote{\url{https://bugzilla.redhat.com/show_bug.cgi?id=1536762}} 
There were no checks for Java exceptions after all JNI calls that might throw them. In JRuby, several problems were also reported mainly related to incompatibilities between languages and missing checks of return values and crashes related to the C language.\footnote{\url{https://www.jruby.org/2012/05/21/jruby-1-7-0-preview1.html}}


Software quality has been widely studied in the literature and was often associated with the presence of design patterns, anti-patterns and code smells in the context of mono-language systems. 
Several studies in the literature have investigated the popularity and challenges of multi-language systems \cite{kochhar2016large,Lee:2009,goedicke2000object,goedicke2002piecemeal,neitsch2012build}, but very few of them studied multi-language patterns and practices \cite{goedicke2000object,goedicke2002piecemeal,neitsch2012build}. 
Kochhar \al \cite{kochhar2016large} claims that the use of several programming languages significantly increases bug proneness. 
They assert that design patterns and design smells are present in multi-language systems and suggest that researchers study them thoroughly. 
Mono-language design smells are conjectured in the literature to hinder software reliability. While a design smell may not definitively identify an error, its presence suggests a potential trouble spot, that is, a place where there is an increased risk of bugs or potential failure in the future. 

However, despite the importance and increasing popularity of multi-language systems, to the best of our knowledge, no approach has been proposed to detect multi-language smells. Also, there is no existing study that empirically evaluates the impacts of multi-language smells on the software fault-proneness.
Through this paper, we aim to fill this gap in the literature. We present an approach to detect multi-language design smells. Based on our approach, we detect occurrences of 15 multi-language design smells in 98 releases of nine open source multi-language projects (\ie \vlc, \conscrypt, \rocksdb, \realm, \javasmt, \pljava, \javacpp, \zstd, and \jpype). We focus on the analysis of JNI systems because they are commonly used by developers and also introduce several challenges \cite{Lee:2009,Tan:2008,abidi2019behind}. 
Our analysis is based on a previously published catalog comprising of anti-patterns and code smells related to multi-language systems \cite{abidi2019anti,abidi2019code}. 
In this paper, we aim to investigate the evolution of multi-language design smells and the relations between these smells and software fault-proneness. 
More specifically, we investigate the prevalence of 15 multi-language design smells in the context of JNI open source projects and evaluate their impact on fault-proneness. 

Our four key contributions are: (1) an approach to automatically detect multi-language design smells in the context of JNI systems, (2) evaluation of the prevalence of those design smells in the selected projects, (3) empirical evaluation of the impacts of multi-language design smells on software fault-proneness, and (4) text-based analysis to identify activities that are more likely to introduce bugs once performed in files with design smells. 



Our results show that in the analyzed systems, 33.95\% of the files involving communication between Java and C/C++ contain occurrences of the studied design smells. Some types of smells are more prevalent than others, \eg \unusedParameters, \scatter, \declaration. 
We bring evidence to researchers that (1) the studied design smells are prevalent in the selected projects and persist within the releases, (2) some types of design smells are more prevalent than others, (3) files with the studied multi-language design smells are more likely to be the subject of bugs than files without these smells, 
(4) some specific smells are more correlated to fault-proneness than others i.e., \unusedParameters, \scatter, \cluster, \hardCodingLib, and \memoryMagement, and (5) \emph{data conversion, memory management, restructuring the code, API usage} and \emph{exception management} activities could increase the risk of inducing bugs once performed in smelly files. We believe that our results could help not only researchers but also practitioners involved in the development of multi-language software systems. We also provide evidence to developers and quality assurance teams of the importance and usefulness of avoiding multi-language design smells. 





\textbf{The remainder of this paper is organized as follows.} 
Section \ref{sec:Background} discusses the background of multi-language systems and the design smells studied in this paper. 
Section \ref{sec:Study} describes our methodology. 
Section \ref{sec:Results} reports our results, while Section \ref{sec:Discussion} discusses these results for better insights and implications. 
Section \ref{sec:Threats} summarises the threats to the validity of our methodology and results. 
Section \ref{sec:RW} presents related work. 
Section \ref{sec:Conclusion} concludes the paper and discusses future works.
Appendix \ref{appendixA} describes the detection rules of the proposed approach.
Appendix \ref{appendixB} presents an overview of the approach validation.
\section{Background}
\label{sec:Background}
To study the impact of multi-language design smells on fault-proneness, we first introduce a brief background on multi-language (JNI) systems. We then discuss different types of multi-language design smells and illustrate them with examples. 

\subsection{Multi-language Systems}
 Nowadays, multi-language application development is gaining popularity over mono-language programming, because of their different inherent benefits.
 Developers often leverage the strengths of several languages to cope with the challenges of building complex systems. 
 By using languages that complement one another, the performance, productivity, and agility may be improved \cite{tomassetti2014empirical,Pfeiffer:2012,Mushtaq:2015}. 


Java Native Interface (JNI) is a foreign function interface programming framework for multi-language systems. JNI enables developers to invoke native functions from Java code and also Java methods from native functions. JNI presents a simple method to combine Java applications with either native libraries and/or applications \cite{Liang:1999, Hunt:1999}. It allows Java developers to take advantage of specific features and functionalities provided by native code. 
We present in Fig. \ref{Figure: HW2} an example of a JNI code extracted from \cite{Liang:1999}. Fig. \ref{jni-ex}(a) presents a Java class that contains a native method declaration \texttt{Print()} and loads the corresponding native library while Fig. \ref{jni-ex}(b) presents the C file that contains the implementation of the native function \texttt{Print()}.

\begin{figure} [!ht] 
\centering
\begin{subfigure}[b]{=0.5\linewidth}{(a) JNI method declaration} 
\centering
\begin{lstlisting}
class HelloWorld {
 
 static {
         System.loadLibrary("HelloWorld");}
 
     private native void print();
 
     public static void main(String[] args) 
     { new HelloWorld().print();
     }
 }
\end{lstlisting}
\end{subfigure}%
\begin{subfigure}[b]{0.5\linewidth}{(b) JNI implementation function} 
\centering
\begin{lstlisting}
 #include <jni.h>
 #include <stdio.h>
 #include "HelloWorld.h"
  
 JNIEXPORT void JNICALL 
 Java_HelloWorld_print(JNIEnv *env, jobject obj)
 {
     printf("Hello World!\n");
     return;
 }
 
\end{lstlisting}
\end{subfigure}%
\caption{JNI HelloWorld Example} \label{jni-ex}
\label{Figure: HW2}
\end{figure}

\subsection{Anti-patterns and Code Smells}

Patterns were introduced for the first time by Alexander in the domain of architecture \cite{alexander1977pattern}. From architecture, design patterns were then introduced in software engineering by Gamma \al \cite{Gamma:1995:DPE:186897}. They defined design patterns as common guidelines and ``good'' solutions based on the developers' experiences to solve recurrent problems. \newtext{Design smells (\ie anti-patterns and code smells), on the other hand, are symptoms of poor design and implementation choices.
They represent violations of best practices that often indicate the presence of bigger problems \cite{brown1998antipatterns,fowler1999refactoring}. 
There exist several definitions in the literature about code smells, anti-patterns, and their distinction  \cite{sharma2018survey,zhang2011code}. However, in this paper we consider design smells, in general, to refer to both code smells and anti-patterns. Several studies in the literature studied the impacts of design smells for mono-language systems and reported that classes containing design smells are significantly more fault-prone and change-prone compared to classes without smells \cite{khomh2009exploratory,romano2012analyzing,soh2016code,yamashita2013developers}.}

\subsection{Multi-language Design Smells} \label{sec:ML-smells}

Design patterns, anti-patterns, and code smells studied in the literature are mainly presented in the context of mono-language programming. While they were defined in the context of object oriented programming and mainly Java programming language, most of them could be applied to other programming languages. However, those variants consider mono-language programming and do not consider the interaction between programming languages. In a multi-language context, design smells are defined as poor design and coding decisions when bridging between different programming languages. They may slow down the development process of multi-language systems or increase the risk of bugs or potential failures in the future \cite{abidi2019anti,abidi2019code}. 

Our study is based on the recently published catalog of multi-language design smells \cite{abidi2019anti,abidi2019code}. 
The catalog was derived from an empirical study that mined the literature, developers' documentation, and bug reports. 
This catalog was validated by the pattern community and also by surveying professional developers \cite{abidi2019behind,abidi2019anti,abidi2019code}. 
Some of those design smells could also apply to the context of mono-language systems, however, in this study we focus only on the analysis of JNI systems. 
In this paper, since we are not analyzing anti-patterns and code smells separately but as the same entity, we will use the term design smells for both anti-patterns and code smells. 
In the following paragraphs, we elaborate on each of the design smells; providing an illustrative example. More details about these smells are available in the reference catalog \cite{abidi2019anti,abidi2019code}. 

 \begin{enumerate}
 
  \item \exception: The exception handling flow may differ from one programming language to the other. In case of JNI applications, developers should explicitly implement the exception handling flow after an exception has occurred \cite{Tan:2008,Li:2009,Kondoh:2008}.\footnote{\label{IBMSite}\url{https://www.ibm.com/developerworks/library/j-jni/index.html}} Since JNI exception does not disrupt the control flow until the native method returns, mishandling JNI exceptions may lead to vulnerabilities and leave security breaches open to malicious code \cite{Tan:2008,Li:2009,Kondoh:2008}. Listing \ref{fig:exceptionOccured} presents an example of occurrences of this smell extracted from IBM site\textsuperscript{\ref{IBMSite}}. In this example, developers are using predefined JNI methods to extract a class field that was passed as a parameter from Java to C code. However, they are returning the result without any exception management. If the class or the field C is not existing, this could lead to errors. A possible solution would be to use the function Throw() or ThrowNew() to handle JNI exception, and also to add a return statement right after one of these functions to exit the native method at a point of error. 

\begin{figure}[!ht]
\center

\lstset{backgroundcolor = \color{gray!4},language=C,basicstyle=\small\ttfamily , showspaces=false, showstringspaces=false,breaklines=true}
\begin{lstlisting} [language=C, caption={Design Smell - Not Handling Exceptions Across Languages}, label={fig:exceptionOccured}]
/* C++ */
jclass objectClass;
jfieldID fieldID;
jchar result = 0;
objectClass= (*env)->GetObjectClass(env, obj);
fieldID= (*env)->GetFieldID(env, objectClass, "charField", "C");
result= (*env)->GetCharField(env, obj, fieldID);
\end{lstlisting}


\end{figure}

 \item \SafeReturnValue: Similar to the previous design smell, in the context of JNI systems, not checking return values may lead to errors and security issues \cite{Li:2009,abidi2019code}. The return values from JNI methods indicates whether the call succeeded or not. It is the developers' responsibility to always perform a check before returning a variable from the native code to the host code to know whether the method ran correctly or not. Listing \ref{fig:returnval} presents an example of occurrences of this smell. If the class \textit{NIOAccess} or one of its methods is not found, the native code will cause a crash as the return value is not checked properly. A possible solution would be to implement checks that handle situations in which problems may occur with the return values. 

\begin{figure}[!ht]
\center

\lstset{backgroundcolor = \color{gray!4},language=C,basicstyle=\small\ttfamily, showspaces=false, showstringspaces=false,breaklines=true}
\begin{lstlisting} [language=C, caption={Design Smell - Assuming Safe Multi-language Return Values},label={fig:returnval}]
/* C++ */
staticvoid nativeClassInitBuffer(JNIEnv *_env){
 jclass nioAccessClassLocal= _env->FindClass("java/nio/NIOAccess");
 nioAccessClass=(jclass) _env->NewGlobalRef(nioAccessClassLocal);
 bufferClass=(jclass) _env->NewGlobalRef(bufferClassLocal);
 positionID= _env->GetFieldID(bufferClass, "position", "I");
\end{lstlisting}


\end{figure}


\item \notSecuringLib: A common way to load the native library in JNI is the use of the method \textit{loadLibrary} without the use of a secure block. In such situation, the code loads a foreign library without any security check or restriction. However, after loading the library, malicious code can call native methods from the library, this may impact the security and reliability of the system \cite{long2013java,abidi2019code}. Listing \ref{fig:SafeLoad}, presents an example of a possible solution by loading the native library within a secure block to avoid malicious attacks. 

\begin{figure}[!ht]
\center
\lstset{backgroundcolor = \color{gray!4},language=Java,basicstyle=\small\ttfamily , stepnumber=1, showspaces=false, showstringspaces=false,breaklines=true}
\begin{lstlisting} [language=Java, caption={Securing Library Loading}, label={fig:SafeLoad}]
/* Java */
 static { AccessController.doPrivileged(
          new PrivilegedAction<Void>() {
          public Void run() {
          System.loadLibrary("osxsecurity");
          return null; } }  ); }
\end{lstlisting}
\end{figure}

\item \hardCodingLib: \newtext{Let us consider a situation in which we have the same code to run on different platforms. We need to customize the loading according to the operating system. However, when those libraries are not loaded considering operating system specific conditions and requirements, but for instance with hard coded names and a try-catch mechanism, it is hard to know which library has really been loaded which could bring confusion especially during the maintenance tasks. Listing \ref{fig:hardcode} provides an example of native libraries loaded without any information about how to distinguish between the usage of those libraries.} 

\begin{figure}[!ht]
\center

\lstset{backgroundcolor = \color{gray!4},language=Java,basicstyle=\small\ttfamily  , stepnumber=1, showspaces=false, showstringspaces=false,breaklines=true}
\begin{lstlisting} [language=Java, caption={Design Smells - Hard Coding Libraries}, label={fig:hardcode}]
/* Java */
public static synchronized Z3SolverContext create(
try { System.loadLibrary("z3"); System.loadLibrary("z3java");
} catch (UnsatisfiedLinkError e1) {    
try {  System.loadLibrary("libz3");
       System.loadLibrary("libz3java");
} catch (UnsatisfiedLinkError e2) {...}
  \end{lstlisting}
  

\end{figure}

\item \relativePath: \newtext{This smell occurs when the library is loaded by using an absolute path to the library instead of the corresponding relative path. Using a relative path, the native library can be loaded and installed everywhere. However, the use of an absolute library path can introduce future bugs in case the library is no longer used. This may also impact the reusability of the code and its maintenance because the library can become inaccessible due to incorrect path. \texttt{System.loadLibrary("osxsecurity")} is an example of this design smell.} 

\item \cluster: \newtext{Too many native methods declared in a single class would decrease readability and maintainability of the code. This will increase the lines of code within that class and thus make the code review
process harder. Many studies discussed good practices about the number of methods to have within the same class, some examples are the rule of 30 introduced by Martin Lippert \cite{lippert2006refactoring}, or the \textit{7 plus/minus 2 rule} stating that a human mind can hold and comprehend from five to 9 objects. Most of the relevant measures are the coupling, cohesion, the single principle responsibility, and the separation of concerns. In this context, a bad practice would be to concentrate multi-language code in few classes, regardless of their role and responsibilities. This may result in a blob multi-language class with many methods and low cohesion. We present in Fig. \ref{Figure:clustring} an example that we extracted from ZMQJNI.\footnote{\url{https://github.com/zeromq/zmq-jni/blob/master/src/main/java/org/zeromq/jni/ZMQ.java}} In this example, native methods related to cryptographic operations are mixed in the same class as the methods used for network communication. This merging of concerns resulted in a blob multi-language class that contains 29 native declaration methods and 78 attributes. In the current study we are considering the case of having an excessive number of calls to native methods within the same class.} 

\begin{figure}
\center

\includegraphics[width=\linewidth]{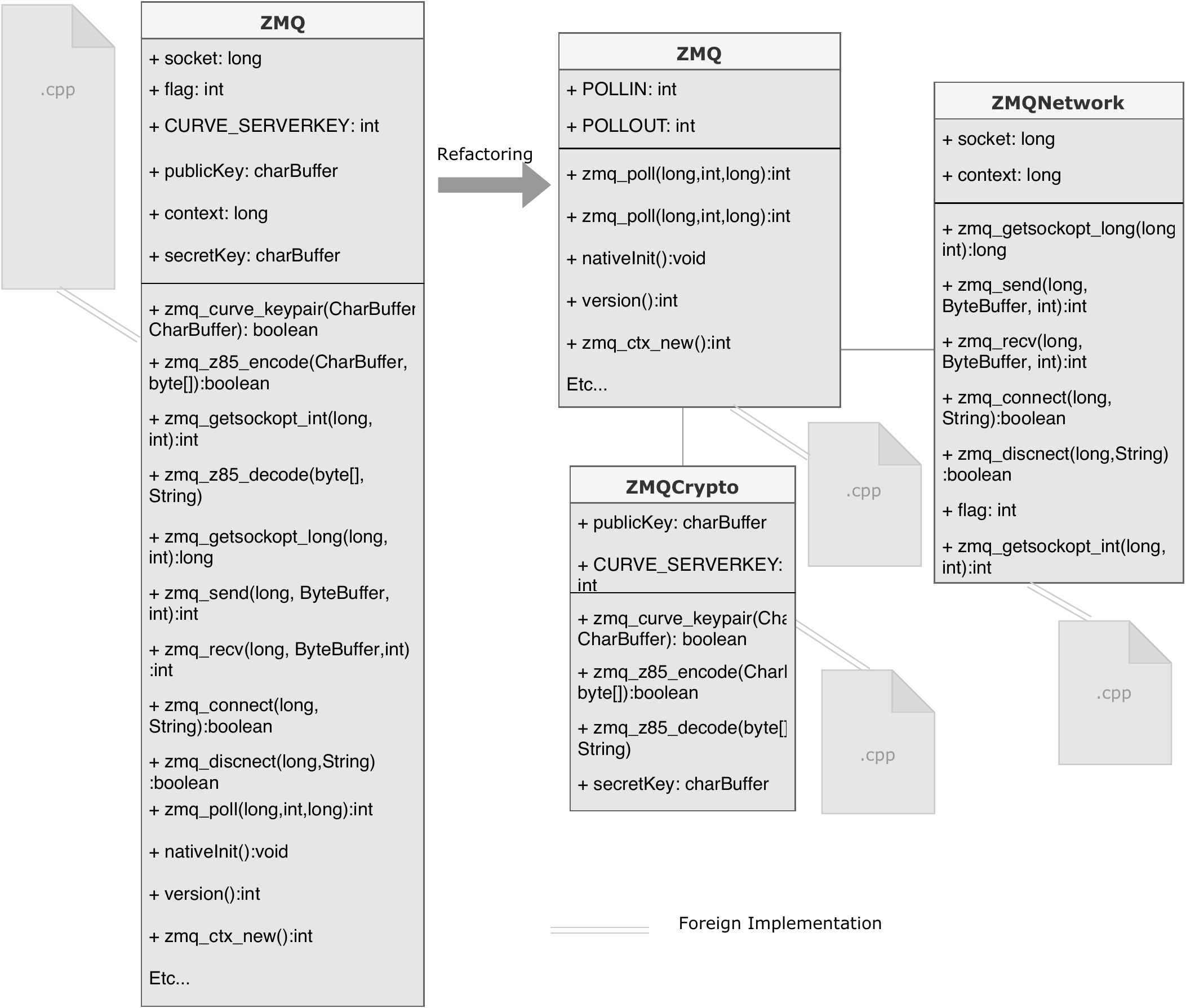}

\caption{Illustration of Design Smell - Too Much Clusterings}
\label{Figure:clustring}

\end{figure}

\item \scatter: Similar to too much clustering, when using multi-language code, developers and managers often have to decide on a trade-off between isolating or splitting the native code. Accessing this trade-off is estimated to improve the readability and maintainability of the systems \cite{abidi2019code}. 
This design smell occurs when classes are scarcely used in multi-language communication without satisfying both the coupling and the cohesion. In Figure \ref{Figure:scattering} extracted from a previous work \cite{abidi2019anti}, we have three classes with only two native methods declaration with duplicated methods. A possible good solution would be to reduce the number of native method declaration by removing the duplicated ones possibly by regrouping the common ones in the same class. This will also reduce the scattering of multi-language participants and concerns by keeping the multi-language code concentrated only in specific classes.


\begin{figure}
\center

\includegraphics[width=\linewidth]{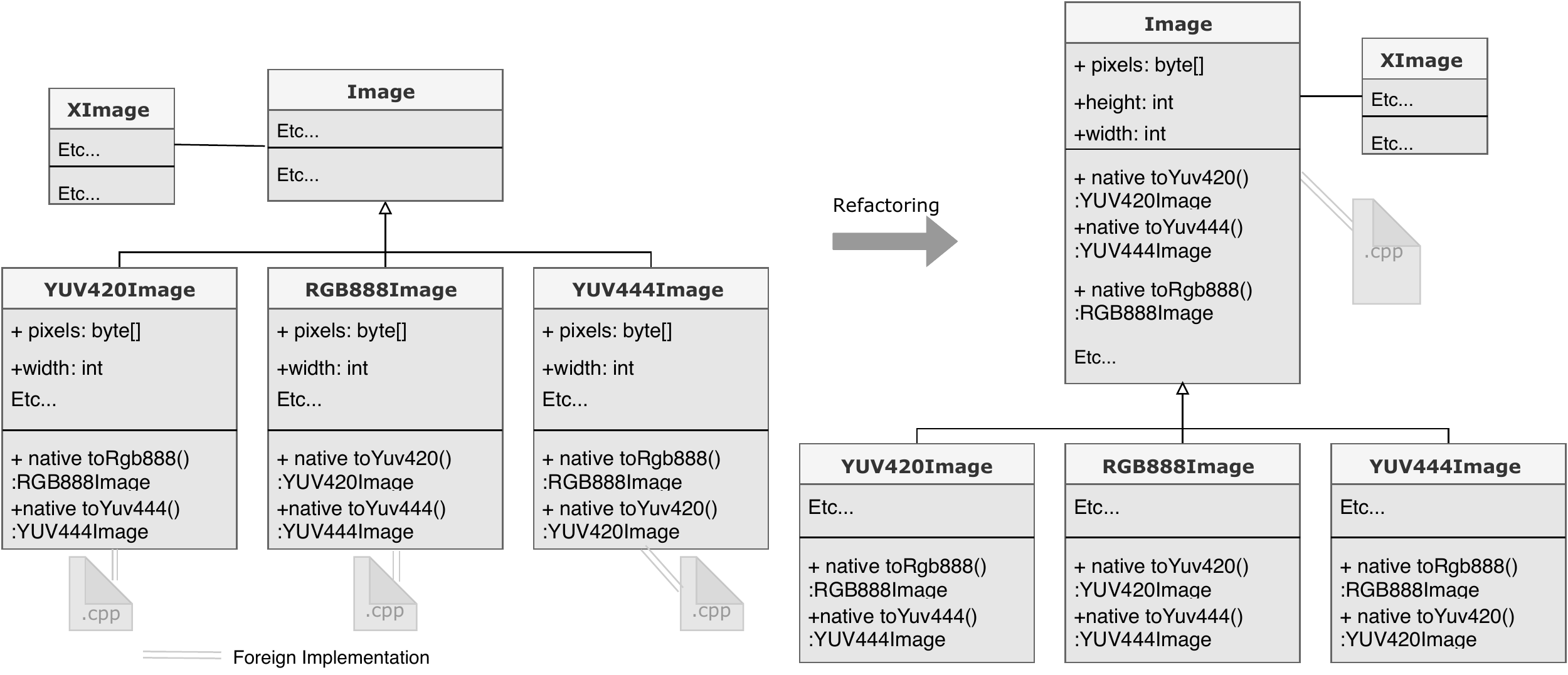}

\caption{Illustration of Design Smell - Too Much Scattering}
\label{Figure:scattering}

\end{figure}

\item \excessivecom: A wrong partitioning in components written in different programming languages leads to many calls in one way or the other. This may add complexity, increase the execution time, and may indicate a bad separation of concerns. Occurrences of this design smell could be observed in systems involving different layers or components. For example, the same object could be used and-or modified by multiple components. An excessive call of native code within the same class, could be illustrated whether by having too many native method calls in the same class or having the native method call within a large range loop. In \textit{Godot}, the function \texttt{process()} is called at each time delta. The time delta is a small period of time that the game does not process anything \ie the engine does other things than game logic out of this time range. The foreign function \texttt{process()} is called multiple times per second, in this case once per frame.\footnote{\url{https://github.com/godotengine/godot-demo-projects/blob/master/2d/pong/paddle.gd}}

\item \localReference: For any object returned by a JNI function, a local reference is created. JNI specification allows a maximum of 16 local references for each method. Developers should pay attention on the number of references created and always deleted the local references once not needed using \texttt{JNIDeleteLocalRef()}. 
Listing \ref{fig:Localref} illustrates an example of this design smell in which local references are created without deleting them.


\begin{figure}[ht]
\center

\lstset{backgroundcolor = \color{gray!4},language=C,basicstyle=\small\ttfamily  , stepnumber=1, showspaces=false, showstringspaces=false,breaklines=true}
\begin{lstlisting} [language=Java, caption={Design Smell - Local References Abuse}, label={fig:Localref}]
/* C++ */
for (i=0; i < count; i++) {
jobject element = (*env)->GetObjectArrayElement(env, array, i);
if((*env)->ExceptionOccurred(env)) { break;}
  \end{lstlisting}
  

\end{figure}

\item \memoryMagement: \newtext{Data types differ between Java and C/C++. When using JNI, a mapping is performed between Java data types and data types used in the native code.\footnote{\url{https://www.developer.com/java/data/jni-data-type-mapping-to-cc.html}} JNI handles Java objects, classes, and strings as reference types. JVM offers a set of predefined methods that could be used to access fields, methods, and convert types from Java to the native code. Those methods return pointers that will be used by the native code to perform the calculation. The same goes for reference types, the predefined methods used allow to either return a pointer to the actual elements at runtime or to allocate some memory and make a copy of that element. Thus, due to the differences of types between Java and C/C++, the memory will be allocated to perform respective type mapping between those programming languages. Memory leaks will occur if the developer forgets to take care of releasing such reference types. Listing \ref{fig:Releasestring} presents an example in which the memory was not released using \texttt{ReleaseString} or \texttt{ReleaseStringUTF}.}

\begin{figure}[ht]
\center

\lstset{backgroundcolor = \color{gray!4},language=C,basicstyle=\small\ttfamily, showspaces=false, showstringspaces=false,breaklines=true, morekeywords={ReleaseStringUTFChars}}
\begin{lstlisting}  [language=C, caption={Refactoring - Memory Management Mismatch}, label={fig:Releasestring}]
/* C++ */
str = env->GetStringUTFChars(javaString, &isCopy);
\end{lstlisting}


\end{figure}


\item \cachingObjects: To access Java objects' fields from native code through JNI and invoke their methods, the native code must perform calls to predefined functions \ie \textit{FindClass(), GetFieldId(), GetMethodId(), and GetStaticMethodId()}. For a given class, IDs returned by \textit{GetFieldId(), GetMethodId(), and GetStaticMethodId()} remain the same during the lifetime of the JVM process. The call of these methods is quite expensive as it can require significant work in the JVM. In such situation, it is recommended for a given class to look up the IDs once and then reuse them. In the same context, looking up class objects can be expensive, a good practice is to globally cache commonly used classes, field IDs, and method IDs. Listing \ref{fig:cachingobj} provides an example of occurrences of this design smell that does not use cached field IDs. 

\begin{figure}[ht]
\center

\lstset{backgroundcolor = \color{gray!4},language=C,basicstyle=\small\ttfamily  , showspaces=false, showstringspaces=false,breaklines=true}
\begin{lstlisting}[language=C, caption={Design Smell - Not Caching Objects' Elements}, label={fig:cachingobj}]
/* C++ */
int sumVal (JNIEnv* env,jobject obj,jobject allVal){
   jclass cls=(*env)->GetObjectClass(env,allVal);
   jfieldID a=(*env)->GetFieldID(env,cls,"a","I");
   jfieldID b=(*env)->GetFieldID(env,cls,"b","I");
   jfieldID c=(*env)->GetFieldID(env,cls,"c","I");
   jint aval=(*env)->GetIntField(env,allVal,a);
   jint bval=(*env)->GetIntField(env,allVal,b);
   jint cval=(*env)->GetIntField(env,allVal,c);
   return aval + bval + cval;}
\end{lstlisting}

\end{figure}

\item \excessiveObject: \newtext{Accessing field's elements by passing the whole object is a common practice in object oriented programming. However, in the context of JNI, since the Object type does not exist in C programs, passing excessive objects could lead to extra overhead to properly perform the type conversion. Indeed, this design smells occurs when developers pass a whole object as an argument, although only some of its fields were needed, and it would have been better for the system performance to pass only those fields except the purpose to pass the object to the native side was to set its elements by the native code using \textit{SetxField} methods, with x the type of the field. Indeed, in the context of object-oriented programming, a good solution would be to pass the object offering a better encapsulation, however, in the context of JNI, the native code must reach back into the JVM through many calls to get the value of each field adding extra overhead. This also increases the lines of code which may impact the readability of the code \cite{abidi2019code}. Listing \ref{fig:ExcessiveObj} presents an example smell of passing excessive objects. The refactored solution of this smell would be to pass the class' fields as a method parameters as described in our published catalog \cite{abidi2019code}.}

\begin{figure}[ht]
\center

\lstset{backgroundcolor = \color{gray!4},language=Java,basicstyle=\small\ttfamily, showspaces=false, showstringspaces=false,breaklines=true}
\begin{lstlisting} [language=Java, caption={Design Smell - Passing Excessive Objects}, label={fig:ExcessiveObj}]
/* C++ */
int sumValues (JNIEnv* env,jobject obj,jobject allVal)
{ jint avalue= (*env)->GetIntField(env,allVal,a);
  jint bvalue= (*env)->GetIntField(env,allVal,b);
  jint cvalue= (*env)->GetIntField(env,allVal,c);
  return avalue + bvalue + cvalue;}
\end{lstlisting}


\end{figure}


\item \implementation: This appears when a method is declared in the host language (Java in our case) and implemented in the foreign language (C or C++). However, this method is never called from the host language. This could be a consequence of migration or refactoring in which developers opted for keeping those methods to not break any related features. 

\item \declaration: \newtext{Similar to \implementation, this design smell occurs when a method is declared in the host language but is never implemented in the native code. This smell and the previous one are quite similar. However, they differ in the implementation part, while for the smell \implementation, the method is implemented but never called, in case of the smell \declaration, the unused method is not implemented and never called in the foreign language. Such methods could remain in the system for a long period of time without being removed because having them will not introduce any bug when executing the program but they may negatively impact the maintenance activities and effort needed when maintaining those classes.} 

\item \unusedParameters: Long list of parameters make methods hard to understand \cite{fontana2012automatic}. It could also be a sign that the method is doing too much or that some of the parameters are no longer used. In the context of multi-language programming, some parameters may be present in the method signature however they are no longer used in the other programming language. Since multi-language systems usually involve developers from different teams, those developers often prefer not to remove such parameters because they may not be sure if the parameters are used by other components. Listing \ref{fig:UnecessayParam} presents an illustration of this design smell where the parameter acceleration is used in the native method signature. However, it is not used in the implemented function. 

\begin{figure}[ht]
\center

\lstset{backgroundcolor = \color{gray!4},language=Java,basicstyle=\small\ttfamily, showspaces=false, showstringspaces=false,breaklines=true}
\begin{lstlisting} [language=Java, caption={Design Smell - Unnecessary Parameters}, label={fig:UnecessayParam}]
/* C++ */
JNIEXPORT jfloat JNICALL Java_jni_distance
  (JNIEnv *env, jobject thisObject,
  jfloat time, jfloat speed,
  jfloat acceleration) {
    return time * speed;}
\end{lstlisting}


\end{figure}

 \end{enumerate}

\section{Study Design}
\label{sec:Study}

\begin{figure*}[ht]
\center
\includegraphics[width=\linewidth, height=3.3cm]{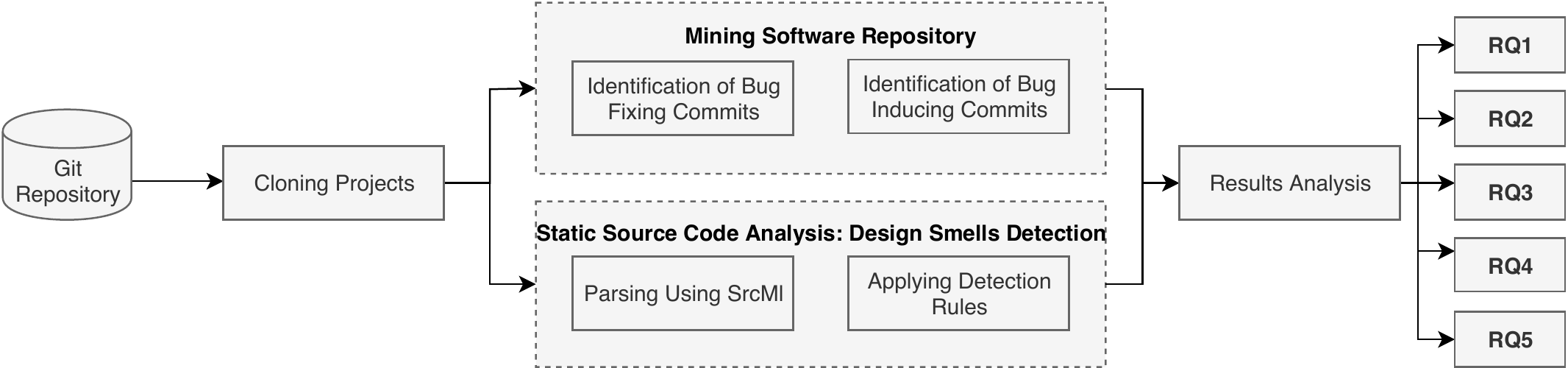}
\caption{Schematic Diagram of the Study}
\label{fig:netho}
\end{figure*}

In this section, we present the methodology we followed to conduct this study. Figure \ref{fig:netho}
provides an overview of our methodology.

\subsection{Setting Objectives of the Study}

We started by setting the objective of our study. Our objective is to investigate the prevalence of multi-language design smells in the context of JNI systems and the relation between those smells with software fault-proneness. We also aim to investigate what kind of activities once performed in smelly files are more likely to introduce bugs.
The quality focus in this study is the occurrence of bugs due to the presence of design smells in JNI systems. 
The perspective is that of researchers, interested in the quality of JNI systems, and who want to get evidence on the impact of design smells on the software fault-proneness. 
Also, these results can be of interest to professional developers performing maintenance and evolution activities on JNI projects and who need to take into account and forecast their effort, since like for mono-language projects, the presence of fault-prone files is likely to increase the maintenance effort and cost. 
These results are also of interest to testers since they need to know which files are more important to test. 
Finally, they can be of interest to quality assurance teams or managers who could use design smells detection techniques to assess the fault-proneness of in-house or to-be-acquired source code, to better quantify the cost-of-ownership of JNI systems. 

We defined our research questions as follows:
\begin{itemize}
\item[RQ1:] \textbf{\RQOne} \newline 
\newtext{Several articles in the literature discussed the prevalence, detection, and evolution of design smells in the context of mono-language systems \cite{khomh2012exploratory,moha2009decor}. Occurrences of design smells may hinder the evolution of a system by making it hard for developers to maintain the system. The detection of smells can substantially reduce the cost of maintenance and development activities. However, most of those research are focusing on mono-language systems. Thus, we decided to fill this gap in the literature and investigate the frequency of design smells in the context of multi-language systems. This research question is preliminary to the remaining questions.
It aims to examine the frequency and distribution of multi-language design smells in the selected projects and their evolution over the releases of the project.
We defined the following null hypothesis: \textit{H$_{1}$: there are no occurrences of the multi-language design smells studied in the literature in the selected projects.}}
  
\item[RQ2:] \textbf{\RQTwo} \newline   
\newtext{
Given that multi-language design smells are prevalent in the studied systems, it is important to know the distribution and evolution of the different types of smells for a better understanding of the implication of their presence for maintenance activities. 
Developers are likely to benefit from knowing the dominating smells to treat them in priority and avoid introducing such occurrences. Consequently, in this research question, we aim to study whether some specific types of design smells are more prevalent than others. We are also interested in the evolution of each type of smells over the releases of the project.
 We aim to test the following null hypothesis: \textit{H$_{2}$: The proportion of files containing a specific type of design smell does not significantly differ from the proportion of files containing other kinds of design smells.}}

\item[RQ3:] \textbf{\RQThree} \newline 
\newtext{
Prior works reported that classes containing design smells in mono-language systems are more prone to faults than other classes \cite{khomh2009exploratory,saboury2017empirical}. Due to components written in different languages, multi-language systems may have more complexities in architecture and inter-component interactions. 
Given the known impacts of design smells on mono-language systems, it is thus important to investigate the impacts of multi-language design smells on the corresponding software systems. To examine this, we aim to investigate whether source files containing multi-language design smells are more likely to experience faults than files without smells. 
We investigate whether files with multi-language design smells are more fault-prone than others by testing the null hypothesis: \textit{H$_{3}$: The proportion of files experiencing at least one bug does not significantly differ between files with design smells and files without.}}

\item[RQ4:] \textbf{\RQFour} \newline 
\newtext{
During maintenance and quality assurance activities, developers are interested in identifying parts of the code that should be tested and–or refactored in priority. Hence, we are interested in identifying design smells that are more fault-prone than others. Thus, we defined the null hypothesis. 
\textit{H$_{4}$: There is no significant difference between the impacts of different kinds of multi-language design smells on the fault-proneness of files containing those smells.}}

\item[RQ5:] \textbf{\RQFive} \newline \newtext{During the maintenance of a project, having knowledge of possible risky activities could help developers and managers to reduce the risk of bugs. They could benefit from that knowledge to capture activities that should be performed with caution in smelly files.
Hence, we are interested in identifying what kinds of activities once performed in smelly files are likely to introduce bugs. Capturing such information could provide insights about what kind of activities could increase the risk of bugs in smelly files.}
\end{itemize}

\begin{table*}
\centering
\caption{Research Objectives and Research Questions}
\label{table:Rq}
{\renewcommand{\arraystretch}{1.3}
\begin{tabular}{p{8cm} p{5cm}}
 \hline
 \rowcolor{gray!35}
 \textbf{Research Objectives} &\textbf{Methodology}\\
 \hline
\textbf{Objective 1:} Detect multi-language design smells & Detection approach (case study of JNI systems)\\
\rowcolor{gray!15}
\textbf{Objective 2:} Investigate the prevalence of multi-language design smells & RQ1 and RQ2\\ 
\textbf{Objective 3:} Study the relationship between multi-language design smells and fault-proneness	& RQ3 and RQ4\\
\rowcolor{gray!15}
\textbf{Objective 4:} Identifying fault-inducing activities & RQ5\\
 \hline
\end{tabular}
}
\end{table*}

\subsection{Data Collection}

In order to address our research questions, we selected 
nine open source projects hosted on GitHub. We decided to analyze those nine systems because they are well maintained, and highly active. 
Another criteria for the selection was that those systems have different size and belong to different domains. 
\newtext{They also have the characteristic of being developed with more than one programming language.
While those systems contain different combinations of programming languages, for this study, we are analyzing the occurrences of design smells for only Java and C/C++ code. For each of the nine selected subject systems, we selected a minimum of 10 releases. For projects with relatively frequent releases and comparatively a small volume of changes per release, we extended our analysis to a few extra releases to cover a longer evolution period for our analysis.
Tables \ref{systemoverview} and \ref{tab:oversyst} summarise the characteristics of the subject systems and releases. 
We also provide the percentage of the Java and C/C++ code in the studied projects in Table \ref{systemoverview}.}  

Among the nine selected systems, \vlc~is a highly portable multimedia player for various audio and video formats. 
\rocksdb~is developed and maintained by Facebook, it presents a persistent key-value store for fast storage. 
It can also be the foundation for a client-server database. \realm~is a mobile database that runs directly inside phones and tablets. 
\conscrypt~is developed and maintained by Google, it is a Java Security Provider (JSP) that implements parts of the Java Cryptography Extension (JCE) and Java Secure Socket Extension (JSSE). 
\javasmt~is a common API layer for accessing various Satisfiability Modulo Theories (SMT) solvers. 
\pljava~is a free module that brings Java Stored Procedures, Triggers, and Functions to the PostgreSQL backend via the standard JDBC interface. 
\javacpp~provides efficient access to native C++ inside Java, not unlike the way some C/C++ compilers interact with assembly language. 
\zstd~present a binding for Zstd native library developed and maintained by Facebook that provides fast and high compression lossless algorithms for Android, Java, and all JVM languages. 
\jpype~is a Python module to provide full access to Java from within Python.

\begin{table*}[t]
 \centering\scriptsize
 \caption{Overview of the Studied Systems}
 \label{systemoverview}
{\renewcommand{\arraystretch}{1.15}
\begin{tabular}{p{1.7cm}lrrrrrr}
\hline
  \rowcolor{gray!35}
\textbf{Systems} & \textbf{Domain} & \textbf{\#Releases} & 
\textbf{\#Commits} &  \textbf{\#Issues}
&\textbf{LOC}& 
\textbf{Java} & \textbf{C/C++} 
\\ \hline
\rocksdb\footnote{\url{https://github.com/facebook/rocksdb/}} & Facebook Database    & 189  
   & 8375  & 1748  & 487853      & 11\%    & 83.1\% \\
\rowcolor{gray!15}
\vlc\footnote{\url{https://github.com/videolan/vlc-android}}    & Media Player and Database   & 176
   & 12697   & 1091 & 125037       & 10.1\%   & 6.7\%  \\
\realm\footnote{\url{https://github.com/realm/realm-java}}    & Mobile Database  & 169     & 8244 & 3886 & 171705   & 82\%  & 8.1\% \\
\rowcolor{gray!15}
\conscrypt\footnote{\url{https://github.com/google/conscrypt}}  & Cryptography (Google)   & 32 & 3874 & 186 & 91765  & 85.3\%  & 14\%   \\
\pljava\footnote{\url{https://github.com/tada/pljava}}  & Database   & 27   & 1236 & 123 & 71910 & 67\%   & 29.7\%  \\
\rowcolor{gray!15}
\javacpp\footnote{\url{https://github.com/bytedeco/javacpp}} & Compiler  & 34  & 658 & 269 & 28713   & 98\% & 0.6\% \\
\zstd\footnote{\url{https://github.com/luben/zstd-jni}}  & Data Compression (Facebook)  & 36 & 423 & 78 & 72824  & 4.3\% & 92.1\%   \\
\rowcolor{gray!15}
\jpype\footnote{\url{https://github.com/jpype-project/jpype}}  & Cross Language Bridge  & 14  & 895 & 305 & 53826  & 7.8\%  & 58\%  \\
\javasmt\footnote{\url{https://github.com/sosy-lab/java-smt}}  & Computation  & 22  & 1822 & 146 & 42049   & 88\% & 4.6\%                               \\\hline                 
\end{tabular}
}
\end{table*}

\begin{table}[t]
\centering\footnotesize
\caption{\label{tab:oversyst} Analyzed Releases in Each Project}
{\renewcommand{\arraystretch}{1.15}
\begin{tabular}{lclc}
	\hline
    \rowcolor{gray!35}
	\textbf{Systems} &  \textbf{\#Releases Analyzed}  & \textbf{Releases} & \textbf{Analysis Periods}\\\hline

\rocksdb & 10  & 5.0.2 - latest release  & 	2017-18-01 - 	2019-14-08 \\
\rowcolor{gray!15}
\vlc & 10  &  3.0.0 - latest release  & 	2018-08-02 - 	2019-13-09\\
\realm & 10  &  0.90.0 - 5.15.0  & 	2016-03-05 - 	2019-04-09\\
\rowcolor{gray!15}
\conscrypt & 11   & 1.0.0.RC11 - 2.3.0  & 	2017-25-09 - 	2019-25-09\\
\pljava & 12 & 1\_2\_0 - latest release   & 	2015-20-11  -  	2019-19-03\\
\rowcolor{gray!15}
\javacpp & 13 & 0.5 - 1.5.1-1  & 	2013-07-04 - 	2019-05-09\\
\zstd &  11 & 0.4.4 - latest release  & 	2015-17-12 - 	2019-19-08\\
\rowcolor{gray!15}
\jpype & 11  & 0.5.4.5 - latest release  & 	2013-25-08 - 	2019-13-09\\
\javasmt & 10  & 0.1 - 3.0.0  & 	2015-27-11 - 	2019-30-08

 \\ \hline 
\end{tabular}
}	
\end{table}

\subsection{Data Extraction}
To answer our research questions, we first have to mine the repositories of the nine selected systems to extract information about the occurrences of smells existing in each file and also the bugs reported for those systems.

\subsubsection{\textbf{Detection of Design Smells}} \label{sec:detection}

\paragraph{\textbf{Detection Approach:}} Because no tools are available to detect design smells in multi-language systems, we extended the Ptidej Tool suite\footnote{\url{http://www.ptidej.net/tools/}} by building a new detection approach for multi-language smells. The approach is closely inspired by DECOR and PtiDej tool Suite \cite{moha2009decor}. 
We used srcML\footnote{\url{https://www.srcml.org/}}, a parsing tool that converts source code into srcML, which is an XML format representation. The srcML representation of source code adds syntactic information as XML elements into the source code text. Listing \ref{lst:srcml} presents the srcML representation of the code snippet presented in \ref{lst:java}. 
The main advantage of srcML, is that it supports different programming languages, and generates a single XML file for the supported programming languages. 
For now, our approach includes only Java, C, and C++, however, it could be extended to include other programming languages in the future. 
SrcML provides a wide variety of predefined functions that could be easily used through the XPath to implement specific tasks. XPath is frequently used to navigate through XML nodes, elements, and attributes. In our case, it is used to navigate through srcML elements generated as an AST of a given project.
The ability to address source code using XPath has been applied to several applications \cite{gottlob2005efficient}.

\newtext{Our detection approach reports smell detection results for a selected system in a CSV file. The report provides detailed information for each smells detected such as smell type, file location, class name, method name, parameters (if applicable). We then used a python script to post-process the results to create a summary file. The summary results contain the total number of occurrences of each type of smell in a specific file or class in a specific release of the selected system. Two members of our research team manually 
validated the results of smell detection for five systems.} 


\paragraph{\textbf{Detection Rules:}} \ 
The detection approach is based on a set of rules defined from the documentation of the design smells. Those rules were validated by the pattern community during the Writers' workshop to document and validate the smells. For example, for the design smell \textit{Local Reference Abuse}, we considered cases where more than 16 references are created but not deleted with the \textit{DeleteLocalRef} function. 
The threshold 16 was extracted from developers blogs discussing best practices and the Java Native Interface specification \cite{Liang:1999}.\footnote{\url{https://www.cnblogs.com/cbscan/articles/4733508.html}}$^{,\thinspace}$\footnote{\url{https://docs.oracle.com/javase/7/docs/technotes/guides/jni/spec/functions.html\#global_local}} We present in the following two examples of rules as well as the thresholds used to define them, and their detection process. All the other rules are available in Appendix \ref{appendixA}.

\begin{figure}[ht]
\center

\lstset{backgroundcolor = \color{gray!4},language=Java,basicstyle=\small\ttfamily, showspaces=false, showstringspaces=false,breaklines=true, label={lst:java}}

\begin{lstlisting}[language=Java, caption={Example of Java Code}]
public class HelloWorld {

    public static void main(String[] args) {
        // Prints "Hello World!" to stdout
        System.out.println("Hello World!");
    }
}
\end{lstlisting}
\vspace{-0.5cm}
\end{figure}

\begin{figure}[h]
\center

\lstset{backgroundcolor = \color{gray!4},language=Java,basicstyle=\small\ttfamily, showspaces=false, showstringspaces=false,breaklines=true, label={lst:srcml}}
\begin{lstlisting}[language=XML, caption={Example of Java Code Converted to SrcML}]
<?xml version="1.0" encoding="UTF-8" standalone="yes"?>
<unit xmlns="http://www.srcML.org/srcML/src" revision="0.9.5" language="Java" filename="HelloWorld.java"><class><specifier>public</specifier> class <name>HelloWorld</name> <block>{

    <function><specifier>public</specifier> <specifier>static</specifier> <type><name>void</name></type> <name>main</name><parameter_list>(<parameter><decl><type><name><name>String
     </name><index>[]</index></name></type> <name>args</name></decl></parameter>)</parameter_list> <block>{
        <comment type="line">// Prints "Hello World!" to stdout</comment>
        <expr_stmt><expr><call><name><name>System</name><operator>.</operator><name>out
        </name><operator>.</operator><name>println</name></name><argument_list>
        (<argument><expr><literal type="string">"Hello World!"</literal>
        </expr></argument>)</argument_list></call></expr>;</expr_stmt>
    }</block></function>

}</block></class></unit>
\end{lstlisting}
\end{figure}



\begin{enumerate}
\item \textbf{Rule for the smell \exception} 
 \newtext{ 
 \begin{multline} 
  (f(y)~|~ f \in \{GetObjectClass, FindClass, GetFieldID, GetStaticFieldID,\\ GetMethodID, GetStaticMethodID\})\\ \nonumber
  \textbf{\textit{AND}}~
  (isErrrorChecked(f(y))=False ~ \textbf{\textit{OR}}~ ExceptionBlock(f(y))=False) 
\end{multline}
 }
  
\newtext{Our detection rule for the smell \exception~ is based on the existence of call to specific JNI methods requiring explicit management of the exception flow. 
The JNI methods (\eg \textit{FindClass}) listed in the rule should have a control flow verification. 
The parameter \textit{y} represents the Java object/class that is passed through a native call for a purpose of usage by the C/C++ side. 
Here, \textit{isExceptionChecked} allows to verify that there is an error condition verification for those specific JNI methods, while \textit{ExceptionBlock} checks if there is an exception block implemented. 
This could be implemented using \texttt{Throw()} or \texttt{ThrowNew()} or a return statement that exists in the method in case of errors.} 

\newtext{If we recheck Listing \ref{fig:exceptionOccured} in Section \ref{sec:Background}, the code illustrated in this example satisfies the rule of using predefined methods to access classes and field Ids. Another condition is that those methods are not followed by an explicit exception block. Thus, this example will be captured by our approach as an occurrence of the design smell \exception.} 
\vspace{0.2cm}   
\item \textbf{Rule for the smell \localReference}
   \newtext{
   \begin{multline}
   (NbLocalReference(f_1(y)) > MaxLocalReferenceThreshold )~ \textbf{AND} \\ (f_1(y)~ |~ f_1  \in  \{ GetObjectArrayElement, GetObjectArrayElement, NewLocalRef, AllocObject, \\ NewObject, NewObjectA, NewObjectV, NewDirectByteBuffer, \\ ToReflectedMethod, ToReflectedField \})~ \textbf{AND} \\ \nonumber (\nexists~ f_2(y)~ |~ f_2  \in  \{ DeleteLocalRef, EnsureLocalCapacity \}) 
   \end{multline}
   }
   
\newtext{The smell \localReference~ is introduced when the total number of local references created inside a called method exceeds the defined threshold and without any call to method \texttt{DeleteLocalRef} to free the local references or a call to method \texttt{EnsureLocalCapacity} to inform the JVM that a larger number of local references is needed. }

\newtext{In the same vein, if we recall the example provided in Listing \ref{fig:Localref}, in which a local reference is created to retrieve an array element. This is implemented inside a loop (\textit{for}). Thus, if the total number for the count is more than 16, this indicates that we are exceeding the authorized number of local references. In this situation, our approach will capture the method exceeding the authorized number of local references and will then check for any possible usage of functions to release the memory. Since this example does not provide any functions to release the memory, this will be detected by our approach as an occurrence of the design smell \localReference.}

\end{enumerate}

\textbf{Validation Approach:} 


To assess the recall and precision of our detection approach, we evaluated the results of our detection approach at the first level by creating dedicated unit tests for the detector of each type of smell to confirm that the approach is detecting the smells introduced in our pilot project. 
We relied on six open source projects used in previous works \cite{abidi2019code,abidi2019anti} on multi-language design smells. For each of the systems, we manually identified occurrences of the studied design smells. 
Two of the authors independently identified occurrences of the design smells in JNI open source projects, and resolved disagreements through discussions with the whole research team. 
Using the ground truth based on the definition of the smell and the detection results, we computed \textit{precision} and \textit{recall} as presented in Table \ref{tab:systvalidation} to evaluate our smell detection approach.
Precision computes the number of true smells contained in the results of the detection tool, while recall computes the fraction of true smells 
that are successfully retrieved by the tool.
From the six selected systems, we obtained a precision between 88\% and 99\%. and a recall between 74\% and 90\%. We calculate precision and recall based on the following equations (\ref{eq1}) and (\ref{eq2}) respectively:

\begin{equation} \label{eq1}
     Precision = \frac{\left \{ existing~true~smells \right \}\bigcap \left \{ detected~smells \right \}} { \left \{  detected~smells\right \}}
\end{equation}

\begin{equation} \label{eq2}
     Recall = \frac{\left \{ existing~true~smells \right \}\bigcap \left \{ detected~smells \right \}} { \left \{  existing~true~smells\right \}}
\end{equation}

\subsubsection{\textbf{Detection of Fault-inducing Commits}} \label{sec:bug}

Our studied systems use Github as the issue tracker. 
We used Github APIs and PyDriller to mine the software repositories and get the list of all the commit logs and resolved issues for the systems \cite{spadini2018pydriller}. 
PyDriller 
provides a set of APIs to extract information from Git repositories. These include important historical information regarding commits, developers, and modifications. 
 PyDriller is very convenient for mining software repositories to analyze changes or bugs. It rely on the SZZ algorithm \cite{Sliwerski:2005} to detect changes that introduce faults. We used PyDriller because this approach was not only evaluated regarding existing tools but also with experiments involving developers \cite{spadini2018pydriller}. 
We started by retrieving all the information related to the projects. 
\newtext{We analyzed all commit messages to identify the fault-fixing commits. We used a set of error related keywords to identify commits related to fault-fixing using a heuristic similar to that presented in the study by Mockus and Votta \cite{mockus2000identifying}. Our list of keywords include ``fix'', ``crash'', ``resolves'', ``regression'', ``fall back'', ``assertion'', ``coverity'', ``reproducible'', ``stack-wanted'', ``steps-wanted'', ``testcase'', ``fail'', ``npe'', ``except'', ``broken'', ``bug'', ``differential testing'', ``error'', ``addresssanitizer'', ``hang'', ``permaorange'', ``random orange'', ``intermittent'', ``steps to reproduce'', ``assertion'', ``leak'', ``stack trace'', ``heap overflow'', ``freez'', ``str:'', ``problem '',  ``overflow'', ``avoid``, `` issue'', ``workaround'',  ``break'', and ``stop''.} 

To retrieve fault-inducing commits, given a commit, 
PyDriller returns the set of commits that previously modified the lines from the files included in the given commit.
It applies the SZZ algorithm to find the commit when the bug was initially introduced as used in some earlier studies \cite{palomba2014they,spadini2018testing,palomba2017scent}. To locate the fault-inducing commits, PyDriller algorithm works as follows: for every file in the commit, it obtains the diff between the files, then obtains the list of all deleted lines. It then blames the file to obtain the commits where the deleted lines were changed. We tagged fault-inducing commits as \textit{buggy}. We used this tag later to distinguish between files containing bugs and files without.

\newtext{
Since Pydriller's SZZ implementation was not previously evaluated, we 
manually examined the bug inducing commits retrieved by 
Pydriller from two of our studied projects, \pljava~ and \zstd. We performed this manual analysis in two steps. 
First, we executed an existing implementation of the SZZ algorithm available on GithHub\footnote{https://github.com/saheel1115/szz} on \pljava~ and \zstd. We compared its reported results 
with the results obtained from Pydriller. For each bug fixing commit, we manually verified if the related bug inducing commit reported by Pydriller matches with the one reported by SZZ. For that, we used two labels (True or False) to distinguish between the bug inducing commits that match with those retrieved by SZZ and those that do not match.
Next, one of the authors manually 
verified if the changes in the bug inducing commits reported by Pydriller were indeed related to the changes performed in the corresponding bug fixing commits. 
We also analyzed the commit messages. 
We labeled each of the bug inducing commits with three tags (True, False, and Unclear). 
We used the tag True in situations in which we were convinced that the change performed in the bug fixing was indeed related to the changes applied in the bug inducing. We assigned False in situations in which it was evident that the changes are not related, and Unclear, in situations in which it was not completely evident to assign a True or False tag. 
We analyzed for \pljava~ and \zstd~ respectively a total of 113 and 96 bug-fixing commits. We performed a cleaning process on those commits and removed the commits related to typos fixing and merge commits. We kept in our validation bug-fixing commits with their corresponding bug-inducing commits. Our final dataset results on 61 bug-fixing commits for \pljava~ and 66 bug-fixing commits for \zstd. 
From our manual validation of fault-inducing commits reported by Pydriller for \pljava~ and \zstd, we found respectively precision values of 78.94\% and 70.83\%. Those values are computed considering only the True and False tags for Java and C/C++ files resulting from our manual validation. From the comparison between Pydriller'szz and the recent implementation of szz, we found for \pljava~ and \zstd, respectively precision values of 85\% and 80\%.  We did not include in our validation the recall because in our study we are considering only JNI code. However, szz is considering the whole project in general without considering multi-language interactions.
So, those results may not generalize to the whole system.
However, the results of our manual validation do not directly contribute to any of our empirical findings, and we did this validation as a complementary step to reduce the threats to validity of our study. 
We also analyzed changes related to multi-language programming. Indeed, in many situations, the Java and C/C++ code are changed within the same commit. This was helpful to validate the bug inducing commits involving Java and C/C++ code. 
}

\begin{table}[t]
\centering\footnotesize
\caption{\label{tab:systvalidation}Validation of the Smell Detection Approach} 
{\renewcommand{\arraystretch}{1.15}
\begin{tabular}{lccccc}
	\hline
    \rowcolor{gray!35}
	\textbf{Systems} &  \textbf{True Positive}  & \textbf{False Positive} & \textbf{False Negative} & \textbf{Recall} & \textbf{Precision}\\\hline

openj9 & 3293  & 137 & 250  & 	93\% & 96\% \\
\rowcolor{gray!15}
rocksdb & 922  & 50 & 136  & 	87\% & 95\% \\
conscrypt & 556  & 29 & 133  & 	80\% & 95\% \\
\rowcolor{gray!15}
pilot project & 32  & 0 & 0  & 	100\% & 100\% \\
pljava  & 511  & 5 & 53  & 	90\% & 99\% \\
\rowcolor{gray!15}
jna & 375  & 50 & 127  & 	74\% & 88\% \\
jmonkey & 2210  & 142 & 185  & 	92\% & 94\%
 \\ \hline
\end{tabular}
}	
\end{table}

\subsection{Analysis Method} 

We present in the following the analysis performed to answer our research questions.

\subsubsection{\textbf{Analyzing the prevalence of design smells}} \label{sec:analysis} We investigate the presence of 15 different kinds of design smells. Each variable s$_{i,j,k}$ reflects the number of times a file $i$ has a smell $j$ in a specific release r$_k$. 

\newtext{For RQ1, since we are interested to investigate the prevalence of multi-language design smells, we aggregate these variables into a Boolean variable s$_{i,k}$ to indicate whether a file $i$ has at least any kind of smells in release $r_k$. 
 We calculate the percentage of files affected by at least one of the studied design smells, s$_{j}$. We use our detection approach to detect occurrences of multi-language design smells following the methodology described earlier. For each file, we compute the value of a variable \textit{$Smelly_{i,r}$} which reflects if the file $i$ has a least one type of smell in a specific release $r$. This variable takes 1 if the file contains at least one design smell in a specific release $r$, and 0 otherwise. Similarly, we also compute the value of variable \textit{$Native_{i,r}$} which takes 1 if the file  $i$ of a specific release $r$ is native and 0 if not. Since our tool is focusing on the combination of Java and C/C++, we compute for each release the percentage of files participating in at least one JNI smells out of the total number of JNI files (files involved in Java and C/C++).} 
 
\newtext{For RQ2, we investigate whether a specific type of design smells is more prevalent in the studied systems than other types of design smells.
For that, we calculate for each system the percentage of files affected by each type of the studied smells $j$. For each file $i$ and for each release $r$, we defined a flag \textit{Smelly}$_{i,j,r}$ which takes the value 1 if the release $r$ of the file $i$ contains the design smell type $j$ and 0 if it does not contain that specific smell. Based on this flag, we compute for each release the number of files participating in that specific smell. We also calculate the percentage of smelly files containing each type of smell.  Note that the same file may contain more than one smell. We investigate the presence of 15 different kinds of smells. We also compute the metric $s_{i,j,k}$ which reflects the number of occurrences of smells of type $j$ in a file $i$ in a specific release $r_k$.
} 



\subsubsection{\textbf{Analyzing the impacts of smells on bugs}} \label{sec:smell_impact}

For RQ3, we focus on each of the smells to study whether the proportion of files containing at least one bug, significantly differs between files containing smells and files without smells. We consider the number of bugs c$_{i,k}$ a file $i$ encountered between releases r$_k$ and r$_{k+1}$, and convert c$_{i,k}$ into a Boolean variable f$_{i,k}$ (true if the file underwent at least one bug, false otherwise).
We rely on Fisher’s exact test \cite{sheskin2003handbook} to check whether the proportion of buggy files varies between two samples (files with and without smells). This test is useful for categorical data that result from the classification of objects. It is used to examine the significance of the association between the two kinds of classification.
We also calculate the \textit{odds ratio} (OR) indicating the likelihood for an event (bug in our case) to occur. 
The odds ratio is calculated (as in Equation (\ref{eq3})) as the ratio of the odds \textit{p} of an event occurring in a sample, \ie the odds that files with some specific smells contain a bug (defined as experimental group), to the odds \textit{q} of the same event occurring in another sample, \ie the odds that files with no smells contain a bug (defined as control group):

\begin{equation} \label{eq3}
OR = \frac{p/(1-p)}{q/(1-q)}
\end{equation}

An OR equal to 1 indicates that the event of interest is equally likely in both samples. While an OR greater than 1 stipulates that the event is more likely to occur in the first sample (files participating in some design smells), having an OR less than 1 indicates that it is more likely to occur in the second sample (control group of files not participating in any design smell).
\newtext{
We use the \textit{fisher{\textunderscore}exact} function of the \textit{stats} module from \textit{scipy} Python package to compute the odds ratio and the \textit{p}-value for statistical significance of the test. By processing the commits and bug information we set different flags for each of the source files. 
As mentioned earlier, the \textit{smelly} flag takes the value 1 if the associated source file contains at least one design smell of any type, and 0 otherwise. The flag \textit{buggy}, takes the value 1 if the associated source file was identified by SZZ algorithm as related to a fault-inducing commit, and 0 otherwise. 
Now, for a given release of a system, we consider all JNI source files for analysis. We count the number of buggy and non-buggy files with design smells. Similarly, we also count the number of buggy and non-buggy files without design smells. With these four values, we form the 2x2 contingency table for Fisher's exact test.}

For RQ4, we investigate the relationship between different types of design smells with fault-proneness.
Unlike using logistic regression for prediction purposes (\cite{khomh2012exploratory,gyimothy2005empirical}), we use it to examine whether some types of design smells are more related to fault-proneness. 
Our analysis approach is similar to the one presented by Khomh \al\cite{khomh2012exploratory} where they investigate the impacts of different types of anti-patterns on change- and fault-proneness using logistic regression model. The multivariate logistic regression is based on the following Equation (\ref{eq4}). 

\begin{equation} \label{eq4}
\pi (X_1,X_2, .... , X_n) = \frac{e^{\beta_0+\beta_1.X_1+~...~+\beta_n.X_n}}{1+e^{\beta_0+\beta_1.X_1+~...~+\beta_n.X_n}}
\end{equation}

Here, \newline
\begin{itemize}
    \item  $X_i$ are the independent variables for the logistic regression model. In our case, $X_i$ represents the number of smells of type $S_i$ in a given source file and $S =\{S_1, S_2, ..., S_n\}$ is the set of the types of smells investigated.
    \item $\beta_i$ are the model coefficients, and 
    \item $0\leq\pi\leq1$ is the value on the logistic regression curve representing the probability of bugs for a file with smells. 
\end{itemize}
\newtext{
In our regression model, independent variables are the number of occurrences of each type of design smells. 
The dependent variable is the flag (\textit{buggy}) representing the presence or absence of bugs. 
Thus, the dependent variable is dichotomous and assumes values either 0 (non-buggy) or 1 (buggy). For each system, we build a regression model and analyze the model coefficients and p-values for individual types of smells. Each row in our data set contains the values of the metrics (number of occurrences) for different smells, file size (LOC), number of previous bug-fix, code churn, and the bug status (1 or 0).} 

\newtext{Each logistic regression model gives the log odds (regression coefficient estimate) of individual independent variables and their corresponding p-values for a particular system. The log odds represent the factors by which the odds of the dependent variable will change for a unit change in values of corresponding independent variables. \newtext{When the logistic regression coefficient is positive ($\beta_i>0$), unit increase of the value of the corresponding independent variable will increase the log odds of the dependant variable by $\beta_i$ assuming that other independent variables are either 0 or remain unchanged. For a negative regression coefficient ($\beta_i<0$), on the other hand, the value of the log odds of the dependant variable will decrease by $\beta_i$ for unit increase in the value of the associated independent variable.} Thus, the higher the positive log odds of an independent variable, the higher is the impact of that independent variable on bug-proneness. We rank the smells based on the model coefficients and the corresponding p-values. We select files that contain at least one smell of any type. 
For a given type of smell, if the model coefficients show higher log odds (LO) of bugs in the majority (in percentage) of the systems, we consider the smell to be related to fault-proneness. It is important to mention that we analyzed the data for correlation among smells and dropped one independent variable from each pair of highly correlated variables. This ensures a non-redundant set of variables for the logistic regression models. From a highly correlated pair, we keep the variable representing smell type with a comparatively higher overall prevalence in the studied systems. 
Because the following metrics are known to be related to fault-proneness \cite{saboury2017empirical,koru2008theory,selim2010studying}, we add the file size, code churn, and the number of the previous occurrence of faults to our model, to control their effect. Here, (i) LOC: number of lines of code in the file at that specific release; (ii) Code Churn: the sum of lines added and removed in the file before that specific release;
(iii) No. of Previous-Bugs: the number of faults fixing related to that file before the particular release \textit{r}.}

\subsubsection{\textbf{Topic Modeling to Identify Fault-inducing Activities}} \label{sec:topics}

For \textbf{RQ5}, we are interested to investigate what kind of activities once performed in smelly files, are more likely to introduce bugs than other activities. We decided to analyze the commit messages that developers described when they performed a change that was captured by the SZZ algorithm as a fault-inducing commit. Having knowledge about those activities, developers could pay more attention to avoid introducing additional bugs.
We collect all the fault-inducing commits messages related to smelly files as described earlier. We then classify those commit messages into different topics of activities based on the keywords mentioned by developers using a mix of automated and manual techniques. We decided to apply both topic modeling strategies and manual text analysis. Similar to previous work \cite{ray2014large,sharma2017cataloging}, we used Latent Dirichlet Allocation (LDA) \cite{blei2010probabilistic}, a well-known
topic modeling algorithm to analyze the text and extract a set of frequently co-occurring words (\ie topics). We treat the commit messages as a corpus of textual documents, that is used as a basis for topic modeling. Given a corpus of $n$ documents $f_{1}$, ..., $f_{n}$, topic modeling techniques automatically discover a set $Z$ of topics, $Z$ = {$z_{1}$, ..., $z_{k}$}. The variable $k$ presents the number of topics. It is an input that controls the granularity of the topics.

\newtext{To generate the topic of activities introducing bugs, we combine both manual and automated approaches to build a categorization of risky activities. Based on developers' commit messages, similar to previous work \cite{chen2012explaining}, we used \textit{MALLET}\footnote{\url{http://mallet.cs.umass.edu/}}, a specific type of LDA implementation to generate a set of topics based on frequently co-occurring words. We removed stop words using MALLET stop words list (\eg a, the, is, this, punctuation marks, numbers, and non-alphabetical characters). We also used Porter stemmer to reduce words to their root words (\eg programmer became program) \cite{porter2001snowball}. Since our objective is to study the activities that could introduce bugs once performed in smelly files, we limited our study to the smelly files in which a bug was introduced (Flag =1). Thus, our data-set resulted in 2707 commit messages. We manually inspected the commit messages to estimate the number of possible topics for each system and also to assign a meaningful name to each topic. Once the number of possible topics was fixed, we used a python script that takes as input the list of all the commits messages in a CSV file and returns the list of commit messages with common keywords that could be used to build the topic. Two of the authors went through all the topics extracted for all the systems, and manually assigned meaningful names to each topic. The name of the topic was decided based on manual inspections of the commit messages and the keywords used to build that topic. We relied on the keywords generated by MALLET but also on frequent keywords captured during the manual analysis. We manually analyzed a total of 500 commits. To resolve the disagreements, two of the authors went through those commit messages and discussed the main topics of activities performed on those commits. Through those analysis,
we aim to capture the possible types of activities that were described in the commit messages of the bug-inducing commits.
} 

\section{Study Results}
\label{sec:Results}

In this section, we report on the results of our study by addressing the five research questions defined in Section \ref{sec:Study}. We focus on the three key research objectives of our study. First, research questions RQ1 and RQ2 investigate the prevalence of the multi-language design smells in software systems. Then, research questions RQ3 and RQ4 evaluate the impacts of the design smells on the fault-proneness of JNI systems. Finally, RQ5 investigates fault-inducing activities. We present additional insights into the findings from the research questions later in Section \ref{sec:Discussion}.

\subsection{\textbf{RQ1: \RQOne{} }}

We use our detection approach to detect occurrences of multi-language design smells following the methodology discussed in Section \ref{sec:Study}. For each file, we compute the value of a variable {$Smelly_{i,r}$} that takes 1 if the file $i$ contains at least one design smell in a specific release $r$, and 0 otherwise. We also compute {$Native_{i,r}$} which takes 1 if the file $i$ in a specific release $r$ is native and 0 if not, following the rules discussed in Section \ref{sec:analysis}. Since our tool is focusing on the combination of Java and C/C++, we compute for each release the percentage of files participating in at least one JNI smell out of the total of JNI files (files involved in Java and C/C++). 

Table \ref{tab:RQ1.1} summarises our results on the percentages of files with JNI smells in each of the studied systems. We report in this table the average number of JNI files participating in, at least one of the studied design smells for each system.
Our results show that indeed, the JNI smells discussed in the literature are prevalent in the nine studied open source projects with average occurrences from 10.18\% in \jpype~system to 61.36\% in \zstd. The percentage of files with smells differ from one project to another. We compute the average of the percentage of smells in all the systems. We find that on average, one-third (33.95\%) of the JNI files in the studied systems contain multi-language design smells.
 
 \begin{table}[t]
\centering\footnotesize
\caption{\label{tab:RQ1.1} Percentage of JNI Files Participating in Design Smells in the Release of 9 Systems}
{\renewcommand{\arraystretch}{1.15}
\begin{tabular}{llcc} \hline
\rowcolor{gray!35}
\textbf{Systems} & \textbf{Releases Analyzed} & \textbf{\% Files with Smells} & \textbf{Smells Density per KLOC} \\ \hline
\zstd & 0.4.4 - latest release &  61.36\% &  8.14
 \\ 

\rowcolor{gray!15}
\javacpp & 0.9 - 1.5.1-1 &  58.97\% & 17.84
 \\ 

\rocksdb & 5.0.2 - latest release &  36.30\%  & 8.54
 \\ 
\rowcolor{gray!15}
\javasmt & 1.0.1 - 3.0.0 &  36.21\% & 26.08 \\ 
\vlc & 3.0.0 - latest release &  30.49\% & 17.67 \\ 
\rowcolor{gray!15}
\conscrypt & 1.0.0.RC2 - 2.3.0 & 30.21\% & 14.05 \\ 
\pljava & REL1\_5\_STABLE - latest release  &  30.13\% & 7.59 \\ 
\rowcolor{gray!15}
\realm & 0.90.0 - 5.15.0 &  11.67\%  & 4.63 \\ 
\jpype & 0.5.4.5 - latest release  & 7.45\% &  7.45
 \\\hline 
 \textbf{Average} & & \textbf{33.95\%} & 12.44
 \\ \hline
	\end{tabular}
}	
\end{table}

\begin{figure}[ht]
\center
\vspace{-0.35cm}
\includegraphics[width=1\linewidth, height= 9cm]{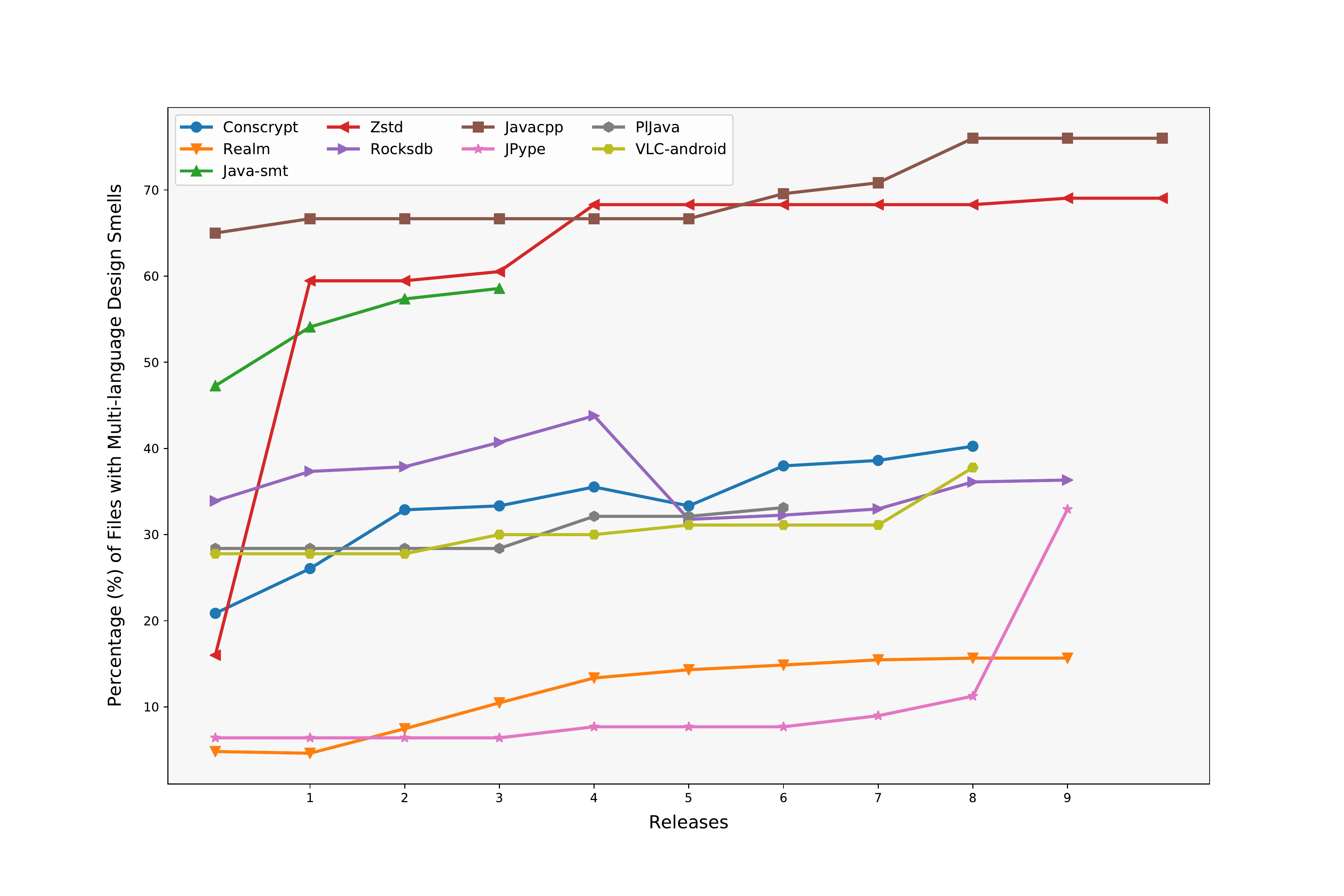} 
\vspace{-0.5cm}
\caption{Evolution of Design Smells in the Releases of the 9 Systems} 
\label{fig:RQ1.1}

\end{figure}

Besides analyzing in each system the percentage of files affected by each of the studied JNI smells, we also investigate their evolution over the releases. 
Figure \ref{fig:RQ1.1} presents an overview of the evolution of the percentage of files participating in multi-language design smells in the releases of each system.
The X-axis in Fig. \ref{fig:RQ1.1} represents the releases analyzed. 
The Y-axis represents the percentage of files affected by at least one of the studied design smells, while the lines are related to each system. 
Results show that these percentages vary across releases in the nine systems with peaks as high as 69.04\%. Some of these systems \ie \realm~and \jpype~contain respectively 4.61\% and 6.41\% in the first releases, but the occurrences of smells increased over time to reach respectively 15.66\% and 32.94\%.
Overall, the number of occurrences of smells are increasing over the releases.
Although, in some cases such as in \rocksdb, the number of occurrences seems to decrease from one release to the next one, (from 43.78\% to 31.76\%). The fact that developers might not be aware of occurrences of such smells and the lack of tools for their detection might explain the observed prevalence. 
The observed decrease in the number of occurrences observed in certain cases could be the result of fault-fixing activities, features updates, 
or any other refactoring activities. In general, as one can see in Fig. \ref{fig:RQ1.1}, these decreases are temporary; the number of occurrences often increase again in the next releases. Overall, the proportions of files with smells are considerably high and the smells persist, thus allowing to reject $H_{1}$.

\begin{tcolorbox}
\textit{\textbf{Summary of findings (RQ1)}}: JNI smells discussed in the literature are prevalent and persistent in open source projects. The number of their occurrences even increases over the releases.
\end{tcolorbox}

\subsection{\textbf{RQ2: \RQTwo{} }} 

Similar to \textbf{RQ1}, we use our approach from Section \ref{sec:Study} to detect the occurrence of the 15 design smells in the nine subject systems. For each file and for each release, we defined a metric \textit{Smelly}$_{i,r}$ which takes the value 1 if the release $r$ of the file contains the design smell type \textit{i} and 0 if it does not contain that specific smell. We compute for each release the number of files participating in that specific smell. Note that the same file may contain more than one smell.

Table \ref{tab:RQ2} shows the distribution of the studied smells in the analyzed open source systems. We calculate the percentage of files containing these smells and compute the average. Since our goal is to investigate if some specific smells are more prevalent than others, we compute the percentage of files containing that specific smell out of all the files containing smells. Our results show that some smells are more prevalent than others, \ie \emph{Unused parameter, Too much scattering, Too much clustering, Unused Method Declaration, Not securing libraries, Excessive Inter-language communication}. In studied releases from \jpype, on average, 89.24\% of the smelly files contain the smell \emph{Unused parameter}. In \javasmt, on average, 94.06\% of the smelly files contain the smell \emph{Unused Parameters}. Our results also show that some smells discussed in the literature and developers' blogs have a low diffusion in the studied systems, \ie \emph{Excessive objects, Not caching objects, Local reference abuse}, while the other smells are quite diffused in the analyzed systems. \conscrypt~ presents 79.60\% occurrences of the design smell \emph{Unused Parameters}. As described in the commit messages in \conscrypt,~this could be explained by the usage of BoringSSL which has many unused parameters. Results presented in Table \ref{tab:RQ2} report a range of occurrences from 0\% to 94.06\%. Some specific types of smells seem to be more frequent than others. On average \unusedParameters~represents 57.36\% of the existing smells, followed by the smell \cluster~with 20.91\%. We also report in Table \ref{tab:RQ22}, the distribution of smells normalized by the number of KLOC.

\begin{table*}[t]
\centering\scriptsize
\caption{Percentage of JNI Files Participating in Design Smells in the Releases of the Studied Systems}\label{tab:RQ2}

{\renewcommand{\arraystretch}{1.15}
\begin{tabular}{p{1.67cm}p{0.48cm}p{0.48cm}p{0.48cm}p{0.49cm}p{0.4cm}p{0.49cm}p{0.30cm}p{0.49cm}p{0.48cm}p{0.32cm}p{0.4cm}p{0.4cm}p{0.45cm}p{0.43cm}p{0.4cm}}	\hline
\rowcolor{gray!35}
	System$\downarrow$/Smells$\rightarrow$ & UP &	UM	& TMS	& TMC &	UMI &	ASR &	EO &	EILC	& NHE &	NCO &	NSL &	HCD &	NURP &	MMM &	LRA	\\\hline
	
\conscrypt &	79.60\% &	4.40\%  & 0\%  &	1.90\%  &	0\%  &	3.99\%  &	0\%  &	1.90\%  &	3.99\%  &	0\%  &	5.71\%  &	0\%  &	3.80\% 	& 3.78\%  &	3.78\% 	\\ 
\rowcolor{gray!15}					
\realm  &	67.68\%  &	3.066\%  & 9.75\%  &	14.86\%  &	2.32\%  &	4.33\%  &	0\%  &	12.58\%  &	5.15\%  &	0\%  &	2.17\%  &	0\%  &	0 \% &	0\%  &	0.79\% 	\\ 
\javasmt &	94.06\%  &	2.96\% 	& 0\%   & 2.96\%  &	0\%  &	0\%  &	0\%  &	0\%  &	0\%  &	0\%  &	2.96\%  &	0\%  &	2.96\%  &	0\%  &	0\% 								\\ 
\rowcolor{gray!15}
\zstd  &	10.46\% 	 &		0.95\% 	 &		13.98\% 	 &		12.36\% 	 &		3.47\% 	 &		17.98\% 	 &		0\% 	 &		23.55\% 	 &		21.45\% 	 &	 0\% 	 &		5.74\% 	 &		3.47\% 	 &		0\% 		 &	2.25\% 	 &		0\% 
	\\ 
\rocksdb		 &	44.55\% 	 &	5.48\% 	 &	34.48\% 	 &	23.47\% 	 &	0\% 	 &	0.67\% 	 &	0\% 	 &	14.35\% 	 &	0.67\% 	 &	0.91\% 	 &	2.85\% 	 &	0.95\% 	 &	0.95\% 	 &	0.79\% 	 &	0.10\% 	 
	\\ 
\rowcolor{gray!15}
\javacpp	& 2.53\%  &	31.70\%  &	74.19\%  &	19.49\%  &	0\%  &	0\%  &	0\%  &	 69.14\%  &	0\%  &	0\%  &	6.48\% 	& 2.51\%  &	0\% 	& 0\% 	& 0\% 			\\ 
\jpype	&	89.24\%  &	0\%  &	0\% 	& 0\% 	&	0\% 	&	1.78\% & 0\% 	&	0.35\% 	&	1.78\% 	&	0\% 	&	0\% 	&	0\% 	&	0\% 	&	8.25\% 	&	1.07\% 	\\ 
\rowcolor{gray!15}
\pljava	& 64.45\% 		& 35.62\% 		& 31.02\% 	&	8.42\% 	&	2.04\% 	& 0\% 	& 0\% &	4.36\% 	&	2.04\% 	&	0\% 	&	0\% 	&	0\% 	&	0\% 	&	2.04\% 	&	0\% 	\\ 
\vlc& 63.67\% 		&	25.71\% 	&		24.74\% 	&	  17.10\% 		&	7.34\% 	&		3.67\% 	&		0.82\% 	&		13.29\% 	&		3.67\% 	&		0\% 	&		3.92\% 	&		0\% 	&		6.01\% 	&		0\% 	&		3.67\% \\ \hline
Median	&	64.45	&	4.4	&	13.98	&	12.36	&	0	&	1.78	&	0	&	12.58	&	2.04	&	0	&	2.96	&	0	&	0	&	0.79	&	0.1 \\
Average	&	57.36	&	12.21	&	20.91	&	11.17	&	1.69	&	3.60	&	0.09	&	15.50	&	4.31	&	0.10	&	3.31	&	0.77	&	1.52	&	1.9	&	1.05
\\ \hline 
\multicolumn{16}{l}{\textbf{Acronyms: }\textbf{Up:} UnusedParameter, \textbf{UM:} UnusedMethodDeclaration, \textbf{TMS:} ToomuchScattering, \textbf{TMC: }Toomuchclustring}\\ 
\multicolumn{16}{l}{\textbf{UMI:} UnusedMethodImplementation, \textbf{ASR:} AssumingSafeReturnValue, \textbf{EO:} ExcessiveObjects}\\
\multicolumn{16}{l}{\textbf{EILC:}excessiveInterlangCommunication, \textbf{NHE:} NotHandlingExceptions, \textbf{NCO:} NotCachingObjects, \textbf{NSL:} NotSecuringLibraries} \\ 
\multicolumn{16}{l}{\textbf{HCD:} HardCodingLibraries, \textbf{NURP:} 	NotUsingRelativePath, \textbf{MMM:} MemoryManagementMismatch, \textbf{LRA:} LocalReferencesAbuse} \\ \hline
	\end{tabular}
}
\end{table*}

\begin{table*}[t]
\centering\scriptsize
\caption{Number of Design Smells per KLOC in the Releases of the Studied Systems}\label{tab:RQ22}

{\renewcommand{\arraystretch}{1.15}
\begin{tabular}{p{1.67cm}p{0.48cm}p{0.48cm}p{0.48cm}p{0.49cm}p{0.4cm}p{0.49cm}p{0.30cm}p{0.49cm}p{0.48cm}p{0.32cm}p{0.4cm}p{0.4cm}p{0.45cm}p{0.43cm}p{0.4cm}}	\hline
\rowcolor{gray!35}
	System$\downarrow$/Smells$\rightarrow$ & UP &	UM	& TMS	& TMC &	UMI &	ASR &	EO &	EILC	& NHE &	NCO &	NSL &	HCD &	NURP &	MMM &	LRA	\\\hline
	
\conscrypt &	6.091 &	7.089 &	0.0	& 0.022 &	0.0 &	0.07 &	0.0	& 0.07 &	0.16 &	0.0 &	0.15 &	0.0 &	0.25 &	0.02 &	0.13 \\

\rowcolor{gray!15}					
\realm  &	1.81 &	0.086 &	0.075 &	0.097 &	0.024 &	0.04 &	0.0	& 2.36 &	0.12 &	0.0 &	0.011 &	0.0	& 0.0 &	0.0 &	0.010
 	\\ 
\javasmt &	8.34 &	16.56 &	0.0	& 0.05 &	0.0	&0.0&	0.0&	0.0	& 0.0&	0.0&	0.2&	0.15&	0.78 &	0.0	& 0.0
\\ 
\rowcolor{gray!15}
\zstd  &	2.0	& 0.22 &	0.11 &	0.23 &	1.09 &	1.08 &	0.0	& 1.60 &	1.15 &	0.0 &	0.078 &	0.07 &	0.0 &	0.50 &	0.0
	\\ 
\rocksdb		 &	1.32 &	0.18 &	0.34 &	0.23 &	0.0	& 0.02 &	0.0 &	5.72 &	0.03 & 0.011 &	0.081 &	0.019 &	0.02 &	0.02 &	0.0
	\\ 
\rowcolor{gray!15}
\javacpp	& 0.05 &	7.06 &	1.93 &	0.5	& 0.0 &	0.0 &	0.0	& 8.06 &	0.0	& 0.0 &	0.20 &	0.04 &	0.0	& 0.0 &	0.0	\\ 
\jpype	&	3.18 &	0.0 &	0.0 &	0.0 &	0.0 &	1.37 &	0.0 &	0.007 &	1.32 &	0.0 &	0.0 &	0.0	& 0.0 &	1.5 &	0.08	\\ 
\rowcolor{gray!15}
\pljava	& 5.10 &	1.7 &	0.41 &	0.06 &	0.02 &	0.0 &	0.0 &	0.04 &	0.11 &	0.0	& 0.0 &	0.0	& 0.0 & 0.14 &	0.0 \\ 
\vlc& 4.18 &	4.75 &	0.46 &	0.4 &	0.55 &	0.1 &	0.010 &	5.47 &	0.1 &	0.0 &	0.37 &	0.0	& 1.25 &	0.0 &	0.05
 \\ \hline 
\multicolumn{16}{l}{\textbf{Acronyms: }\textbf{Up:} UnusedParameter, \textbf{UM:} UnusedMethodDeclaration, \textbf{TMS:} ToomuchScattering, \textbf{TMC: }Toomuchclustring}\\ 
\multicolumn{16}{l}{\textbf{UMI:} UnusedMethodImplementation, \textbf{ASR:} AssumingSafeReturnValue, \textbf{EO:} ExcessiveObjects}\\
\multicolumn{16}{l}{\textbf{EILC:}excessiveInterlangCommunication, \textbf{NHE:} NotHandlingExceptions, \textbf{NCO:} NotCachingObjects, \textbf{NSL:} NotSecuringLibraries} \\ 
\multicolumn{16}{l}{\textbf{HCD:} HardCodingLibraries, \textbf{NURP:} 	NotUsingRelativePath, \textbf{MMM:} MemoryManagementMismatch, \textbf{LRA:} LocalReferencesAbuse} \\ \hline
	\end{tabular}
}
\end{table*}

\begin{figure}[ht]
\center
\vspace{-0.45cm}
\includegraphics[width=1.1\linewidth]{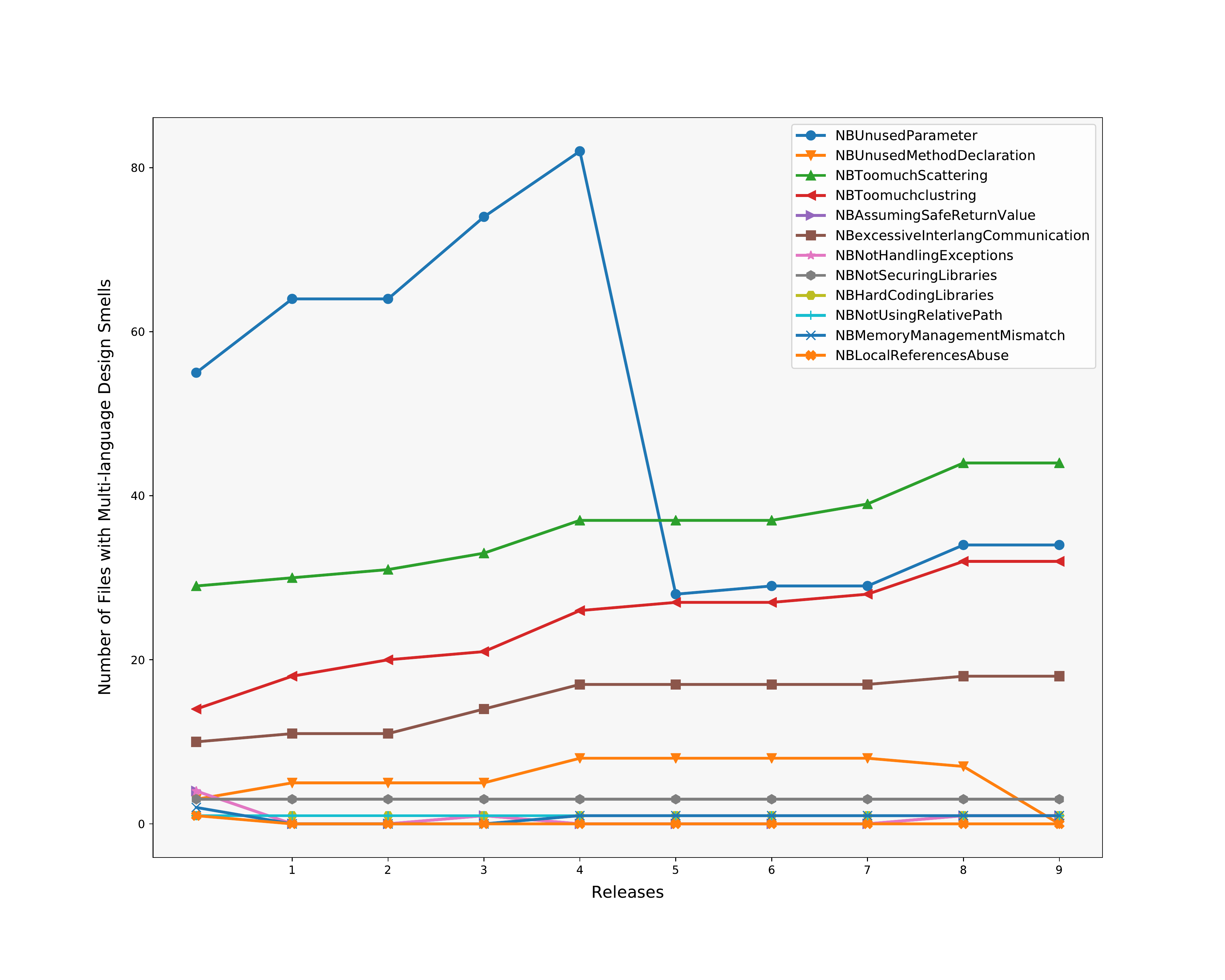}
\vspace{-0.5cm}
\caption{Evolution of the Different Kinds of Smells in \rocksdb~ Releases}
\label{fig:RQ2}
\end{figure}

\begin{figure}[ht]
\center
\vspace{-0.5cm}
\includegraphics[width=1.1\linewidth]{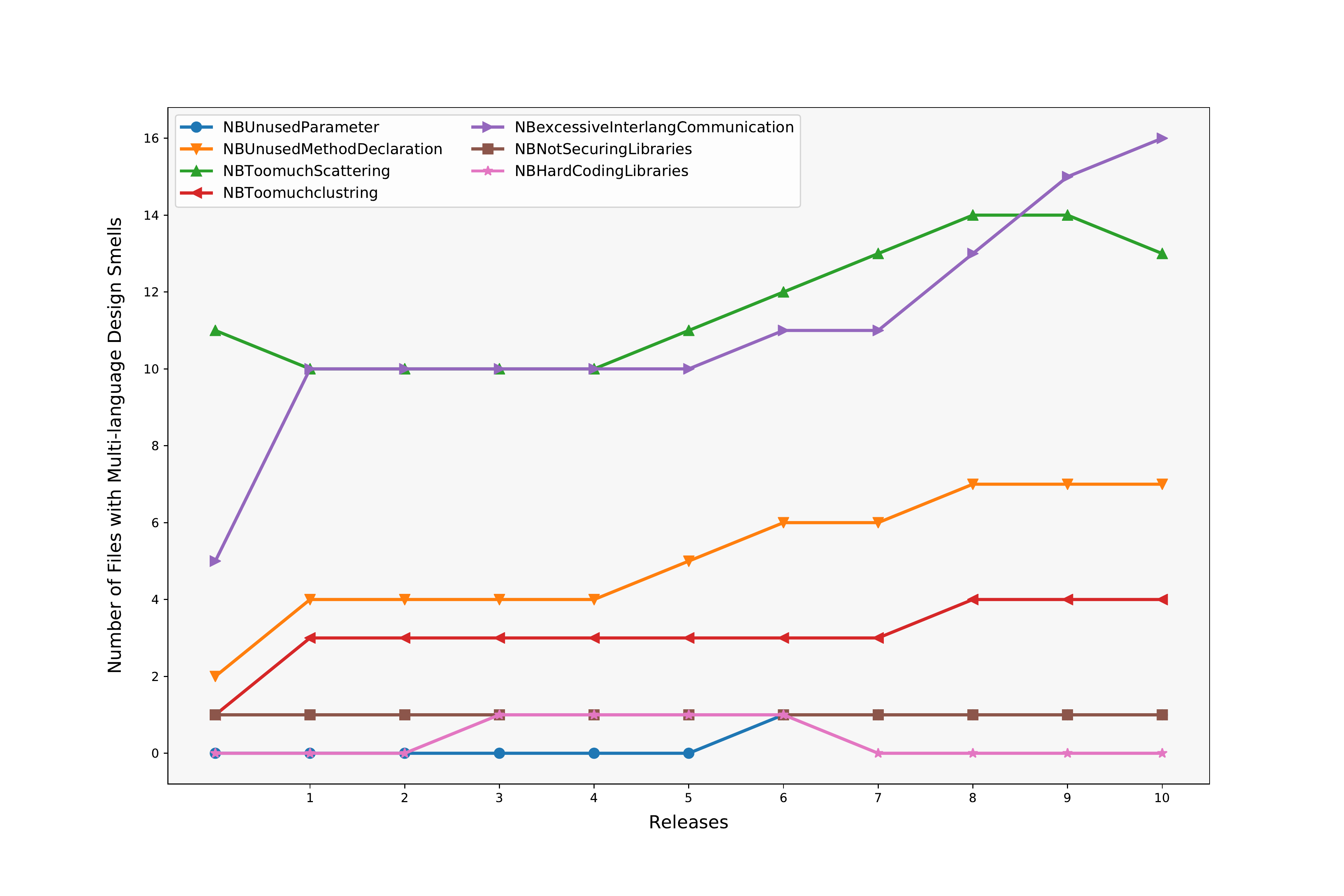}
\vspace{-0.5cm}
\caption{Evolution of the Different Kinds of Smells in \javacpp~ Releases}
\label{fig:RQ22}
\end{figure}

\begin{figure}[ht]
\center
\vspace{-0.5cm}
\includegraphics[width=1.1\linewidth]{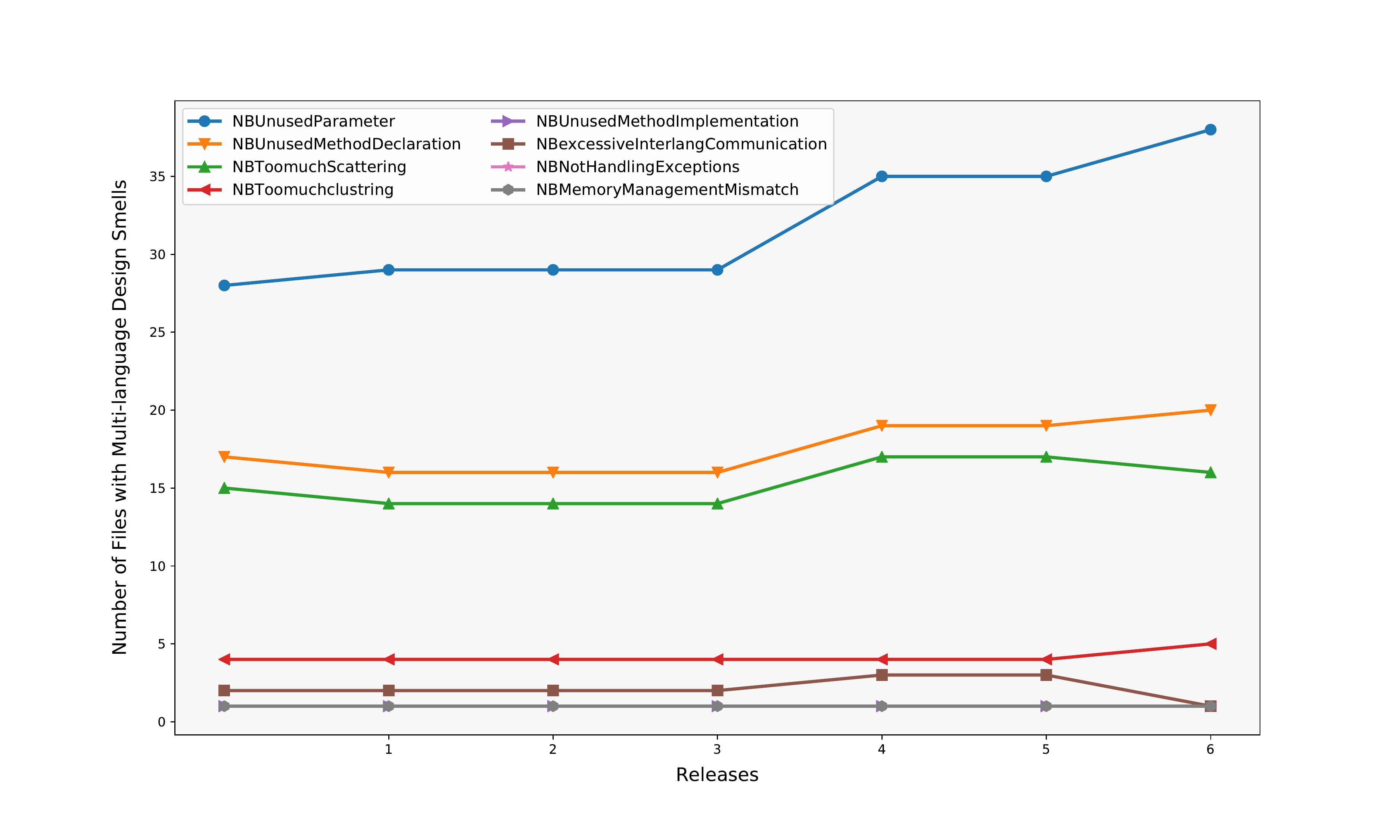} 
\vspace{-0.5cm}
\caption{Evolution of the Different Kinds of Smells in \pljava~ Releases}
\label{fig:RQ222}
\end{figure}

\begin{figure}[ht]
\center
\vspace{-0.5cm}
\includegraphics[width=1.1\linewidth]{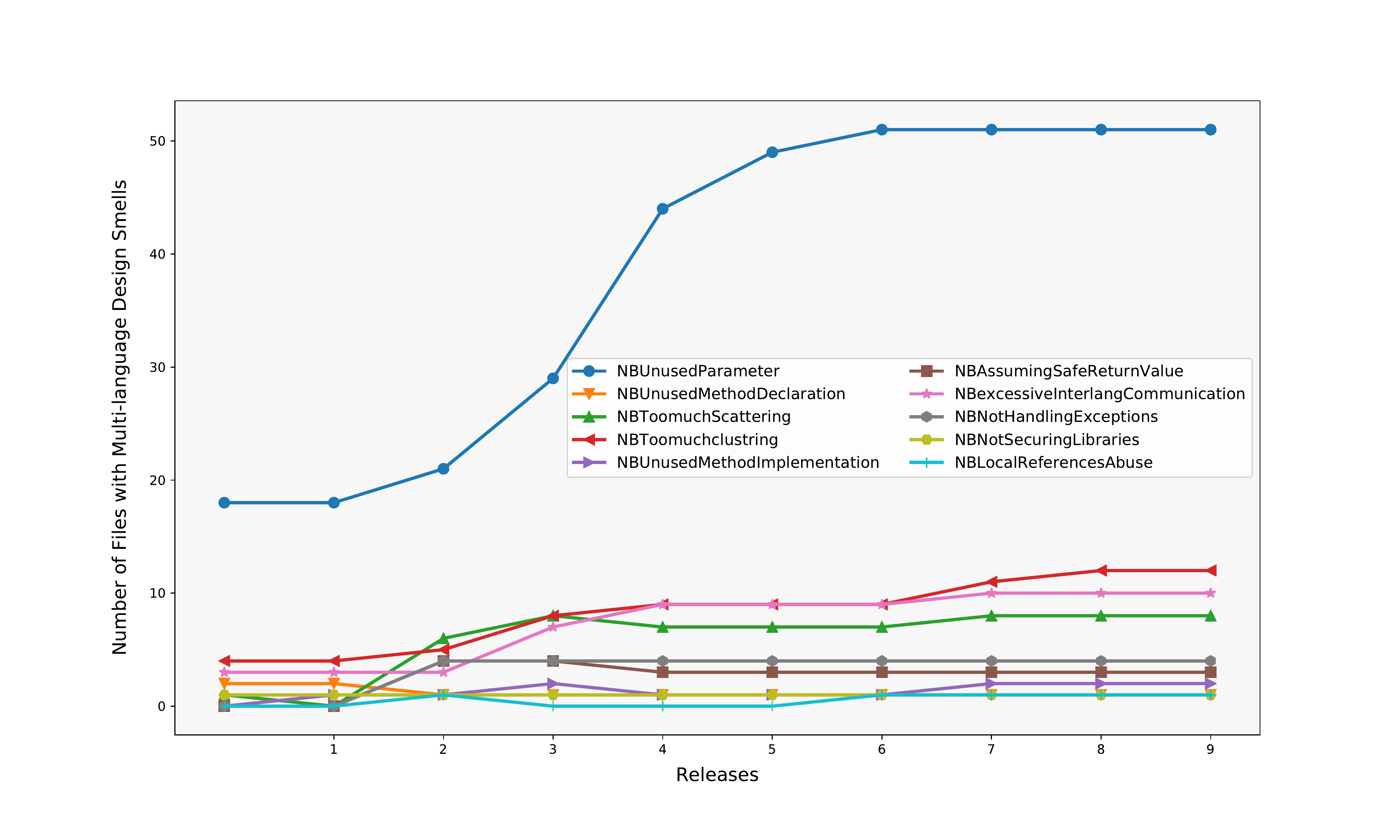} 
\vspace{-0.5cm}
\caption{Evolution of the Different Kinds of Smells in \realm~ Releases}
\label{fig:RQrealm}
\end{figure}

\begin{figure}[ht]
\center
\vspace{-0.4cm}
\includegraphics[width=1.1\linewidth]{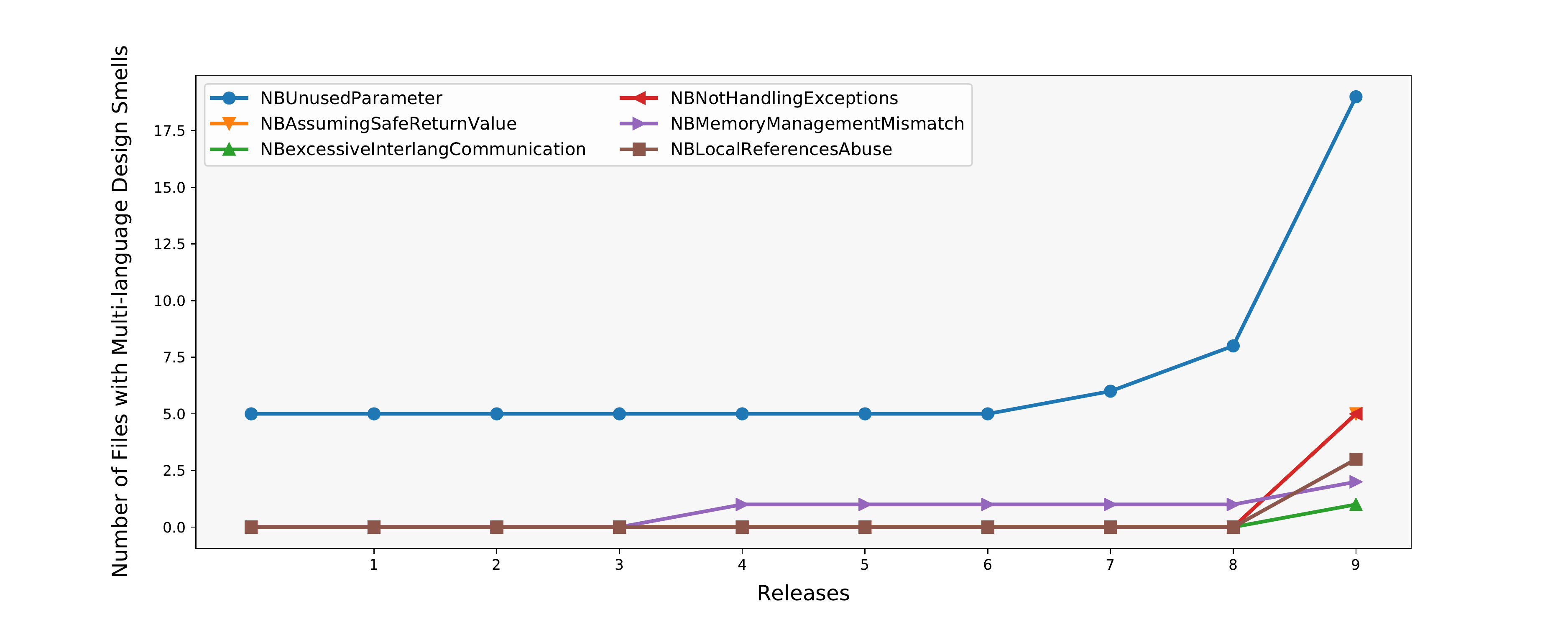} 
\vspace{-0.5cm}
\caption{Evolution of the Different Kinds of Smells in \jpype~ Releases}
\label{fig:RQjpype}
\end{figure}

\begin{figure}[ht]
\center
\vspace{-0.4cm}
\includegraphics[width=1.1\linewidth]{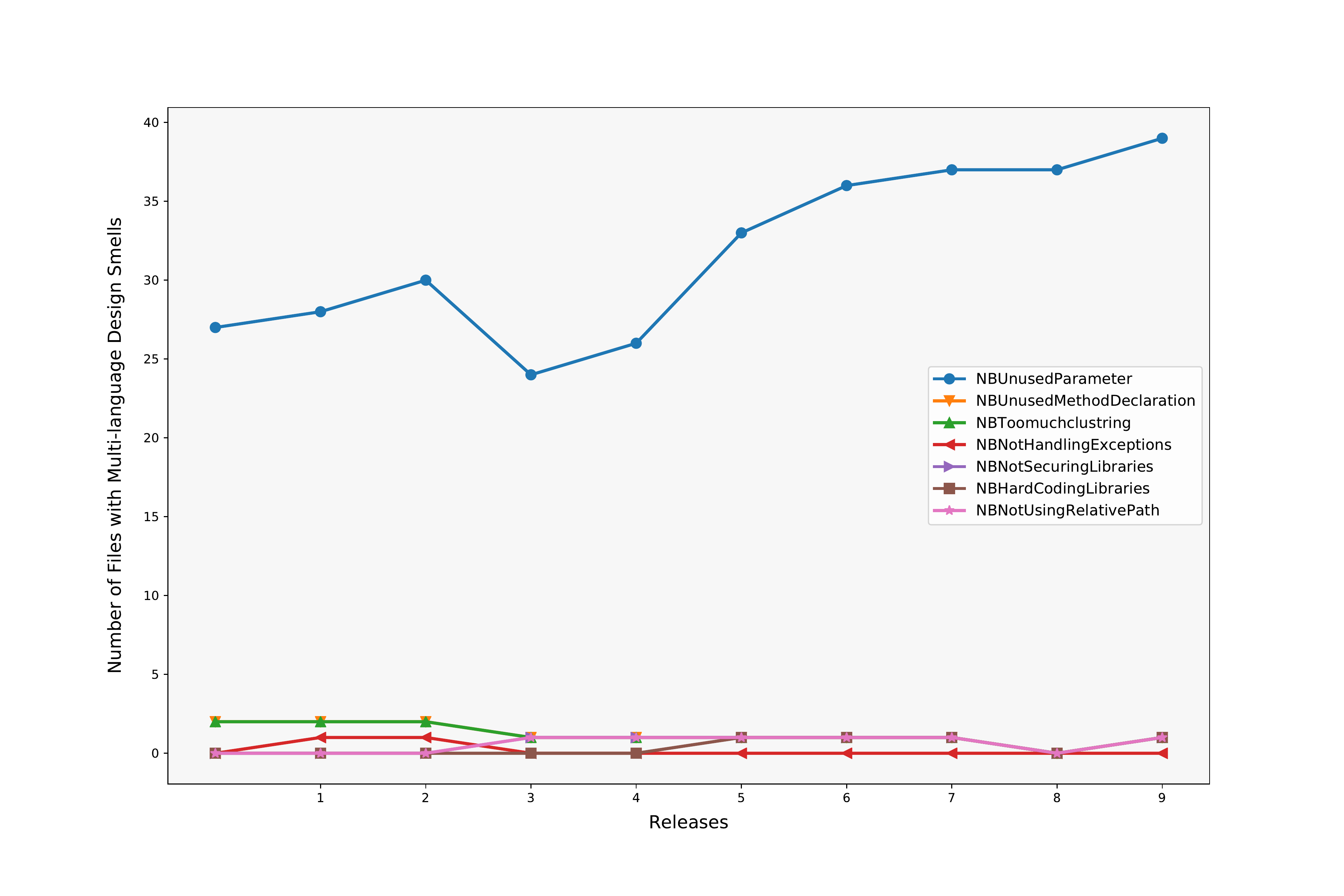} 
\vspace{-0.5cm}
\caption{Evolution of the Different Kinds of Smells in \javasmt~ Releases}
\label{fig:RQjavasmt}
\end{figure}

For each system, in addition to analyzing the percentage of files affected by each type of smell, we also investigate the evolution of the smell over the releases. Figures \ref{fig:RQ2}, \ref{fig:RQ22}, \ref{fig:RQ222}, \ref{fig:RQrealm}, \ref{fig:RQjpype}, and \ref{fig:RQjavasmt} provide an overview of the evolution of smells respectively in \rocksdb, \javacpp, \pljava, \realm, \jpype, and \javasmt~releases. The X-axis in these figures represents the releases analyzed. 
The Y-axis represents the number of files in that specific system affected by that kind of design smells, while the lines are related to the different types of smells we studied. Depending on the system, some smells seem more prevalent than the others. In \javacpp, \scatter, and \excessivecom~seem to be the predominant ones, while \unusedParameters~is less frequent in this system. However, in general, for other systems including \rocksdb~and \realm, \unusedParameters~seems to be dominating. Results show that most of the smells generally persist within the project. 
The smells tend to persist in general or even increase from one release to another. 

Although, in some specific cases, for example, the design smell \unusedParameters~in \rocksdb, presented a peak of 82 and decreased to 28 in the next release. 
However, the number of files containing this smell increased in the next releases and reached to 34 in the last release analyzed. We studied the source code files containing some occurrences of the design smell \emph{unused parameters} between releases (5.11.2 and 5.14.3) of \rocksdb~to understand the reasons behind the peak and the decrease. We found that some method parameters were unused on \rocksdb~(5.11.2) and have been refactored during the next releases by removing occurrences of this smell and also due to project migration features. Another example of refactoring of the code smell \unusedParameters~from one release to another was observed in \conscrypt, where they refactored \unusedParameters~occurrences due to errors generated by those occurrences in the release \textit{1.0.0.RC14} (\emph{``commit message: Our Android build rules generate errors for unused parameters. We cant enable the warnings in the external build rules because BoringSSL has many unused parameters''}). 
From our results, we can clearly observe that occurrences of JNI smells are not equally distributed. We conclude that the proportions of files with specific smells vary significantly between the different kinds of smells. We, therefore, reject hypothesis $H_{2}$.



\begin{tcolorbox}
\textit{\textbf{Summary of findings (RQ2)}}: Some JNI smells are more prevalent than others, \eg \unusedParameters, \scatter, \declaration~while others are less prevalent, \eg \excessiveObject~and \cachingObjects. 
Most of the smells persist with an increasing trend from one release to another in most of the systems.
\end{tcolorbox}

\subsection{\textbf{RQ3: \RQThree{} }}

\newtext{Prior works show that design smells increase the fault-proneness of Java applications\cite{khomh2009exploratory,saboury2017empirical}. 
Since JNI systems introduce other kinds of design smells and those smells are prevalent as observed in research questions RQ1 and RQ2, we are interested in studying the impacts of those design smells on the fault-proneness of JNI systems. 
For that, we applied Fisher's exact test \cite{sheskin2003handbook} to check whether the proportion of bugs varies between two samples (files with and without smells) as discussed in Section \ref{sec:smell_impact}.
The columns \textit{Smelly-buggy (SB), Buggy-Notsmelly (BNS), Smelly-NotBuggy (SNB), NotBuggy-NotSmelly (NBNS)} in Tables \ref{tab:fisher1} and \ref{tab:fisher2} contain the values of the contingency tables for the Fisher's exact test; each row corresponding to a single release. 
The numbers reported in the cells of these columns 
are the total number of JNI source code files (for the specific release) with or without smells and with or without bugs, depending on the column. 
More specifically, these columns present respectively: the total number of source code files with smells and are buggy (SB), the total number of source code files containing bugs without occurrences of smells (BNS), smelly source code files that are not buggy (SNB), and source code files that do not present any occurrences of smells or bugs (NBNS).} 
The value of the odds ratio (OR) greater than 1 from Fisher's exact test indicates that files with design smells have higher odds of being buggy compared to files without design smells. The values $OR<1$ indicate that files with design smells have lower odds of having faults, while $OR=1$ refers to no impact of design smells on fault-proneness of the source files. The \textit{p}-value shows the probability of observing the odds ratio by chance, and thus lower values ($<0.05$) of \textit{p}-value confirm the significance of the impacts of design smells on fault-proneness. \newtext{In addition to significant p-values, we examine the confidence intervals of the odds ratios. A confidence interval specifies the range where the true odds ratio lies in. A significant p-value value ($<0.05$) of an odds ratio ($>1.0$) with confidence interval not containing 1 confirms a true relationship between design smells and fault-proneness. We marked the p-values of such cases with (*) in Table \ref{tab:fisher1} and Table \ref{tab:fisher2}.}


\newtext{Tables \ref{tab:fisher1} and \ref{tab:fisher2} report the results of applying Fisher's exact test and present the values of odds ratios for the studied systems. Each row of those tables shows, for each system and each release, the odds for a file containing at least one type of design smells to be involved in a bug inducing change. 
}
In most of the analyzed releases, Fisher’s exact test indicates a significant difference of proportions between fault-prone JNI files with and without design smells. In some systems (\eg \rocksdb, \javacpp, and \javasmt), odds ratios for specific releases are less than one, or the \textit{p}-value is not statistically significant. However, in general, the values of odds ratios are high (in general greater than 2) in most cases. For \zstd, we found odds ratios always higher than 13. Having this high odds could be explained by the large number of smells contained in that system, as described in Table \ref{tab:RQ1.1}, but also by the nature of smells existing in this system. \newtext{The higher values of statistically significant odds ratios in most cases and the confidence intervals of those significant odds ratios being above the value 1 in some cases show that multi-language design smells are related to fault-proneness. However, this relationship varies with systems and further investigation is necessary to generalize.}  

From analyzing fault-fixing commit messages, we identified some commits reporting a refactoring for specific smells \eg \emph{``removing unused parameter'', ``implementing the handling of exception''}. This could explain cases where  \textit{Smelly-Buggy} values decrease from one release to the other, while the overall number of occurrences of smells are in general increasing from one release to the other, as shown in Fig. \ref{fig:RQ1.1}. For example, in \rocksdb, \textit{Smelly-Buggy} values are decreasing from one release to the other, while \textit{Smelly-NonBuggy} is increasing. This could be explained by the nature of the smells and the refactoring applied. Since one file can contain more than one type of smell, the refactoring of some specific types of smells could decrease the risk of bugs while leaving the file still smelly. This suggests that some specific smells could be more correlated with bugs than others. These hypotheses motivate us to further investigate the relationship between specific types of smells and fault-proneness (\textbf{RQ4}) and also the activities that once performed in smelly files could lead to bugs (\textbf{RQ5}).

\begin{table}[t]
\centering\footnotesize
\caption{\label{tab:fisher1} Fisher's Exact Test Results for the Fault-proneness of Files with and without Design Smells (1)
}
\begin{tabular}{llccccccc}
	\hline
\rowcolor{gray!35}
System & Releases & SB& BNS &  SNB &  NBNS &  Odds Ratios & \textit{p}-values & Confidence Interval\\\hline

\multirow{10}{*}{\begin{sideways}\rocksdb\end{sideways}}
& rocksdb-5.0.2&82&85&17&108&6.1287& < \textbf{0.01*} 
& (1.2184,2.4076)
\\
& \cc{rocksdb-5.4.6}&\cc{89}&\cc{90}&\cc{23}&\cc{98}&\cc{4.2135}&\cc{< \textbf{0.01} 
}
& \cc{(0.8979,1.9787)}\\
& rocksdb-5.6.2&90&80&24&107&5.0156& < \textbf{0.01*} 
& (1.0771,2.1480)
\\
&\cc{rocksdb-5.9.2}&\cc{97}&\cc{84}&\cc{30}&\cc{101}&\cc{3.8876}&\cc{ < \textbf{0.01}}
& \cc{(0.8563,1.8592)}
\\
& rocksdb-5.11.2&99&86&42&95&2.6038&< \textbf{0.01}
& (0.4929,1.4211)
\\
&\cc{rocksdb-5.14.3}&\cc{50}&\cc{101}&\cc{38}&\cc{88}&\cc{1.1464}&\cc{0.607}
& \cc{(-0.3729,0.6462)}\\
& rocksdb-5.17.2&51&92&39&97&1.3787&0.249 & (-0.1840,0.8263)\\
&\cc{rocksdb-5.18.3}&\cc{50}&\cc{101}&\cc{43}&\cc{88}&\cc{1.0131}&\cc{1.0} & \cc{(-0.4848,0.5109)}\\
& rocksdb-6.1.1&49&94&55&90&0.8529&0.541 & (-0.6404,0.3224)\\
&\cc{rocksdb-latest release}&\cc{49}&\cc{101}&\cc{56}&\cc{83}&\cc{0.7190}&\cc{0.181} & \cc{(-0.8108,0.1511)}\\
\hline 
\multirow{10}{*}{\begin{sideways}\pljava\end{sideways}}
& pljava-1\_4\_3&0&36&0&100&-&1.0 & - \\
&\cc{pljava-rel1\_5\_stable}&\cc{33}&\cc{38}&\cc{13}&\cc{78}&\cc{5.2105}&\cc{< \textbf{0.01}}
& \cc{(0.9008,2.4005)}\\
& pljava-1\_5\_0b3&32&33&14&83&5.7489&< \textbf{0.01*}
& (1.0026,2.4954)\\
&\cc{pljava-1\_5\_0}&\cc{33}&\cc{37}&\cc{13}&\cc{79}&\cc{5.4199}&\cc{< \textbf{0.01}}
& \cc{(0.9388,2.4413)}\\
& pljava-1\_5\_1b1&32&38&14&78&4.6917&< \textbf{0.01}
& (0.8077,2.2839)\\
&\cc{pljava-1\_5\_1b2}&\cc{39}&\cc{36}&\cc{14}&\cc{76}&\cc{5.8809}&\cc{< \textbf{0.01*}}
& \cc{(1.0436,2.4998)}\\
& pljava-1\_5\_2&38&34&15&78&5.8117&< \textbf{0.01*}
& (1.0392,2.4806)
\\
&\cc{pljava-latest release}&\cc{39}&\cc{35}&\cc{16}&\cc{76}&\cc{5.2928}&\cc{< \textbf{0.01}}
& \cc{(0.9600,2.3727)}\\
\hline 

\multirow{10}{*}{\begin{sideways}\realm\end{sideways}}
& realm-java-0.90.0&21&89&2&365&43.0617&< \textbf{0.01*}
& (2.2938,5.2315)
\\
&\cc{realm-java-1.2.0}&\cc{20}&\cc{169}&\cc{2}&\cc{285}&\cc{16.8639}&\cc{< \textbf{0.01*}}
& \cc{(1.3592,4.2912)}\\
& realm-java-2.3.2&33&177&3&269&16.7175&< \textbf{0.01*}
& (1.6194,4.0135)
\\
&\cc{realm-java-3.7.2}&\cc{43}&\cc{165}&\cc{8}&\cc{271}&\cc{8.8280}&\cc{< \textbf{0.01*}}
& \cc{(1.3988,2.9570)}\\
& realm-java-4.4.0&48&166&18&262&4.2088&< \textbf{0.01}
& (0.8616,2.0127)\\
&\cc{realm-java-5.4.0}&\cc{50}&\cc{165}&\cc{21}&\cc{261}&\cc{3.7662}&\cc{< \textbf{0.01}}
& \cc{(0.7804,1.8718)}\\
& realm-java-5.7.1&52&164&22&260&3.7472&< \textbf{0.01}
& (0.7856,1.8565)\\
&\cc{realm-java-5.9.0}&\cc{54}&\cc{161}&\cc{23}&\cc{260}&\cc{3.7915}&\cc{< \textbf{0.01}}
& \cc{(0.8066,1.8589)}\\
& realm-java-5.11.0&54&161&24&259&3.6195&< \textbf{0.01}
& (0.7668,1.8059)\\
&\cc{realm-java-5.15.0}&\cc{54}&\cc{162}&\cc{24}&\cc{258}&\cc{3.5833}&\cc{< \textbf{0.01}}
& \cc{(0.7569,1.7957)}\\
\hline 

\multirow{10}{*}{\begin{sideways}\vlc\end{sideways}}
& vlc-android-3.0.92&19&23&8&40&4.1304&< \textbf{0.01}
& (0.4460,2.3907)\\
&\cc{vlc-android-3.1.6}&\cc{22}&\cc{22}&\cc{6}&\cc{40}&\cc{6.6666}&\cc{< \textbf{0.01}}
& \cc{(0.8552,2.9390)}\\
& vlc-android-3.1.0&22&22&6&40&6.6666&< \textbf{0.01}
& (0.8552,2.9390)\\
&\cc{vlc-android-3.0.13}&\cc{18}&\cc{24}&\cc{7}&\cc{41}&\cc{4.3928}&\cc{< \textbf{0.01}}
& \cc{(0.4720,2.4879)}\\
& vlc-android-latest release&21&23&13&33&2.3177&0.081
& (-0.0322,1.7134)\\
&\cc{vlc-android-3.0.11}&\cc{19}&\cc{24}&\cc{6}&\cc{41}&\cc{5.4097}&\cc{< \textbf{0.01}}
& \cc{(0.6411,2.7352)}\\
& vlc-android-3.0.0&19&22&6&43&6.1893&< \textbf{0.01}
& (0.7709,2.8747)\\
&\cc{vlc-android-3.0.96}&\cc{19}&\cc{23}&\cc{8}&\cc{40}&\cc{4.1304}&\cc{< \textbf{0.01}}
& \cc{(0.4460,2.3907)}\\
& vlc-android-3.1.2&22&22&6&40&6.6666&< \textbf{0.01}
& (0.8552,2.9390)\\
\hline 

\multirow{10}{*}{\begin{sideways}\jpype\end{sideways}}
&\cc{jpype-0.5.4.5}&\cc{5}&\cc{28}&\cc{0}&\cc{45}&\cc{-}&\cc{< \textbf{0.02}}
& \cc{-} \\
& jpype-0.5.5.1&5&28&0&45&-&< \textbf{0.02}
& - \\
&\cc{jpype-0.5.5.4}&\cc{5}&\cc{28}&\cc{0}&\cc{45}&\cc{-}&\cc{< \textbf{0.02}}
& \cc{-} \\
& jpype-0.5.6&5&29&0&44&-&< \textbf{0.02}
& - \\
&\cc{jpype-0.5.7}&\cc{6}&\cc{28}&\cc{0}&\cc{44}&\cc{-}&\cc{< \textbf{0.01}}
& \cc{-} \\
& jpype-0.6.0&6&28&0&44&-&< \textbf{0.01}
& - \\
&\cc{jpype-0.6.1}&\cc{6}&\cc{28}&\cc{0}&\cc{44}&\cc{-}&\cc{< \textbf{0.01}}
& \cc{-} \\
& jpype-0.6.2&6&28&1&43&9.2142&< \textbf{0.05}
& (0.0509,4.3906)\\
&\cc{jpype-0.6.3}&\cc{6}&\cc{28}&\cc{3}&\cc{43}&\cc{3.0714}&\cc{0.158}
& \cc{(-0.3432,2.5875)}\\
& jpype-latest release&23&42&5&15&1.6428&0.430
& (-0.6362,1.6291)\\
\hline 
\multicolumn{9}{l}{* = significant p-values for odd ratios with confidence intervals not containing 1} \\ \hline
\end{tabular}

\end{table}

\begin{table}[t]
\centering\footnotesize
\caption{\label{tab:fisher2} Fisher's Exact Test Results for the Fault-proneness of Files with and without Design Smells (2)
}
\begin{tabular}{llccccccc}
	\hline
\rowcolor{gray!35}
System & Releases & SB& BNS &  SNB &  NBNS &  Odds Ratios & \textit{p}-values & Confidence Interval\\\hline
\multirow{10}{*}{\begin{sideways}\javacpp\end{sideways}}
& javacpp-0.5&0&9&0&5&-&1.0
& - \\
&\cc{javacpp-0.9}&\cc{10}&\cc{4}&\cc{3}&\cc{3}&\cc{2.5}&\cc{0.612} & \cc{(-1.0599,2.8926)}\\
& javacpp-1.1&0&10&0&4&-&1.0
& - \\
&\cc{javacpp-1.2}&\cc{9}&\cc{5}&\cc{5}&\cc{2}&\cc{0.72}&\cc{1.0}
& \cc{(-2.2994,1.6424)}\\
& javacpp-1.2.1&12&5&2&2&2.4&0.574
& (-1.3449,3.0958)\\
&\cc{javacpp-1.2.7}&\cc{7}&\cc{4}&\cc{7}&\cc{3}&\cc{0.75}&\cc{1.0} & \cc{(-2.1148,1.5395)}\\
& javacpp-1.3&10&1&4&6&15.0&< \textbf{0.05}
& (0.2942,5.1218)\\
&\cc{javacpp-1.3.2}&\cc{11}&\cc{6}&\cc{3}&\cc{1}&\cc{0.6111}&\cc{1.0} & \cc{(-2.9646,1.9797)}\\
& javacpp-1.4&12&5&4&2&1.2&1.0
& (-1.8101,2.1747)\\
&\cc{javacpp-1.4.2}&\cc{14}&\cc{4}&\cc{3}&\cc{3}&\cc{3.5}&\cc{0.306} & \cc{(-0.6955,3.2011)}\\
& javacpp-1.4.4&11&2&8&4&2.75&0.378
& (-0.9147,2.9379)\\
&\cc{javacpp-1.5}&\cc{14}&\cc{5}&\cc{5}&\cc{1}&\cc{0.56}&\cc{1.0}& \cc{(-2.9573,1.7977)}\\
& javacpp-1.5.1-1&10&4&9&2&0.5555&0.660
& (-2.5093,1.3337)\\
\hline 

\multirow{10}{*}{\begin{sideways}\zstd\end{sideways}}

&\cc{zstd-jni-0.4.4}&\cc{4}&\cc{0}&\cc{0}&\cc{21}&\cc{-}&\cc{< \textbf{0.01}} & \cc{-} 
\\
& zstd-jni-1.3.0-1&13&0&9&15&-&< \textbf{0.01}
& - \\
&\cc{zstd-jni-1.3.2-2}&\cc{15}&\cc{2}&\cc{7}&\cc{13}&\cc{13.9285}&\cc{< \textbf{0.01}} & \cc{(0.8958,4.3721)}
\\
& zstd-jni-1.3.3-1&16&2&7&13&14.8571&< \textbf{0.01} & (0.9649,4.4320)
\\
&\cc{zstd-jni-1.3.4-1}&\cc{20}&\cc{1}&\cc{8}&\cc{12}&\cc{30.0}&\cc{< \textbf{0.01*}}
& \cc{(1.2025,5.5998}\\
& zstd-jni-1.3.4-8&20&1&8&12&30.0&< \textbf{0.01*}
& (1.2025,5.5998)\\
&\cc{zstd-jni-1.3.5-3}&\cc{20}&\cc{1}&\cc{8}&\cc{12}&\cc{30.0}&\cc{< \textbf{0.01*}}
& \cc{(1.2026,5.5999)}\\
& zstd-jni-1.3.7&20&1&8&12&30.0&< \textbf{0.01*}
& (1.2026,5.5998)\\
&\cc{zstd-jni-1.3.8-1}&\cc{20}&\cc{1}&\cc{8}&\cc{12}&\cc{30.0}&\cc{< \textbf{0.01*}}
& \cc{(1.2026,5.5998)}\\
& zstd-jni-1.4.0-1&22&1&7&12&37.7142&< \textbf{0.01*}
& (1.4198,5.8403)\\
&\cc{zstd-jni-latest release}&\cc{22}&\cc{1}&\cc{7}&\cc{12}&\cc{37.7142}&\cc{< \textbf{0.01*}}
& \cc{(1.4198,5.8403)}\\
\hline 

\multirow{10}{*}{\begin{sideways}\conscrypt\end{sideways}}
& conscrypt-1.1.1&42&52&12&46&3.0961&< \textbf{0.01}
& (0.3758,1.8844)\\
&\cc{conscrypt-1.0.0.RC14}&\cc{4}&\cc{64}&\cc{0}&\cc{54}&\cc{-}&\cc{0.128} & \cc{-} \\
& conscrypt-1.0.1&38&55&11&43&2.7008&< \textbf{0.02}
& (0.2128,1.7743)\\
&\cc{conscrypt-2.1.0}&\cc{47}&\cc{53}&\cc{17}&\cc{42}&\cc{2.1908}&\cc{< \textbf{0.05}}
& \cc{(0.0975,1.4711)}\\
& conscrypt-1.0.2&38&55&11&43&2.7008&< \textbf{0.02}
& (0.2128,1.7742)\\
&\cc{conscrypt-1.4.2}&\cc{6}&\cc{0}&\cc{55}&\cc{97}&\cc{-}&\cc{< \textbf{0.01}}
& \cc{-} \\
& conscrypt-1.2.0&45&52&15&46&2.6538&< \textbf{0.01}
& (0.2697,1.6823)\\
&\cc{conscrypt-1.0.0.RC11}&\cc{37}&\cc{55}&\cc{11}&\cc{43}&\cc{2.6297}&\cc{< \textbf{0.02}}
& \cc{(0.1844,1.7493)}\\
& conscrypt-1.0.0.RC2&23&20&6&90&17.25&< \textbf{0.01*}
& (1.8270,3.8686)\\
&\cc{conscrypt-1.0.0.RC8}&\cc{26}&\cc{59}&\cc{11}&\cc{46}&\cc{1.8428}&\cc{0.172} & \cc{(-0.1922,1.4148)}\\
\hline 

\multirow{5}{*}{\begin{sideways}\javasmt\end{sideways}}
& java-smt-0.60&0&23&0&7&-&1.0 & - \\
&\cc{java-smt-1.0.1}&\cc{21}&\cc{20}&\cc{5}&\cc{9}&\cc{1.89}&\cc{0.3667} & \cc{(-0.6165,1.8896)}\\
& java-smt-2.0.0-alpha&22&16&11&12&1.5&0.5966 & (-0.6357,1.4467)\\
&\cc{java-smt-2.2.0}&\cc{30}&\cc{19}&\cc{9}&\cc{10}&\cc{1.7543}&\cc{0.4132} & \cc{(-0.5061,1.6304)}\\
& java-smt-3.0.0&19&17&22&12&0.6096&0.3414 & (-1.4556,0.4658)\\
\hline
\multicolumn{9}{l}{* = significant p-values for odd ratios with confidence intervals not containing 1} \\ \hline
\end{tabular}

\end{table}


We, therefore, conclude that, in most cases, there is a relation between multi-language design smells and fault-proneness in the context of JNI systems: a greater proportion of JNI files participating in design smells experienced bugs compared to other classes. We therefore reject $H_{3}$. The rejection of $H_{3}$ and the statistically significant odds ratios provide a \textit{posteriori} concrete evidence of the impact of multi-language design smells on fault-proneness in the context of JNI files. 


\begin{tcolorbox}
\textit{\textbf{Summary of findings (RQ3)}}:
Our results suggest that files with occurrences of the studied smells are more likely to be associated with faults than files without these smells and this relationship is statistically significant in most cases.
\end{tcolorbox}

\begin{landscape}
\begin{table*}
\caption{Log Likelihood of Different Smells from the Logistic Regression models for Bug-proneness of the Studied Systems}\label{tab:LR-details}
\scalebox{0.62}{
{\renewcommand{\arraystretch}{1.15}
\begin{tabular}{lrcrccrccrccrccrccrccrccrcc} \hline
\multirow{2}{*}{\textbf{Design Smells} $\downarrow$  / \textbf{Systems} $\rightarrow$}  & \multicolumn{2}{c}{\conscrypt} & \multicolumn{2}{c}{\javasmt}  & \multicolumn{2}{c}{\javacpp} & \multicolumn{2}{c}{\jpype}   & \multicolumn{2}{c}{\pljava}  & \multicolumn{2}{c}{\realm} & \multicolumn{2}{c}{\rocksdb} & \multicolumn{2}{c}{\vlc}   & \multicolumn{2}{c}{\zstd}  \\ \cline{2-19}

& Coeff. & Rank & Coeff. & Rank  & Coeff. & Rank  & Coeff. & Rank  & Coeff. & Rank & Coeff. & Rank  & Coeff. & Rank  & Coeff. & Rank  & Coeff. & Rank  \\ \hline

\begin{tabular}[c]{@{}l@{}}Excessive Inter-language\\  Communication\end{tabular} & \textbf{-9.538e+15} &          & NA &        & \textbf{-1.765e+14} &        & -5.380e+01 &        & -4.649e+01 &         & 4.495 & 2        & 1.640e+01 & 4        & \textbf{-7.653e+13} &       & \textbf{-4.265e+13} &        \\
\rowcolor{gray!15}
Too Much Clustering  & \textbf{3.280e+15}  & 1        & -6.864e+03 &        & \textbf{3.280e+15}  & 1        & NA  &        & -5.488e+02 &         & -9.514e+01 &        & 8.593e+02 & 3        & \textbf{1.265e+15} & 2        & \textbf{2.949e+14} & 3        \\
Too Much Scattering  & NA &        & NA &        & \textbf{2.304e+14} & 4        & NA &        & 1.492e+02 & 1        & 9.310e+01 & 1        & 1.324e+03  & 2        & \textbf{3.438e+15} & 1        & \textbf{1.077e+15} & 2        \\
\rowcolor{gray!15}
Unused Method Declaration   & \textbf{-6.275e+13} &        & 2.846e+01 & 1        & \textbf{9.799e+13} & 6        & NA &        & 4.883e+01 & 2        & -6.092e+01 &        & -6.785e+01 &         & \textbf{-6.359e+12} &         & \textbf{-2.448e+13} &         \\
Unused Method Implementation  & NA &        & NA &        & NA &        & NA &        & -3.446e+02 &        & -2.288e+02 &        & NA &        &\textbf{-5.237e+14} &        & \textbf{9.511e+13} & 5        \\
\rowcolor{gray!15}
Unused Parameter     & \textbf{8.145e+13}  & 3        & 8.497 & 2        & \textbf{7.421e+14} & 2        & -5.527e-02 &     & -7.090e-02 &        & 4.828  & 3        & -2.286e+01 &        & \textbf{3.363e+13} &3        & \textbf{2.092e+13} &7        \\
Excessive Objects    & NA &        & NA &        & NA &        & NA &        & NA &        & NA &        & NA &        & NA &        & NA &        \\
\rowcolor{gray!15}
Not Handling Exceptions & \textbf{2.089e+14} & 2        & NA &        & NA &        & 1.993e+01 & 1       & -1.999e+02 &        & -9.445e+01 &        & -4.270e+01 &        & \textbf{-2.951e+15} &        & \textbf{2.373e+14} &  4      \\
Not Caching Objects  & NA &        & NA &        & NA &        & NA &        & NA &       & NA &        & NA &        & NA &        & NA &        \\
\rowcolor{gray!15}
Not Securing Libraries & \textbf{-2.915e+13}  &         & 1.093 &3        & \textbf{3.308e+14} & 3        & NA &        & NA &        & -9.201e+01 &       & -2.498e+03 &        & \textbf{-8.185e+14} &         & \textbf{-1.569e+15} &        \\
Hard Coding Libraries  & NA &        & -8.665e+01 &        & \textbf{1.235e+14} &5        & NA &        & NA &        & NA &        & 4.936e+03 & 1        & NA &        & \textbf{2.135e+15} & 1        \\
\rowcolor{gray!15}
Memory Management Mismatch  & \textbf{-1.180e+15} &        & NA &        & NA &        & 1.162e-01 & 2       & NA &        & NA &        & -1.393e+03 &        & NA &        & \textbf{4.684e+13} & 6       \\
Local References Abuse & \textbf{-1.301e+15} &        & NA &        & NA &        & -2.462e+02 &        & NA &       & -1.891e+02 &        & -3.474e+04 &       & \textbf{-8.620e+15} &        & NA &   \\ \hline
\rowcolor{gray!15}
LOC & \textbf{-8.459e+11} &        &-4.280e-03 &        &\textbf{1.159e+11}  &        &8.468e-05  &        & 5.328e-04 &       & 2.488e-03 &        &8.713e-01  &       & \textbf{1.091e+11} &        &\textbf{5.331e+10}  &   \\ 
Previous bug-fix & \textbf{6.363e+14} &        & 1.864e+02 &        & \textbf{1.293e+15}  &        & 5.728e+01 &        & 3.010e+02 &       &  9.480e+01&        &1.448e+03  &       & \textbf{7.873e+14} &        &\textbf{7.437e+14}  &   \\ \hline \hline 
\rowcolor{gray!15}
Null deviance &2034.3 & &5.2010e+02 & & 349.15& & 1.0902e+03& &2.3975e+03 & &6.6378e+03 & &4048.24 &  &1245.10 & &589.62 & \\
Residual deviance & 1513.8 &  & 2.7256e-09& & 2595.14& & 3.5622e-10&  & 4.8079e-10& &8.8774e-10 & & 8.5475& &360.44 &  &2595.14 & \\
\rowcolor{gray!15}
AIC & 1535.8 & & 16&  & 2615.1 & &16 & &20 & & 24& &34.55 & & 384.44& & 1107.3& \\ \hline
\end{tabular}
}}
\end{table*}
\end{landscape}

\subsection{\textbf{RQ4: \textbf{\RQFour} }}
Findings from RQ3 suggest that source code files with smells in JNI systems are often more prone to faults than files without smells. Although these findings give a general impression of the impacts of smells on the fault-proneness of JNI systems, it is important to know which smell(s) are more related to faults. When we are able to identify some specific smells to be more related to faults, we can prioritize those smells during the maintenance of the JNI systems. 

\newtext{As presented in our methodology described in Section \ref{sec:smell_impact}, we apply multivariate logistic regression to examine whether some types of design smells are more related to fault-proneness. In our logistic regression models, independent variables are the number of occurrences of each type of design smells. 
The dependent variable is a dichotomous flag (\textit{buggy}) that assumes values either 0 (non-buggy) or 1 (buggy). For each system, we build a logistic regression model and analyze the model coefficients and p-values for individual types of smells. To address multicollinearity among the independent variables, we drop one of the variables from each highly correlated pair of variables from the models.} From our analysis, we observed two pairs of smells highly correlated- (\textit{Not Handling Exceptions, Assuming Safe Return Value}) and (\textit{Not Securing Libraries, Not Using Relative Path}) \newtext{with correlation (Spearman's) coefficients of 0.91 and 0.60, respectively}. We keep \textit{Not Handling Exceptions} and \textit{Not Securing Libraries} as they are more prevalent in the systems compared to the other smell in each correlated pair. Similarly, we drop the variable code churn from the model as we found it to be highly correlated (0.99) with the file size (LOC). \newtext{We chose Spearman's rank correlation as it is non-parametric and does not require data to be normally distributed.}


We rank the independent variables based on the logistic regression model coefficients (log odds) and the corresponding p-values. Table \ref{tab:LR-details} presents the model coefficients and their ranking for each system. The coefficients with significant p-values (<0.01) are presented in boldface. 
To evaluate the relationships of individual types of smells with bug-proneness, we summarize the data from Table \ref{tab:LR-details} into Table \ref{tab:LR-Smell-wise} to identify the top five smell types that are more related to bugs. For each smell type, Table \ref{tab:LR-Smell-wise} presents the percentage of systems where the smell type has positive log odds, the number of times the smell type is in the top five in the ranking of positive log odds, and the number of systems where the log odds are statistically significant. For each smell, we consider only the system where we are able to calculate the model coefficients and thus we exclude the systems where we do not get the coefficient values due to singularities. For the smell \cluster~ for example, in Table \ref{tab:LR-details}, for eight (8) out of nine (9) systems (\ie except \jpype) we have values for model coefficients. 
Out of these eight systems, for five (5) systems (\conscrypt, \javacpp, \rocksdb, \vlc, and \zstd~ \ie 5/8 (62.5\%) times) \cluster~ has positive values for log odds. All of these five times (systems) the log odds were ranked in the top five, having significant p-values in four systems (\conscrypt, \javacpp, \vlc, and \zstd). For all smell types in Table \ref{tab:LR-details}, we present such summary in Table \ref{tab:LR-Smell-wise}. We then report the top five smell types \newtext{(underlined in Table \ref{tab:LR-Smell-wise})} based on the percentage of systems in which the smell have positive log odds, number of times the positive log odds were in the top five ranking, and the number of systems in which the log odds have significant p-values respectively as shown in Table \ref{tab:LR-Smell-wise}. \newtext{For the control variables LOC and the number of previous bug-fix, 
we observed positive log odds in most of the systems. So, these coefficients for the control factors agree with the known impacts of these two variables on fault-proneness. We also observed negative log odds for the smells from the logistic regression models for the studied systems. The negative regression coefficients might be interpreted as an 
indication that the corresponding smells are negatively related to fault-proneness. However, this scenario varies across the studied systems.}  

\begin{center}
	\begin{table*}
		\centering
		\caption{Fault-proneness of Different Types of Smells Based on Logistic Regression Analysis}\label{tab:LR-Smell-wise}
		\begin{tabular}{lccc} \hline
		\multirow{2}{*}{Smell Types} & \multicolumn{3}{c}{Number and Percentage of Systems}\\\cline{2-4}
             & LO > 0 & LO in Top 5 & \newtext{(LO>0 and p<0.01)} \\ \hline
			\excessivecom& 25\%(2/8)& 2 & 0\\
		    \rowcolor{gray!15}
			\cluster& \underline{62.5\%}(5/8)& 5 & 4\\
			\scatter& \underline{100\%}(6/6)& 6 & 3\\
		    \rowcolor{gray!15}
			\declaration& 37.5\%(3/8)& 2 & 1\\
			\implementation& 25\%(1/4) & 1& 1\\
		    \rowcolor{gray!15}
			\unusedParameters& \underline{66.6\%}(6/9)& 5 & 4\\
			\exception& 42.8\%(3/7) & 3& 2\\
			\rowcolor{gray!15}
			\notSecuringLib& 28.5\%(2/7)& 2& 1\\
			\hardCodingLib& \underline{75\%}(3/4) & 3 & 2\\
		    \rowcolor{gray!15}
			\memoryMagement& \underline{50\%}(2/4) & 1& 1\\
			\localReference& 0\%(0/5) & 0 & 0\\
			\rowcolor{gray!15}
			\excessiveObject&NA&NA&NA \\
			\cachingObjects& NA& NA& NA\\\hline
			\multicolumn{4} {l}{\newtext{\textbf{LO} = Log Odds of the corresponding smell from the logistic regression model.}} \\ 
			\multicolumn{4} {l}{\newtext{\textbf{NA} = Corresponding Log odds are not available from the LR models due to singularities}} \\\hline
		\end{tabular}
	\end{table*}
\end{center}

\begin{center}
	\begin{table*}
		\centering
		\caption{Fault-proneness of Different Types of Smells Based on Logistic Regression Model for All Systems}\label{tab:LRAll}
		\begin{tabular}{lcc} \hline
		    Smell Types & Log Odds & \textit{p}-values\\\hline
			\excessivecom& 2.985e-01& 1.11e-07 (<0.01)\\
		    \rowcolor{gray!15}
			\cluster& 4.262e+00& 0.000898 (<0.01)\\
			\scatter& 8.359e+00& 1.82e-15 (<0.01)\\
		    \rowcolor{gray!15}
			\declaration&9.078e-01 & < 2e-16 (<0.01)\\
			\implementation& -1.255e+01 & 0.894915 \\
		    \rowcolor{gray!15}
			\unusedParameters& 5.695e-01& < 2e-16 (<0.01)\\
			\exception& 3.248e+00 & 0.000217 (<0.01)\\
			\rowcolor{gray!15}
			\notSecuringLib& -4.469e-01& 0.813090 \\
			\hardCodingLib& 1.841e+00 & 0.546194\\
		    \rowcolor{gray!15}
			\memoryMagement& -9.255e+00 & 0.998258\\
			\localReference& -1.172e+01 & 0.999968\\
			\rowcolor{gray!15}
			\excessiveObject&NA&NA \\
			\cachingObjects& NA& NA\\\hline
			\multicolumn{3} {l}{\newtext{\textbf{NA} = Corresponding Log odds are not available from the LR models due to singularities}} \\\hline
		\end{tabular}
	\end{table*}
\end{center}
As shown in Table \ref{tab:LR-details}, the log odds of the independent variables vary across the systems. In four of the systems (\conscrypt, \javacpp, \vlc, and \zstd), we observe that the log odds for the smells are statistically significant (<0.01). These four systems reject the hypothesis $H_4$, meaning that different smells have different impacts on fault-proneness. However, we cannot generalize it to other systems to have a concrete conclusion. Thus, given the varying log odds of the smells \newtext{from our regression models for individual systems}, we conclude that the relationships between different types of multi-language smells and fault-proneness are system dependent. 

Given that we have limited evidence to draw a firm conclusion on the strength of the relationships between different types of smells and fault-proneness, we focus on identifying smells that are relatively more related to faults based on the ranking of the values of the log odds and their significance. 
In Table \ref{tab:LR-Smell-wise}, the smell type \cluster~ has positive log odds in 62.5\% (5/8) of the systems. Each time, log odds were among the top 5 and was statistically significant in three systems. Similarly, \scatter, \unusedParameters, \hardCodingLib~ and \memoryMagement~ are among the top five smells with positive log odds in 100\%(6/6), 66.6\%(6/9), 75\%(3/4), and 50\%(2/4) systems, respectively. These smells are likely to have a strong relation with fault-proneness. Besides, the smells \exception, \declaration~ and \notSecuringLib~ have significant positive log odds for 2 (\conscrypt, \zstd), 1 (\javacpp), and 1 (\javacpp) system(s), respectively; indicating some degree of relation with fault-proneness. \excessivecom~ and \localReference~which have no significant positive log odds are less likely to be associated with faults. 

\newtext{We also build a single logistic regression model for all the systems combined to evaluate how the findings from individual systems generalize. We presented the regression results for the smells in Table \ref{tab:LRAll}. We observed that smells \excessivecom, \cluster, \scatter, \declaration, \unusedParameters, and \exception~have positive log odds with significant p-values (<0.01). This is an indication that these smells have statistically significant relationships with fault-proneness. This finding corroborates 
our findings from the analysis of individual systems for most cases. However, we did not observe significant relationships between multi-language smells and fault-proneness for the remaining smell types (in the models for all systems combined). 
Now, if we consider positive log odds with significant p-values in the  logistic regression model for all systems combined (Table \ref{tab:LRAll}) and the percentage of positive log odds for regression models for individual systems (Table \ref{tab:LR-Smell-wise}), we observe that the smell types \cluster, \scatter, \unusedParameters, \exception, and \hardCodingLib~are  
the most related to fault-proneness. However, this relationship varies 
with systems.}  
One important point to note is the fact that smells suggested by our empirical results to be more related to fault-proneness constitute roughly over 80\% of the smells in the studied systems (details in Table \ref{tab:RQ2}). This further shows that it is important to detect and remove these smells from the systems as soon as possible. 

\newtext{To study the relationships between multi-language smells and fault-proneness, we also investigate the correlation (Spearman's) 
between the number of smells of individual types in a file and the number of bugs associated with the corresponding file.
We observed that \relativePath~(0.48), \exception~(0.32), \excessivecom~(0.30), \localReference~(0.24), \hardCodingLib~(0.19), and \cluster~(0.18) are the top smells based on the correlation with faults, although most of these and the remaining correlations are weak. Also, we mentioned that \relativePath~was dropped from our logistic regression models because of its high 
correlation with \notSecuringLib. So, we cannot draw a firm conclusion on the impacts of smells on fault-proneness based on these correlation results.} 

To have better insights into the identified relationships between the different types of JNI smells and bugs and to understand the bug-smell contexts in the studied systems, we further manually investigated a random sample of commit messages associated with bugs. 
From analyzing these commit messages, we found some commit messages clearly suggesting that some specific smells are often related to bugs. For example, in the release \textit{1.0.0.RC14} of \conscrypt, a commit message is clearly specifying errors related to the code smell \unusedParameters~ (\emph{``Our Android build rules generate errors for unused parameters. We cant enable the warnings in the external build rules because BoringSSL has many unused parameters''}). The same goes for \memoryMagement, in \realm, a commit message was discussing errors related to memory management \emph{``DeleteLocalRef when the ref is created in loop (\#3366) Add wrapper class for JNI local reference to delete the local ref after using it''}. Another example from \conscrypt~ discussing bugs related to native memory management \emph{``This fixes a memory leak in NativeCrypto\_i2d\_PKCS7. It never frees derBytes''}. The smell \exception~ is also discussed as related to the bug 3482 in \realm~ (\emph{``Add cause to RealmMigrationNeededException (\#3482)''}). \vlc~ also presents bugs related to the smell \exception~  \emph{``rework exceptions throwing from jni''}. Similarly, other commit messages were also describing bugs related to \declaration. \emph{``There were a bunch of exceptions that are being thrown from JNI methods that aren't currently declared''}, and \emph{``Fix latent bug in unused method''} present examples extracted respectively from \conscrypt~ and \pljava. Thus, we see that our identified smells are often related to bugs which highlight practical contexts and usability of our findings. 

We also analyzed some quality attributes of our models such as \textit{Null deviance, Residual deviance, and Akaike's Information Criterion (AIC)} as presented in Table \ref{tab:LR-details}. We observe that there are larger differences between null deviance and the residual deviance for the models of all the systems indicating a good fit of the regression models. We observe lower values for AIC for most of the regression models indicating the simplicity of the models with comparatively higher values for \conscrypt, \javacpp, \vlc, and \zstd.

\begin{tcolorbox}
\textit{\textbf{Summary of findings (RQ4)}}: We conclude that although not always significant, there exists a relation between types of smells and the fault-proneness. The relationship is not consistent for all types of smell and across all the systems. 
Smell types \textit{Too much Scattering, Too Much Clustering, Unused Parameters, Hard Coding Libraries, and Not Handling Exceptions} are observed to be more related to faults compared to other smells, and thus should be prioritized during maintenance.
\end{tcolorbox}

\subsection{\textbf{RQ5: \textbf{\RQFive} }} \label{sec:RsltTopic}

Since the risk of having bugs could differ from one activity to the other, we decided to investigate what kind of activities once performed in smelly files could increase the risk of bug occurrences. Having knowledge of risky activities developers and maintainers could reduce the risk of bugs in smelly files. 
To study the activities that could introduce bugs in smelly files, we collected the fault-inducing commits messages and performed a topic modeling, combining a mix of manual and automatic approaches as described in Section \ref{sec:Study}. 

Table \ref{tab:topics} lists 12 activities that are more likely to introduce bugs in smelly files. For each activity the table lists the systems from which the activity was extracted. For each activity, we also present examples of keywords used to build the topic for that activity. For example, the activity memory management, was extracted using a set of keywords including: buffer, memory, leak, flush, reference, local, memtable from \rocksdb, \realm, \conscrypt, and \jpype. \emph{``Add more numbers to float-conversion test, add new unit-test for float-conversion''}, is an example of a commit message describing data conversion activity when the bug was introduced, extracted from \javasmt. Another example of commit message that introduced bugs extracted from \rocksdb: \emph{``Another change is to only use class BlockContents for compressed block, and narrow the class Block to only be used for uncompressed blocks, including blocks in compressed block cache''}. This commit message is related to compression activities. From \zstd, the following commit messages \emph{``expose faster API to allow re-using of dictionaries''} refers to the usage of APIs. Those activities were extracted from commit messages of fault-inducing commits. Developers performed those activities in files containing occurrences of multi-language design smells when the bug was introduced. 

\begin{table}
\centering\footnotesize
\caption{\label{tab:topics} Activities Introducing Bugs in Smelly Files}
{\renewcommand{\arraystretch}{1.15}
\begin{tabular}{p{0.3cm}p{3cm}p{6.5cm}p{2.9cm}}
	\hline
    \rowcolor{gray!35}
	\textbf{No} &  \textbf{Activities}&  \textbf{Example of Keywords} &  \textbf{Systems} \\\hline
 1 & Compression tasks & compression, decode, memory, blocks, encode, compaction, streaming, frames, block, dictionary. & \rocksdb, \zstd\\
 \rowcolor{gray!15}
2  & Data conversion & parser, type, container, basic, declaration, write, invalid, convert, string, coverage.  & \rocksdb, \realm, \pljava, \javacpp, \conscrypt, \javasmt, \zstd\\
3 & Memory management & buffer, messagesize, memory, leak, local, reference, flush, memtable, allocation, garbage. & \rocksdb, \realm, \conscrypt, \pljava, \jpype, \javacpp \\
\rowcolor{gray!15}
4 & Restructuring the code & add, update, remove, code, reorder, native, move, public, improve, change. &\rocksdb, \vlc, \conscrypt, \jpype, \pljava\\
5 & Database management & stored, db, database, persistence, key, data, visible, size, file, timestamp.  &\rocksdb, \pljava\\
\rowcolor{gray!15}
6 & API usage & external, library, api, include, expose, public, integrate, allow, streaming, wrapper.  & \vlc, \realm, \conscrypt, \rocksdb, \pljava, \zstd,\javasmt\\
7 & Feature migration 	 & upgrade, support, migrate, integrate, create, legacy, simplify, add, format, update.  &\realm, \javacpp, \rocksdb, \javasmt, \zstd\\
\rowcolor{gray!15}
8 & Network management & sslsocket, encrypt, socket, nativessl, token, hostname, protocol, platform, sslSession, activesession. &  \conscrypt, \rocksdb\\
9 & Exception management & occur, handle, check, exception, throw, return, fix, pointer, illegal, runtime.   &\conscrypt, \javacpp, \jpype, \javasmt\\
\rowcolor{gray!15}
10  & Threads management & thread, pull, execution, reflect, client, transaction, monitor, notify, mutex, log. & \pljava, \rocksdb, \realm\\
11 & Performance management  & time, wrap, performance, execution-time, regression, cache, shared, resources, bundle, increase.  &  \zstd, \rocksdb\\
 \rowcolor{gray!15}
12 & Compiler management & compiler, resolve, failure, check, warnings, support, JNI\_ABORT, error, illegal, dynamic. & \jpype, \rocksdb \\
 \hline 
\end{tabular}
}	
\end{table}

From analyzing the commit messages and topics of activities, we found that 
activities related to data conversion, memory management, API usage, code restructuring, and exception management are the most common activities that could increase the risk of bugs when performed in smelly files. Activities related to the compiler management, threads management, and compression tasks could also induce bugs in smelly files. 

To understand why these activities seem to be risky, we decided to investigate further these activities at the source code level. For example, zstd.java is a native class from \zstd~  system. This class contains 1351 lines of code with 75 native methods and exhibits the smell \cluster. This class combines methods performing distinct responsibilities, \ie compression, decompression, computation, data access, and utility methods. As per commit messages identified as introducing bugs, the developer was reordering the statics methods, adding JNI wrappers, and performing compression tasks ``Reorder the static methods, All compression first then all decompression then the rest,  inputs checking + utility methods'', ``Add Java wrappers and C implementations of compress/decompress using direct ByteBuffer''. This class contains two types of smells, \cluster~and \excessivecom. The nature of those two smells is by definition adding complexity to the code by making the readability of such classes hard. Thus, restructuring the code of a large native class could have increased the risk of introducing bugs. Indeed, applying changes on multi-language code could bring some confusion if the developer is not familiar with the components involved in the multi-language interaction. Similarly, activities related to compression are declared in Java side and mainly implemented in the C/C++ side (jni\_zstd.c). Developers should have knowledge of both implementations to correctly perform a change, especially that this class contains \excessivecom~between Java and C/C++ when performing compression activities. Another example is illustrated by Listing \ref{fig:exmpExceptionbug}. It presents a function extracted from the C file jni\_zdict.c from \zstd~system. A developer was ``adding support for legacy dictionary trainer'' in this smelly function when the bug was introduced. However, in the context of JNI, it is important to always perform checks to ensure that the native execution was performed correctly. As described in Section \ref{sec:ML-smells}, when checking JNI exceptions, we should add a return statement just after throwing the excepting to interrupt the execution flow and exit the method in case of errors. The \texttt{ThrowNew()} functions do not interrupt the control flow of the native method. In case an error occurred when retrieving the jclass, the exception will not be thrown in the JVM until the native method returns. Developers should be aware of how to implement the exception in the context of JNI systems to avoid introducing bugs related to mishandling JNI exceptions. Activities related to the conversion of types could also introduce bugs as expressed by a commit message extracted from \javacpp~; \ie ``Provide `BytePointer' with value getters and setters for primitive types other than `byte' to facilitate unaligned memory accesses''.

Another example of bugs related to the management of the memory is extracted from \pljava, c source code file JNICalls.c, ``Eliminate threadlock ops in string conversion''. Both of those files exhibit the smell \memoryMagement. Activities related to data and type conversion could increase the risk of bug because when converting types from Java to C/C++, the conversion will raise two categories of types; primitive types and reference types. Primitive types are simple to convert, we usually add j in front of the type \eg \texttt{int} become \texttt{jint}, \texttt{float} become \texttt{jfloat}, etc. However, for the reference types \ie Class, Object, String, developers should use the predefined method to correctly perform the conversion. However, it happens that they forget to release the memory after such conversion which could introduce additional bugs including memory leaks. Listing \ref{fig:exmpmemory} presents an example extracted from \pljava~as introducing bugs. In this example, the method \texttt{GetObjectArrayElement} is used to capture a Java array. However, the memory is not released after usage as done in Listing \ref{fig:exmpExceptionbug}. From the above examples, we conclude that some specific types of activities are relatively more frequently associated with bugs, especially in the context of multi-language design smells. Developers should be cautious while performing those activities. 


\begin{figure}[ht]
\center

\lstset{backgroundcolor = \color{gray!4},language=C,basicstyle=\small\ttfamily, showspaces=false, showstringspaces=false,breaklines=true, morekeywords={ReleaseStringUTFChars}}
\begin{lstlisting}  [language=C, escapechar=@, caption={Example of Bug in Smelly Method 1/2}, label={fig:exmpExceptionbug}]
/* 
...
*/
 jsize num_samples = (*env)->GetArrayLength(env, sampleSizes);
    jint *sample_sizes_array = (*env)->GetIntArrayElements(env, sampleSizes, 0);
    size_t *samples_sizes = malloc(sizeof(size_t) * num_samples);
    if (!samples_sizes) {
        jclass eClass = (*env)->FindClass(env, "Ljava/lang/OutOfMemoryError;");
       @\bh@(*env)->ThrowNew(env, eClass, "native heap");@\eh@
    }
    for (int i = 0; i < num_samples; i++) {
        samples_sizes[i] = sample_sizes_array[i];
    }
   (*env)->ReleaseIntArrayElements(env, sampleSizes, sample_sizes_array, 0);

\end{lstlisting}
\end{figure}

\begin{figure}[ht]
\center

\lstset{backgroundcolor = \color{gray!4},language=C,basicstyle=\small\ttfamily, showspaces=false, showstringspaces=false,breaklines=true, morekeywords={ReleaseStringUTFChars}}
\begin{lstlisting}  [language=C, escapechar=@, caption={Example of Bug in Smelly Method 2/2}, label={fig:exmpmemory}]
/*
... 
*/
	jsize idx;
	jboolean foundNull = JNI_FALSE;
	BEGIN_JAVA
	idx = (*env)->GetArrayLength(env, array);
	while(--idx >= 0)
	{	@\bh@if((*env)->GetObjectArrayElement(env, array, idx) != 0)@\eh@
			continue;
		foundNull = JNI_TRUE;
		break;
	}
	END_JAVA
	return foundNull;
}
\end{lstlisting}
\end{figure}



\begin{tcolorbox}
\textit{\textbf{Summary of findings (RQ5)}}: Activities related to data conversion, memory management, code restructuring, API usage, and exception management are the most common activities that could increase the risk of bugs once performed in smelly files, and thus should be performed carefully.
\end{tcolorbox}

\section{Discussion}
\label{sec:Discussion}

This section discusses the results reported in Section \ref{sec:Results}.

\subsection{Multi-language Design Smells}
\paragraph{\textbf{Detection of Smells}} We used srcML parser due to its ability to provide a single xml file combining source code files written in more than one programming language. Languages supported in the current version of srcML include Java, C, C++, and C\#.\footnote{\url{https://www.srcml.org/about.html}}  However, this could be extended to include other programming languages \cite{collard2013srcml}. \newtext{The detection approach presents some limitations. The recall and precision vary depending on the type of design smells and mainly on the naming convention used to implement the JNI projects. For the smell \declaration, we are missing some occurrences due to the syntax used in the C implementation that is not completely following the JNI naming convention (\eg \pljava~ \textit{jobject pljava\_DualState\_key}). For \localReference, we are not considering situations in which predefined methods could be used to limit the impact of this design smell,  \ie 
PushLocalFrame\footnote{\url{https://docs.oracle.com/javase/7/docs/technotes/guides/jni/spec/functions.html\#PushLocalFrame}}, and PopLocalFrame.\footnote{\url{https://docs.oracle.com/javase/7/docs/technotes/guides/jni/spec/functions.html\#PopLocalFrame}} These methods were excluded because by a manual validation when defining the smells, we found that those methods do not always prevent occurrences of the design smells and inclusion of those may result in false negatives. Our detection approach also presents some limitations in the detection of \relativePath, particularly in situations where the path could be retrieved from a variable or concatenation of strings. However, this was not captured as a common practice in the analyzed systems. We refined our detection rules to favor the recall over precision, as was done for smells detection approaches for mono-language systems \cite{moha2009decor,gueheneuc2008demima}. However, by refining some rules as explained earlier for the smell \localReference, and mainly due to some situations that are not coherent with the standard implementation of JNI code, we ended up having on average a better precision. The same goes for the smell \memoryMagement. Indeed, we implemented a simple detection approach that could be applied to detect the smell following the definition and rule presented in this paper.
Thus, this could not be generalized to all memory allocation issues. The detection approach relies on rules specific to the JNI usage. Thus, other native methods that could be implemented without considering JNI guidelines could lead to false positives and false negatives.  
To reduce threats to the validity of our work, 
we manually verified instances of smells reported by our detection approach on 
six open source projects along with our pilot project and measured the recall and precision of our detection approach 
as described in Section \ref{sec:Study}.}

\paragraph{\textbf{Distribution of JNI Smells}} From our results we found that most of the studied smells specific to JNI systems are prevalent in the selected projects. Results from the studied systems reflect a range from 10.18\% of smelly files in \jpype~system to 61.36\% of smelly files in \zstd. On average, 33.95\%  of the JNI files in the studied systems contain multi-language design smells. Multi-language systems offer numerous benefits, but they also introduce additional challenges. Thus, it is expected to have new design smells specific to such systems due to their heterogeneity. The prevalence of multi-language smells in the selected projects highlights the need for empirical evaluation targeting the analysis of multi-language smells and also the study of their impact on software maintainability and reliability. We also analyzed the persistence of these smells. Our results show that overall the number of smells usually increases from one release to the other. Such systems usually involve several developers working in the same team and who might not have a good understanding of the architecture of the whole project. Thus, the number of smells may increase if no tools are available to detect those smells and-or to propose refactored solutions. 

We observed situations in which the number of smells could decrease from one release to the next one. From investigating the commit message, we observed that some smells were refactored from one release to the other. Most of them due to the side effect of other refactoring activities, but also due to specific refactoring activities, \eg removing \unusedParameters, unused methods, implementing the handling of native exceptions, etc. This suggests that some developers might be aware of the necessity to remove those smells. However, since no tools are available to automatically detect such occurrences, it is hard for a developer to manually identify all the occurrences. However, we plan in another study to investigate the developers' perceptions and opinions about those smells as well as their impacts on software quality.

\paragraph{\textbf{Distribution of specific kinds of smells}}
We investigated in \textbf{RQ2}, if some specific smells are more prevalent than others. We found that the smells are not equally distributed within the analyzed projects. We also investigated their evolution over the studied releases. Our results show that the studied smells either persist or even mostly increase in number from one release to another. We observed some cases in which there was a decrease from one release to the other, and where smells occurrences were intentionally removed (\rocksdb, \conscrypt) by refactoring. Those systems are emerging respectively from Facebook and Google. In \realm, we also observed the awareness of developers about the bad practice of not removing local references (commit message: \emph{``DeleteLocalRef when the ref is created in loop (\#3366) Add wrapper class for JNI local reference to delete the local ref after using it''}). This could explain the decrease of smells occurrences in some situations. However, since no automatic tool is available, it could be really hard to identify all the occurrences, especially since such systems usually include different teams, which could explain the increase and decrease of multi-language design smells occurrences. 

Our results show that \unusedParameters~is one of the most frequent smells in the analyzed projects. This could be explained by the nature of the smell. This smell is defined when an unnecessary variable is passed as a parameter from one language to another. Since multi-language systems are emerging from the concept of combining heterogeneous components and they generally involve different developers who might not be part of the same team, it could be a challenging task for a developer working only on a sub-part of a project to clearly determine whether that specific parameter is used by other components or not. Thus, developers will probably tend to opt for keeping such parameters for safety concerns. The same goes for \scatter~and \declaration, these smells are defined respectively by occurrences in the code of native methods declarations that are no longer used, and separate and spread multi-language participants without considering the concerns. The number of these smells seems to increase over the releases as shown in Fig. \ref{fig:RQ22}. Under time pressure the developers might not take the risk to remove unused code, especially since in the case of JNI systems, such code could be used in other components. Similarly, the high distribution and increase of \scatter~could be explained in situations where several developers are involved in the same projects, bugs related to simultaneous files changes may occur. When features are mixed together, a change to the behavior of one may cause a bug in another feature. Thus, developers might try to avoid these breakages by introducing scattered participants. 
Similarly, the design smell \notSecuringLib~is prevalent in the analyzed systems. We believe that developers should pay more attention to this smell. Malicious code may easily access such libraries. Occurrences of this smell can introduce vulnerabilities into the system, especially JNI systems that have been reported by previous studies to be prone to vulnerabilities \cite{Tan:2008,Lee:2009}. Several problems may occur due to the lack of security checking. An unauthorized code may access and load the libraries without permission. This may have an adverse impact especially in industrial projects that are usually developed for sale or are available for online use, or other safety-critical systems.

\subsection{Smells and Faults}

\paragraph{\textbf{Relation Between Smells and Faults}}
In \textbf{RQ3}, we analyzed the relation between smells and fault-proneness. We used Fisher’s exact test and the odds ratios to check whether the proportion of buggy files varies between two samples (with and without design smells). From our results, we found that in general odds ratios are higher than one. This confirms previous insights from mono-language studies in which researchers claimed that design smells could increase the risk of faults \cite{khomh2009exploratory,jaafar2013mining}. We cannot claim causation as we do not know whether such faults could have been caused by other factors.
Although, our results suggest that files with JNI systems are more likely to be associated with faults than files without.
In \zstd, we found higher ORs than those of other systems from \textit{13.9285} to \textit{37.7142}; this could be explained by the nature of smells involved in this system as reported in Table \ref{tab:RQ2}. Some types of smells could be more related to bugs than other types. Out of all the 98 releases analyzed, we found eight releases with ORs less than one, however, none of them was with a significant \textit{p}-value. In \javasmt~and \javacpp, \textit{p}-values are not statistically significant (higher than 0.05) in most releases. 

 
From studying bug-fix commit messages, we observed that the impact is also smell-dependent. Occurrences of some types of smells seem more related to bugs than others, which motivates us to perform the \textbf{RQ4}. Some occurrences of smells related to bugs have been refactored from one release to the next one. In many cases, we find a description in the commit message indicating refactoring for removing specific smells that caused the bugs (commit message: \eg \emph{``There were a bunch of exceptions that are being thrown from JNI methods that aren't currently declared'', ``cleaning up JNI exceptions (\#252)'', ``removed a few unused JNI methods''}).  
Mono-language smells have been widely studied in the literature and were reported to negatively impact systems by making classes more change-prone and fault-prone.  
Multi-language systems could introduce additional challenges compared to mono-language systems. Those challenges are mainly related to the incompatibilities of programming languages and the heterogeneity of components. Thus, the design smells occurring on those systems are expected to increase the challenges related to the maintenance of these systems. Even if some smells \eg \declaration, \implementation~could not be directly related to bugs, they seem to increase the maintenance efforts because some of them are intentionally removed by developers. Thus, we believe that developers should be cautious about files with JNI smells, because they are more likely to be subject to faults and thus may incur additional maintenance efforts. 
Developers should also pay attention to avoid introducing occurrences of such design smells when dealing with JNI systems. 

\paragraph{\textbf{Relation between Specific Smells and Faults}}
Results from \textbf{RQ4} show that some smells seem more related to faults than others: \unusedParameters, \cluster, \scatter, \hardCodingLib, and \exception. The smell types \memoryMagement~and \notSecuringLib~are also found to be related to bugs. We believe that files containing these smells should be considered in priority for testing and-or refactoring. The smell \exception~was previously reported as related to bugs \cite{Tan06safejava,Lee:2009}. In fact, we discussed a bug related to this smell early in Section \ref{sec:Intro}. A bug related to this smell was reported in \conscrypt, developers were not checking for Java exceptions after all JNI calls that might throw them. The management of exceptions is not automatically ensured in all the programming languages. Incompatibility between the programming languages may lead to bugs and challenges related to the conversion, management of memory, and other mismatches between programming languages. In JNI projects, developers should explicitly implement the exception handling flow. Similarly, bugs in files containing the smells \unusedParameters~and \cluster~could be explained as the impacts of the noises that these two smells could introduce. Indeed, unused code or huge files with JNI code could impact 
code maintainability and the comprehension of JNI systems, which may lead to the introduction of bugs. From our detection approach, we identified files containing more than 200 method declarations that are not necessarily related in terms of responsibilities and that do not follow the principle of separating the concerns. We believe that faults could be easily introduced in such files, 
especially when dealing with JNI code; a developer might not be an expert on all the languages used and the inter-language interfaces. Developers should be concerned about the types of smells that are more likely to introduce bugs. The code containing these smells should be prioritized for testing and refactoring. 

\subsection{Risky Activities}
From \textbf{RQ5}, we found that activities related to data conversion, memory management, restructuring the code, API usage, and exception management are among the activities that could increase the risk of bugs when performed on smelly files. This is not surprising as several articles and developers' blogs discussed bugs related to the management of the memory in JNI systems \cite{Lee:2009,Tan:2008}. Some of these activities are directly related to design smells discussed in this paper, \eg \memoryMagement, \localReference. Developers who do not follow good practices to avoid such design smells could perform activities that could increase the risk of future bugs in those files. In the context of JNI systems, it is the developers' responsibility to take care of the management of memory because of the incompatibility between Java and C/C++. The same goes for data conversion when using JNI. We should consider specific rules to convert and access data between Java and C/C++. Primitive types could be easy to convert from Java to C/C++. However, reference types are more complex and require additional knowledge on what kind of methods to use to apply a proper conversion. Several studies also discussed issues related to exceptions in JNI context. Unlike Java, C/C++ does not support the automatic handling of exceptions. Developers could introduce bugs if they do not have enough knowledge about how to implement the exception handling flow in JNI context. Such incompatibility between programming languages could introduce bugs and other maintenance challenges including checking exceptions, buffer overflows, and memory leaks \cite{Kondoh:2008}. Following formal guidelines and being aware of the practices to follow could help to improve the quality of those systems \cite{abidi2019anti,abidi2019code,Lee:2009,Tan:2008}. 
We also noticed that in some systems, developers started paying more attention to this smell to avoid bugs related to the management of exceptions \conscrypt: \emph{``This works towards issue \#258. So the exception can be routed out properly, this moves the SSL\_get0\_peer\_certificates call to after doHandshake completes in ConscryptFileDescriptorSocket''}. Another example from \realm, developers started paying more attention to the smell \localReference~\emph{``DeleteLocalRef when the ref is created in a loop (\#3366) Add wrapper class for JNI local reference to delete the local ref after using it''}. We believe that further investigations should be performed to better understand the reasons for bug introduction in the presence of this smell. 

\subsection{ Implications of the Findings} 
\newtext{Based on our results we formulate some recommendations and highlight the implications of our findings that could help not only researchers but also developers and anyone 
considering using more than one programming language in a software system: }

\newtext{Our main goal was to investigate the existence of multi-language design smells and their impact on software quality. We found that multi-language code smells frequently occur within the selected projects and that they may increase the risk of bugs occurrence. Our results also highlight that the frequency and impact differ from one smell to the other. We also studied the activities that could introduce bugs once performed in smelly files.}

\newtext{Some of the implications of this study could be derived directly from the outcome of our research questions. First, researchers could find interest in studying why and how some specific types of smells are more frequent than others and the reasons behind their increase over time. They could also investigate the reasons why 
some specific types of smells are more related to bugs than others. The same goes for the activities, they could investigate further reasons behind the introduction of bugs when those specific activities are performed. They could also explore the existence of other activities that could introduce bugs.  
Second, practitioners could also take advantage of the outcome of this paper to reduce the maintenance cost of multi-language systems. In fact, 
most of the smells discussed in this paper (even those that are not always related to bugs) could introduce additional challenges and increase the effort of maintenance activities. Having knowledge of their existence and the potential impact could help to improve the quality of multi-language systems, and avoid their introduction in systems during evolution activities. 
In fact, as reported earlier, we found multiple commit messages in which developers explicitly mentioned issues caused by the occurrence of a smell studied in this paper. 
Studying each type of smell separately also allowed us to capture their impact individually. The insights from this study could help developers to prioritize multi-language smells for maintenance and refactoring activities. The same goes for the activities introducing bugs. Being aware of those activities could help developers avoid issues when 
performing them. Finally, the catalog of design smells studied in this paper is not exhaustive and presents only a small sample of possible multi-language smells and practices. Therefore, researchers and developers could further investigate smells and practices in multi-language software development. Our focus in this paper was on the JNI systems, and the researchers could also investigate other combination of programming languages. 
Additionally, they can also examine the impact of design smells on other quality attributes.}

\newtext{
We recommend that developers pay more attention to the design patterns and design smells discussed in the literature that could be applied to the context of multi-language systems. Our results highlight the need for more empirical studies on the impact of 
multi-language smells on maintainability and program comprehension. We recommend to developers to be cautious when editing files containing design smells 
\unusedParameters, \emph{To Much Clustering, Too much Scattering, Not Handling Exception, Hard Coding Libraries} since their occurrence seems to increase the risk of fault introduction.}

\section{Threats To Validity}
\label{sec:Threats}
In this section, we shed light on some potential threats to the validity of our methodology and findings following the guidelines for empirical studies \cite{yin2002applications}. 

\paragraph{\textbf{Threats to Construct Validity}} These threats concern the relation between the theory and the observation. 
In this study, these threats are mainly due to measurement errors. \newtext{Most of the studied projects rely on Github issues to report bugs. 
Therefore, we identified fault-fixing commits by mining the Github commit logs using a set of keywords extracted from the literature \cite{zafar2019towards,mockus2000identifying,ray2014large}. We used a set of keywords similar to those previously used in studies focusing on bug prediction. However, this technique may not capture 
all the commits related to fault-fixing if the commit messages were not representative enough of the developer's intention or were not containing any of those keywords. Nevertheless, this methodology was successfully used in multiple previous empirical studies \cite{selim2010studying,saboury2017empirical,khomh2012exploratory,zafar2019towards}. Moreover, in \cite{CastelluccioAK19}, the authors report that this technique can achieve a precision of 87.3\% and a recall of 78.2\%.
Another threat to construct validity is related to the accuracy of the SZZ heuristic 
used to identify fault-inducing commits. Although this heuristic does not achieve a 100\% accuracy, it has been successfully employed and reported to achieve good results in multiple empirical studies from the literature \cite{rodriguez2018if,rodriguez2018reproducibility,8870178}. We also did a manual validation of the bug inducing commits as described in Section \ref{sec:bug} by inspecting the changes of a small sample of bug inducing commits.
For our smell detection approach, we applied simple rules. We adapted our detection approach to ensure a balanced trade-off between the precision and the recall. For some smells, \eg \memoryMagement, we considered specific situations in which the smell occurs following simple rules and the definition presented earlier in Section \ref{sec:ML-smells}. Thus, this is not currently covering all possible issues related to memory management. However, the approach could be extended to include other contexts and types of memory issues following other rules. }

\newtext{When analyzing the smelliness of files that experienced bugs, we considered the whole file as participating in the design smell. Hence, the smell present in the file could be in different code lines than the bug. 
There is a similar threat in our analysis of the activities introducing bugs. We rely on commit messages provided by developers to identify the activities. We are aware that in some cases, developers might not have provided all the details of the activities performed or might have used some abbreviations. However, we mitigated this threat by combining both manual and automatic approaches to capture the possible activities that were performed. We are aware that the retrieved topics may not be 100\% accurate. 
However, we followed the coding methodology applied in previous studies \cite{8816780,jelodar2019latent} and two of the authors manually validated a subset of the commit messages. Through the manual analysis, we found that some commit messages describe more than one activity in the same commit (\eg Commit extracted from \zstd: ``Align the JNI names with the new streaming API, Move to the new streaming API, Use the ZBUFF based streaming compression'') while they assigned by the automatic approach to a single category. Although, the category to which they are assigned is based on the frequencies of the related keywords. We mitigated this threat by performing a manual validation over 500 commit messages. We also investigated some examples of activities at the source code level in the smelly code as described in Section \ref{sec:RsltTopic}. The list of the activities may not be exhaustive and do not present a 100\% recall and precision. However, in this paper we are reporting our observation on the activities that once performed in smelly files could introduce bugs without any empirical comparison of the risk introduced by each activity. 
However, we consider this as our future work in which we plan to perform a full manual validation approach to capture individual activities and the risk of introducing bugs related to each of them. 
}

\paragraph{\textbf{Threats to Internal Validity}}
We do not claim causation and only relate the presence of multi-language design smells with the occurrences of faults. We report our observations based on empirical results and explain these observations with manually analyzed examples from the studied systems to better contextualize our findings. We are aware that smells can depend on each other and we select the subset of non-correlated smells while building the logistic regression models. However, the variations in the distribution of smells, and some smells being very infrequent can have negative impacts on the regression models. As our model for each system considers all releases of a particular system than individual releases separately, it helps compensate for the infrequent classes by boosting the per-class data size. 
\newtext{Our study is an internal validation of multi-language design smells that we previously defined and cataloged. Thus, this may present a threat to validity. However, this threat was mitigated by publishing our catalog in a pattern conference. The paper went through rounds of a shepherding process. In this process, an expert on patterns provided three rounds of meaningful comments to refine and improve the patterns. The catalog then went through the writers' workshop process, in which five researchers from the pattern community had two weeks before the writers' session to carefully read the paper and provide detailed comments for each defined smell. The catalog was then discussed during three sessions of two hours each. During these sessions, each smell was examined in detail along with their definition and concrete examples. The conference chair also provided additional comments to validate the catalog. In addition, the results of this paper have shown that the studied smells are related to bugs. From the commit messages, we also found that some smells were explicitly discussed by developers who contributed in the smelly files. For example, one developer discussed exception handling as \textit{``There were a bunch of exceptions that are being thrown from JNI methods that aren't currently declared''}. Therefore, we believe that the studied smells should be considered with caution by developers since they may hinder the software maintenance and may lead to bugs.}

\paragraph{\textbf{Threats to External Validity}} These threats concern the possibility to generalize our results. 
We studied nine JNI open source projects with different sizes and domains of application. We focused on the combination of Java and C/C++ programming languages. Nevertheless, further validation of a larger number of systems with other sets of languages would give more opportunities to generalize the results. We studied a particular yet representative subset of multi-language design smells. Future works should consider analyzing other sets of design smells. 
\paragraph{\textbf{Threats to Conclusion Validity}} These threats are related to the relationship between the treatment and the outcome. We were careful to take into account the assumptions of each statistical test. 
We mainly used non-parametric tests that do not require any assumption about the data set distribution.

\paragraph{\textbf{Threats to Reliability Validity}} We mitigate the threats by providing all the details needed to replicate our study in section \ref{sec:Study}. 
We analyzed open source projects hosted in GitHub.

\section{Related Work}
\label{sec:RW}

We now discuss the literature related to this work.
\subsection{Multi-language Systems}
Several studies in the literature discussed multi-language systems. One of the very first studies, if not the first, was by Linos \al\ \cite{linos1995polycare}. They presented \textit{PolyCARE}, a tool that facilitates the comprehension and re-engineering of complex multi-language systems. \textit{PolyCARE} seems to be the first tool with an explicit focus on multi-language systems. They reported that the combination of programming languages and paradigms increases the complexity of program comprehension.
Kullbach \al\cite{kullbach1998program} also studied program comprehension for multi-language systems. They claimed that program understanding for multi-language systems presents an essential activity during software maintenance and that it provides a large potential for improving the efficiency of software development and maintenance activities. 
Linos \al\cite{linos2003tool} later argued that no attention has been paid to the issue of measuring multi-language systems' impact on program comprehension and maintenance. They proposed \textit{Multi-language Tool (MT)}; a tool for understanding and managing multi-language programming dependencies. Kontogiannis \al\cite{kontogiannis2006comprehension} stimulated discussion around key issues related to the comprehension, reengineering, and maintenance of multi-language systems. 
They argued that creating dedicated multi-language systems, methods, and tools to support such systems is expected to have an impact on the software maintenance process which is not yet known. 
Kochhar \al\cite{kochhar2016large} investigated the impact on software quality of using several programming languages. They reported that the use of multi-programming languages significantly increases bug proneness. They claimed that design patterns and anti-patterns were present in multi-language systems and suggested that researchers study them thoroughly. Kondoh \al\cite{Kondoh:2008} presented four kinds of common JNI mistakes made by developers. They proposed \textit{BEAM}, a static-analysis tool, that uses a typestate analysis, to find bad coding practice pertaining to error checking, virtual machine resources, invalid local references, and JNI methods in critical code sections. 
Tan \al\cite{Tan:2008} studied JNI usages in the source code of part of JDK v1.6. They examined a range of bug patterns in the native code and identified six bugs. The authors proposed static and dynamic algorithms to prevent these bugs. 
Li and Tan \cite{Li:2009} highlighted the risks caused by the exception mechanisms in Java, which can lead to failures in JNI implementation functions and affect security. 
They defined a pattern of mishandled JNI exceptions. 

\subsection{Impacts of Patterns and Smells} 
Several studies in the literature have studied the impact of design smells on software quality but mainly for mono-language systems.

Khomh \al\cite{khomh2009exploratory} analyzed nine releases of Azureus and 13 releases of Eclipse to investigate if the classes with occurrences of design smells are more change-prone than classes without those occurrences. They concluded that the classes with occurrences of design smells are more likely to be the subject of changes than classes without those occurrences. Olbrich \al\cite{olbrich2009evolution} proposed an approach that analyses the evolution of design smells and study their impact on the frequency and size of changes. They study two design smells: God Class and Shotgun Surgery. They used an automated approach based on detection strategies to detect the occurrences of design smells. They identified different phases in the cycle of design smells evolution during the different phases of the system development. They also found that components infected by design smells exhibit different behavior. Abbes \al\cite{abbes2011empirical} investigated the impact of occurrences of anti-patterns in the developers' understandability of systems while performing comprehension and maintenance tasks. They conducted three experiments to collect data about the performance of developers and study the impact of Blob and Spaghetti Code anti-patterns and their combinations. They concluded that the occurrence of one anti-pattern does not significantly impacts comprehension while the combination of the two anti-patterns negatively impact program comprehension. This finding was corroborated by Politowski et al. \cite{POLITOWSKI2020106278}. Linares \al\cite{linares2014domain} studied the potential relationship between the occurrence of design smells and quality attributes as well as the possible relation between design smells and application domains. They analyzed 1,343 Java Mobile applications in 13 different application domains. They concluded that anti-patterns negatively impact software metrics in Java Mobile applications, in particular, fault-proneness. They observed that there is a difference in the metric values between classes containing occurrences of smells and classes without smells. They also found that some smells are more frequently present in a domain of application while other smells are more present in other domains. Soh \al\cite{soh2016code} performed a study with six developers, three maintenance tasks, and four equivalent functions in Java. They used the Eclipse Mimec plugin and Thinkaloud sessions to analyze the effort spent by different developers when performing different maintenance activities (editing, reading, navigating, searching, static navigation, executing, and other activities). They concluded that design smells differently impact the effort needed to perform the different activities. They also found that the effort needed for reading, navigating, and editing is affected by three smells: ``Feature Envy'', ``God Class'', and ``ISP Violation''.

\subsection{Patterns and Smells Detection Approaches}

\newtext{
Van Emden et al. \cite{van2002java} proposed the JCosmo tool that supports the visualization of the code layout and design defects locations. They used primitives and rules to detect occurrences of anti-patterns and code smells while parsing the source code into an abstract model.}
\newtext{
Marinescu et al. \cite{marinescu2004detection} proposed an approach for design defects detection based on detection strategies. The approach captures deviations from good design principles and heuristics to help developers and maintainers in the detection of design problems.}
\newtext{
Lanza et al. \cite{lanza2007object} presented the platform iPlasma for software modeling and analysis of object oriented software systems to detect occurrences of design defects. The platform applies rules based on metrics from C++ or Java code.}
\newtext{
Moha et al. \cite{moha2007p} introduced DECOR which detects design defects in Java programs. DECOR is based on a domain-specific language that generates the design defect detection algorithms.} 
\newtext{
Khomh et al. \cite{khomh2009bayesian} proposed a Bayesian approach to detect occurrences of design defects by converting the detection rules of DECOR into a probabilistic model. Their proposed approach has two main benefits over DECOR: (i) it can work with missing data and (ii) it can be tuned with analysts’ knowledge. Later on, they extended this Bayesian approach as BDTEX \cite{khomh2011bdtex}, a Goal Question Metric (GQM) based approach to build Bayesian Belief Networks (BBNs) from the definitions of anti-patterns. They assessed the performance of BDTEX on two open-source systems and found that it generally outperforms DECOR when detecting Blob, Functional Decomposition, and Spaghetti code anti-patterns. 
}

\newtext{
Kessentini et al. \cite{kessentini2011design} proposed an automated approach to detect and correct design defects. The proposed approach automatically finds detection rules and proposes correction solutions in term of combinations of refactoring operations.}
\newtext{
Rasool et al. \cite{rasool2017lightweight} proposed an approach to detect occurrences of code smells that supports multiple programming languages. They argued that most of the existing detection techniques for code smells focused only on Java language and that the detection of code smells considering other programming languages is still limited. They used SQL queries and regular expressions to detect code smells occurrences from Java and C\# programming languages. In their approach, the user should have knowledge about the internal architecture of the database model to use the SQL queries and regular expressions. In addition, each language needs a specific regular expression.}
\newtext{Fontana et al. \cite{fontana2016comparing} conducted a study applying machine learning techniques for smell detection. They empirically created a benchmark for 16 machine learning algorithms to detect four types of code smells. The analysis was performed on 74 projects belonging to the \texttt{Qualitas Corpus} dataset. They found that \texttt{J48} and \texttt{Random Forest} classifiers attain the highest accuracy.}
\newtext{Liu et al. \cite{liu2018deep} proposed a smell detection approach based on Deep Learning to detect Feature Envy. The proposed approach relies on textual features and code metrics. 
It relies on deep neural networks to extract textual features.}
\newtext{Barbez et al. \cite{barbez2020machine} proposed a machine learning based method SMAD that combines several code smells detection approaches based on their detection rules. 
The core of their approach is to extract metrics based on existing approaches and use those metrics as features to train the classifier for smell detection.
The proposed approach supports the detection of the smells of type God Class and Feature envy. Their approach outperforms other existing methods in terms of recall and Matthews Correlation Coefficient (MCC).}
\newtext{Palomba et al. \cite{7503704} proposed TACO, an approach that relies on textual information to detect code smells at different levels of granularity. They evaluated their approach on ten open source projects and found that the proposed approach outperforms existing approaches.}
\newtext{
While there are some studies in the literature that document the good and bad practices related to multi-language systems,\cite{Tan:2008,Tan06safejava,goedicke2001message,goedicke2000object,goedicke2002piecemeal}  
to the best of our knowledge, this is the first study that automatically detects occurrences of multi-language design smells in the context of JNI systems and evaluates their impact on software fault-proneness. Other studies in the literature are focusing on the detection and analysis of design smells in 
mono-language systems. 
}

\section{Conclusion}

\label{sec:Conclusion}

In this paper, we present an approach to detect multi-language design smells and empirically evaluate the impacts of these design smells on fault-proneness. We performed our empirical study on 98 releases of nine open source JNI systems.  
Those systems provide a great variety of services to numerous different types of users. They introduce several advantages, however, as the number of languages increases so does the maintenance challenges of these systems. 
Despite the importance and increasing popularity of multi-language systems, studying the prevalence and impact of patterns and smells within these systems is still under-investigated. In this paper, we studied the impact of multi-language design smells on the software fault-proneness. 
We investigated the prevalence and impact of 15 design smells on fault-proneness. We showed that the design smells are prevalent in the selected projects and persist across the releases. Some types of smells are more prevalent than others. Our results suggest that files with JNI smells are more likely to be subject of bugs than files without those smells. We also report that some specific smells, are more likely to be of a concern than others, \ie \unusedParameters, \scatter, \cluster, \hardCodingLib, and \exception. These smells seem more related to faults, thus we suggest that practitioners consider them in priority for testing and-or refactoring. 
This empirical study supports, within the limits of its threats to validity, the conjecture that multi-language design smells are prevalent in the selected projects and that similar to mono-language smells, JNI smells may have a negative impact on software reliability. From analyzing fault-inducing commits we found that data conversion, memory management, code restructuring, API usage, and exception management activities could increase the risk of bug introduction when performed on smelly files. We believe that the results of this study could help not only researchers but also practitioners involved in building software systems using more than one programming language.
Our future work includes (i) replicating this study with a larger number of
systems for further generalization of our results; (ii) studying
the impact of design smells on change-proneness, (iii) investigating the occurrences of other patterns and defects related to multi-language systems. 
\vspace{0.35cm}

\bibliographystyle{IEEEtran}

\bibliography{references.bib}

\appendix
\section{Appendix} \label{appendixA}
\newtext{
We present in the following the smell detection rules of the proposed approach. These rules are applied on the srcML elements generated as an XML representation of a given project as described in Section \ref{sec:detection}. Since the smells described in this paper are multi-language smells, the following rules detect the occurrences of smells by using the XPath queries in the srcML representation of the source code that contains Java and C/C++ native code.} 


\begin{enumerate}
\item \newtext{ \textbf{Rule 1: \exception} }
  \newtext{     
 \begin{multline} 
 (f(y)~|~ f \in \{GetObjectClass, FindClass, GetFieldID, GetStaticFieldID,\\ GetMethodID, GetStaticMethodID\})\\ \nonumber
  \textbf{\textit{AND}}~
  (isExceptionChecked(f(y))=False ~ \textbf{\textit{OR}}~ ExceptionBlock(f(y))=False) 
\end{multline}
}
\newtext{Our detection rule for the smell \exception~is based on the existence of call to specific JNI methods requiring an explicit management of the exception flow. 
The JNI methods (\eg FindClass) listed in the rule should have a control flow verification. 
The parameter \textit{y} presents the Java object/class that is passed through a native call for a purpose of usage by the C/C++ side. 
Here, \textit{isExceptionChecked} allows to verify that there is an error condition verification for those specific JNI methods, while \textit{ExceptionBlock} checks if there is an exception block implemented. 
This could be implemented using Throw() or ThrowNew() or a return statement that exists in the method in case of errors. }\\

    \item \newtext{  \textbf{Rule 2: \SafeReturnValue}}
    \vspace{-0.15cm}
    \newtext{ 
     \begin{multline}
      ~~~~~x:= f(y)| ~f \in\{ FindClass, GetFieldID, GetStaticFieldID, GetMethodID, GetStaticMethodID \} \\
    \textbf{\textit{AND}}~  
~ isErrrorChecked(x)=False ~  \nonumber \textbf{\textit{AND}}~ IsReturn(x)=True  
   \end{multline}
   }
   \newtext{
    This rule is quite similar to the previous rule. However, it considers the return value from the native code. Indeed, the JNI methods called in this context are used for specific calculation and the result then needs to be passed as a method return value to the Java side. Here, \textit{x} presents the native variable used within the method to receive the returned value and perform computation on the Java side. \textit{isErrrorChecked(x)} allows to verify if there is an error condition verification applied to the variable \textit{x} that will be returned back to the Java code (\textit{IsReturn(x)=True}). The use of the variable \textit{x} as a return value by a native method without any check of its correctness will introduce smell of type \SafeReturnValue~ given other conditions hold.}\\

    \item \newtext{ \textbf{Rule 3: \notSecuringLib}}

  \newtext{   \[  \textit{ IsNative(Lib) = True \textbf{\textit{AND}} loadedWithinAccessBlock (Lib) = False}\]}

    \newtext{This rule implies that in the Java code, a native library is used (\textit{IsNative(Lib) = True}) and that this library is loaded outside a block \textit{AccessController.doPrivileged} without a try and catch statements for safe handling of potential exceptions. This introduces smell of type \notSecuringLib.}\\
    
    \item  \newtext{ \textbf{Rule 4: \hardCodingLib}}
    
    \vspace{0.15cm}
   \newtext{   \textit{ IsNative(Lib)= True \textbf{\textit{AND}} AccessiblePath(Lib)=False \textbf{\textit{AND}} OsBlock(m) = True}}
    \vspace{0.15cm}
    
   \newtext{This rule implies that in the Java code, a native library (\textit{Lib}) is used in a native method $m$ and that the path used for accessing that library is an absolute path  while the code loading the library depends on the operating systems. Here, the access to libraries is hard coded for specific operating system rather than implementing a platform independent access mechanism for libraries. This limits the portability of the code and may cause issues in accessing the libraries for different operating systems.}\\
  
    \item  \newtext{ \textbf{Rule 5: \relativePath}}
     
      \newtext{ \textit{ IsNative(Lib)= True \textbf{\textit{AND}} RelativePath(Lib) = False}}

     \newtext{This rule implies that in the Java code, a native library is used. However, the native library loaded from an absolute path and not from a relative path.}\\
        
    \item  \newtext{ \textbf{Rule 6: \cluster}}
  
      \newtext{ \textit{NbNativeMethods (C) >= MaxMethodsThreshold}  \textbf{\textit{AND}}  \textit{IsCalledOutside(m) = True}}

     \newtext{ This rule detects cases where the total number of native methods (\textit{NbNativeMethods}) within any class $C$ is equal to or higher than a specific threshold while those methods $m$ are used by other classes and not only the one where they are declared (\textit{IsCalledOutside(m) = True}). In our case, we used the default values for the threshold eight. However, all the thresholds could be easily adjusted as discussed earlier in Section \ref{sec:detection}. }\\
    
    \item  \newtext{ \textbf{Rule 7: \scatter}}
     \newtext{ 
       \begin{multline}
    ~~~~~~NBNativeClass (P) >= MaxClassThreshold~  \\
       \nonumber \textbf{AND}~ ( NbNativeMethods (C) < MaxMethodsThreshold~ \textbf{AND}~  C \in P)
     \end{multline}
    }

    \newtext{The smell of type \scatter~ occurs when the total number of native classes in any package $P$  (\textit{NBNativeClass(P)}) is more than a specific threshold (\textit{MaxClassThreshold}) for the number of maximum native classes. In addition, each of those native classes $C$ contains a total number of native methods (\textit{NbNativeMethods(C)}) less than a specific threshold (\textit{MaxMethodsThreshold}) \ie the class does not contain any smell of type \cluster. We used default values for the threshold three for the minimum number of classes with each a maximum of three native method each.}\\
    

    \item  \newtext{ \textbf{Rule 8: \excessivecom}}

    \newtext{
    \textit{(NBNativeCalls(C,m) > MaxNbNativeCallsThreshold )  \textbf{OR} \\ 
    (NbNativeCalls(m(p)) > MaxNativeCallsParametersThreshold ) \textbf{OR} \\
    ((NBNativeCalls (m) > MaxNbNativeCallsMethodsThreshold ) \textbf{AND} IsCalledInLoop(m) = True) } 
    }\\

    \newtext{The smell \excessivecom~ is detected based on the existence of at least one of the three possible scenarios. First, in any class $C$ the total number of calls to a particular native method $m$ exceeds the specified threshold (\textit{NBNativeCalls(C,m) > MaxNbNativeCallsThreshold}). Second, the total number of calls to the native methods $m$ with the same parameter $p$ exceeds the specific threshold (\textit{MaxNativeCallsParametersThreshold}). Third, the total number of calls to a native method $m$ within a loop is more than the defined threshold (\textit{(MaxNbNativeCallsMethodsThreshold}).} \\
 
 
    \item \newtext{\textbf{Rule 9: \localReference}}
  
   
   \newtext{
   \begin{multline}
   (NbLocalReference(f_1(y)) > MaxLocalReferenceThreshold )~ \textbf{AND} \\ (f_1(y)~ |~ f_1  \in  \{ GetObjectArrayElement, GetObjectArrayElement, NewLocalRef, AllocObject, \\ NewObject, NewObjectA, NewObjectV, NewDirectByteBuffer, \\ ToReflectedMethod, ToReflectedField \})~ \textbf{AND} \\ \nonumber (\nexists~ f_2(y)~ |~ f_2  \in  \{ DeleteLocalRef, EnsureLocalCapacity \}) 
   \end{multline} }
\newtext{The smell \localReference~ is introduced when the total number of local references (\textit{NbLocalReference(f$_1$(y)}) created inside a called method exceeds the defined threshold and without any call to method \texttt{DeleteLocalRef} to free the local references or a call to method \texttt{EnsureLocalCapacity} to inform the JVM that a larger number of local references is needed.} \\


  
    \item \newtext{\textbf{Rule 10: \memoryMagement}}
\newtext{
 \begin{multline}
  (mem \leftarrow f_1(y) ~|~ f_1  \in \{GetStringChars, GetStringUTFChars, GetBooleanArrayElements, \\ GetByteArrayElements, GetCharArrayElements, GetShortArrayElements, \\ GetIntArrayElements,  GetLongArrayElements, GetFloatArrayElements, \\ GetDoubleArrayElements, GetPrimitiveArrayCritical, GetStringCritical \}) ~\\
   \textbf{AND}~~   
 \nonumber (\nexists~ f_2(mem)~ |~ f_2  \in \{ReleaseGetStringChars, ReleaseGetStringUTFChars, \\ ReleaseGetBooleanArrayElements,  ReleaseGetByteArrayElements, \\ ReleaseGetCharArrayElements, ReleaseGetShortArrayElements, \\ ReleaseGetIntArrayElements,  ReleaseGetLongArrayElements, \\ReleaseGetFloatArrayElements,  ReleaseGetDoubleArrayElements, \\ ReleaseGetPrimitiveArrayCritical, ReleaseGetStringCritical \})  
\end{multline}
 }
\newtext{As discussed earlier, JNI offers predefined methods to manage the access of reference types that are converted to pointers. These methods are used to create pointers and to allocate the corresponding memory. The rule described here allows to detect the native implementation in which the memory was allocated by calling one of these allocation methods, however, the memory allocated was never released. The rule detects situations in which `get' methods are used to allocate memory for specific JNI elements that are not released after usage by calling the corresponding `release' methods.}\\

    \item \newtext{\textbf{Rule 11: \cachingObjects}}
\newtext{
\begin{multline}
((Parameter(m,p)=Object)~ \textbf{AND} ~ \\ ((NbCalls(C,m) >= MaxNbCallsThreshold) ~\textbf{OR}~  (IsLoop(m)= True\\ ~\textbf{AND}~ NoOfIterations >= MaxCountThreshold))\\
~\textbf{AND}~ (IsCalled(m,f_n(y)) = True) \\
~\textbf{AND}~  (f_n(y) | f_n  \in \{GetFieldID, GetMethodID, GetStaticMethodID \}))\\ ~\textbf{OR}~ 
 ((Parameter(m,p)=Object) ~\textbf{AND}~ (IsCalledInMethod(m, f_n)=True \\~\textbf{AND}~ NbCalls(f_n(y)) >= MaxNbCallsThreshold) ~\textbf{AND}~ \\ \nonumber (f_n(y) | f_n  \in \{GetFieldID, GetMethodID, GetStaticMethodID \}))
\end{multline}
 }

\newtext{This rule allows to detect occurrences of the smell \cachingObjects~ based on two situations. The first one is where the total number in which ids related to the same object $p$ are looked up for the same class $C$ through JNI allocation methods is greater than or equal to a specific threshold or the method is called within a loop. Indeed, the ids returned for a given class $C$ remain the same for the lifetime of the JVM execution. Considering that we have a native method $m$ and one of its parameter $p$ is a Java object (\textit{Parameter(m,p)=Object}), this type is considered in the native code as a reference type. Thus, unlike primitive types, its element could not be accessed directly by the native code but should be accessed through the usage of the methods defined in (\textit{IsCalled(m,$f_n(y)$}) = True). In this first scenario, the total number of calls from the Java code to a native method $m$ that is defined in a class $C$ exceeds a specific threshold (\ie \textit{NbCalls(C,m) >= MaxNbCallsThreshold}) or the method is called within a loop. In the second scenario, the number of times the same id for an object $p$ is looked up inside the same method $m$ (\textit{IsCalledInMethod(m, $f_n$)=True}) more than a given threshold even if the method $m$ is called only once (\textit{NbCalls($f_n(y)$) >= MaxNbCallsThreshold}). This last scenario includes the total number of calls to the predefined methods (\textit{NbCalls($f_n(y)$}) independent of the total number of calls to the method itself.}\\
 
 
    \item \newtext{ \textbf{Rule 12: \excessiveObject}}
 \newtext{
 \begin{multline}
  (Parameter(m,p)=Object) ~\textbf{AND}~ (IsCalledInMethod(m, f_1)=True)  ~\textbf{AND}~ \\ (NbCalls(f_1(y)) >= MaxNbCallsThreshold)~\textbf{AND}~ \\ (f_1(y) | f_1  \epsilon  \{GetObjectField, GetBooleanField, GetByteField, GetCharField, GetShortField,\\ GetIntField, GetLongField, GetFloatField, GetStaticObjectField  \}) 
    ~\textbf{AND}~ \\ 
 (\nexists~ f_2(y) ~|~ f_2  \in  \{SetObjectField, SetBooleanField, SetByteField, SetCharField, \nonumber \\SetShortField, SetIntField, SetLongField, SetFloatField, SetStaticObjectField \})  
 \end{multline}
 }

\newtext{This rule identifies situations in which a JNI object is passed as a parameter (\textit{Parameter(m,p)=Object}) to the native code. In this context the total number of calls to allocation methods to retrieve its field id in the same method is higher than a specific threshold (\ie \textit{NbCalls($f_1(y)$) >= MaxNbCallsThreshold}), without a call to corresponding set functions to set the object fields by the native code. However, as described in the specification of the smell in Section \ref{sec:ML-smells}, having the total number of calls to allocation methods higher than the threshold is not considered as a smell only in situations where the purpose of those calls was to set the object fields by the native code.}\\

    \item \newtext{\textbf{Rule 13: \implementation}}
  
    \vspace{0.15cm}
   \newtext{ \textit{IsNative(m) = True ~\textbf{AND} IsDeclared(m) = True ~\textbf{AND} IsImplemented(m) = True \\ \textbf{AND} IsCalled(m)=False}}
    \vspace{0.15cm}
    
 \newtext{ This rule allows to capture the native functions $m$ (\textit{IsNative(m) = True}) implemented in the C/C++ (\textit{IsImplemented(m) = True}), declared in Java with the keyword \textit{native} but never used in the Java code (\textit{IsCalled(m)=False}). It looks for the native methods that are declared using the keyword \textit{native} with a header in the Java code and looks for the corresponding native implementation nomenclature.}\\
    
    \item \newtext{\textbf{Rule 14: \declaration:}}
        
    \vspace{0.15cm}
     \newtext{\textit{IsNative(m)=True ~\textbf{AND} IsDeclared(m)=True ~\textbf{AND} IsImplemented(m)=False}}
    \vspace{0.15cm}
     
    \newtext{Native functions declared in Java with the keyword \textit{native} (\textit{IsDeclared(m) = True}) that are not implemented in C/C++ (\textit{IsImplemented(m)=False}). 
    This rule allows to retrieve the native methods that are declared with a header in the Java code using the keyword \textit{native} and checks for the corresponding implementation nomenclature. However, those methods were never used or even implemented in the C/C++ code.}\\
  
    \item \newtext{\textbf{Rule 15: \unusedParameters}}
 
    \newtext{
    \begin{multline}
      ~~~~~(IsNative(m(p))=True ~\textbf{\textit{AND}}~ IsDeclared(m(p))=True ~\textbf{\textit{AND}}~ IsImplemented(m(p))=True \\
      \nonumber ~\textbf{\textit{AND}}~
      IsParameterUsed(p) = False
    \end{multline}
    }

    \newtext{This rule reports the method parameters that are used in the Java native method declaration header using the keyword \textit{native} (\textit{IsDeclared(m(p))=True}). However the parameter is never used in the body of the implementation of the methods, apart from the first two arguments of JNI functions in C/C++. The rule checks if the parameter $p$ is used in the corresponding native implementation (\textit{IsParameterUsed(p) = False}).}

\end{enumerate}

\section{Appendix} \label{appendixB}

We present in Table \ref{tab:valsmells} the results of the evaluation of the performance of our design smell detection approach.

\newtext{The pilot project as described in Section \ref{sec:Study} was the project we developed with the occurrences of the smells along with the clean code without any smell to test and validate our approach. This explains the 100\% precision and 100\% recall for all the smells. For the other projects, the precision and recall were evaluated through the investigation of the occurrence of the smell itself and the multi-language files involved on that smell. Most of the rules are trivial and the corresponding smells could easily be detected by our approach. Therefore, the reasons for false positives and false negatives are mainly related to the alternative implementation choices of the multi-language code that do not follow JNI specification guidelines and therefore are not currently covered by our approach. Indeed, our approach consider the JNI implementation with the appropriate naming convention as described in the JNI specification (\eg using the native keyword in the Java native method declaration, using JNIENV, JNIEXPORT, JNICALL, and Java\_ClassName\_methodname) \cite{liang1999java}. Thus, our detection approach could only be considered for JNI systems that follows the JNI specification guideline. The validation on some systems were done earlier and thus on the older versions of the systems. Thus, the validation results only reports on the smell types available in the analyzed version.} 
\newtext{
As described in the Section \ref{sec:Discussion}, our approach may present some limitations for the smell \localReference~ in situations in which some specific methods are used to ensure the memory capacity. However, as per our manual analysis when defining the smells, those methods are not considered relevant to detect the smell and are not usually used. However, we are aware that in similar situation, the approach may result in false positives. For the smells \unusedParameters~ and \declaration, when evaluating the recall and precision we noticed that our approach was not always able to correctly match the java and corresponding native  implementation. This was mainly due the syntax used in the C implementation that is not completely following the JNI specification for the naming convention (\eg \pljava~ \textit{jobject pljava\_DualState\_key}). For the smells \SafeReturnValue~ and \exception, the false negatives in \conscrypt~ were related an intermediate step that made the detection harder. In this intermediate step, the native value was checked before returning it to the Java code. The same goes for the smell \memoryMagement, our rules allows the detection of a specific type of memory issues and do not cover other types of issues related to the memory.} 

\begin{landscape}
\begin{table*}
\centering\footnotesize
\caption{\label{tab:valsmells} Validation for Each Type of Smells
}
\scalebox{0.55}{
\begin{tabular}{ll|llll|llll|llll|llll|llll|llll|llll|cc|} \hline
\multicolumn{2}{c} \textbf{}                  & \multicolumn{4}{|c|}{\textbf{pilotproject}}                        & \multicolumn{4}{|c|}{\textbf{conscrypt}}   & \multicolumn{4}{|c|}{\textbf{pljava}}      & \multicolumn{4}{|c|}{\textbf{openj9}}      & \multicolumn{4}{|c|}{\textbf{rocksdb}}     & \multicolumn{4}{|c|}{\textbf{jmonkey}}     & \multicolumn{4}{|c|}{\textbf{jna}}         & \multicolumn{2}{c|}{\textbf{Average}}      \\ \hline
\multicolumn{2}{c}{Design Smells} & \multicolumn{1}{|c}{FP} & \multicolumn{1}{c}{Precision} & FN & \multicolumn{1}{c|}{Recall} & \multicolumn{1}{|c}{FP} & \multicolumn{1}{c}{Precision} & FN & \multicolumn{1}{c|}{Recall} & \multicolumn{1}{|c}{FP} & \multicolumn{1}{c}{Precision} & FN & \multicolumn{1}{c|}{Recall} & \multicolumn{1}{|c}{FP} & \multicolumn{1}{c}{Precision} & FN & \multicolumn{1}{c|}{Recall} & \multicolumn{1}{|c}{FP} & \multicolumn{1}{c}{Precision} & FN & \multicolumn{1}{c|}{Recall} & \multicolumn{1}{|c}{FP} & \multicolumn{1}{c}{Precision} & FN & \multicolumn{1}{c|}{Recall} & \multicolumn{1}{|c}{FP} & \multicolumn{1}{c}{Precision} & FN & \multicolumn{1}{c|}{Recall} & \multicolumn{1}{c} \textbf{Precision} & \textbf{Recall} \\ \hline
1 & \begin{tabular}[c]{@{}l@{}}Excessive Inter-language\\  Communication\end{tabular} & 0  & 100\%  & 0 & 100\% & 0  & 100\%  & 0 & 100\% & 4  & 95\% & 3 & 96\%  & 9 & 96\% & 37  & 85\%  & 24 & 96\% & 91  & 86\%  & 8 & 95\% & 75  & 65\%  & 5 & 84\% & 22  & 54\%  & 94\% & 81\% \\

2 & Too Much Clustering & 0 & 100\% & 0 & 100\% & 0 & 100\% & 0 & 100\% & 0 & 100\% & 0 & 100\% & 0 & 100\% & 0 & 100\% & 2 & 96\% & 4 & 92\%  & 0 & 100\% & 0 & 100\% & 0 & 100\% & 0 & 100\% & 99\%	& 98\% \\

3 & Too Much Scattering & 0 & 100\% & 0 & 100\% & - & - & - & - & 0 & 100\% & 0 & 100\% & 0 & 100\% & 0 & 100\% & 5 & 92\% & 5 & 92\%  & 0 & 100\% & 0 & 100\% & 0 & 100\% & 0 & 100\% & 98\% & 98\%  \\

4 & Unused Method Declaration & 0 & 100\% & 0 & 100\% & 6 & 98\% & 0 & 100\% & 1 & 98\% & 42  & 67\%  & 29  & 95\% & 40  & 94\%  & 2 & 88\% & 0 & 100\% & 12  & 95\% & 86  & 72\%  & - & - & - & - & 94\% & 86\% \\

5 & Unused Method Implementation & 0 & 100\% & 0 & 100\% & - & - & - & - & 0 & 100\% & 0 & 100\% & - & - & - & - & - & - & - & - & - & - & - & - & - & - & - & - & - & - \\

6 & Unused Parameter  & 0 & 100\% & 0 & 100\% & 15  & 95\% & 132 & 67\%  & 0 & 100\% & 1 & 99\%  & 95 & 95\% & 76 & 96\%  & 17  & 93\% & 31  & 88\%  & 59  & 96\% & 13  & 99\%  & 43  & 88\% & 64  & 83\%  & 94\% & 88\%\\

7 & Assuming Safe Return Value  & 0 & 100\% & 0 & 100\% & 3 & 0\% & 0 & 100\% & - & - & - & - & 0 & 100\% & 4 & 66\%  & 0 & 100\% & 0 & 100\% & 32  & 88\% & 7 & 97\%  & - & - & - & -  & 72\%  & 90\% \\

8 & Excessive Objects & 0 & 100\% & 0 & 100\% & - & - & - & - & - & - & - & - & - & - & - & - & - & - & - & - & 0 & 100\% & 0 & 100\% & - & - & - & - & - & -\\

9 & Not Handling Exceptions & 0 & 100\% & 0 & 100\% & 5 & 0\% & 0 & 100\% & 0 & 100\% & 0 & 100\% & 3 & 98\% & 86  & 63\%  & 0 & 100\% & 0 & 100\% & 31  & 89\% & 0 & 100\% & 0 & 100\% & 1 & 83\%  & 81\% & 91\%\\
10 & Not Caching Objects & 0 & 100\% & 0 & 100\% & - & - & - & - & - & - & - & - & - & - & - & - & - & - & - & - & - & - & - & - & - & - & - & - & - & - \\

11 & Not Securing Libraries  & 0 & 100\% & 0 & 100\% & 0 & 100\% & 0 & 100\% & - & - & - & - & 0 & 100\% & 0 & 100\% & 0 & 100\% & 3 & 73\%  & 0 & 100\% & 0 & 100\% & 0 & 100\% & 2 & 71\% & 100\% & 88\%
 \\

12 & Hard Coding Libraries & 0 & 100\% & 0 & 100\% & - & - & - & - & - & - & - & - & - & - & - & - & 0 & 100\% & 0 & 100\% & - & - & - & - & 0 & 100\% & 0 & 100\% & - & - \\

13 & Not Using Relative Path & 0 & 100\% & 0 & 100\% & 0 & 100\% & 0 & 100\% & - & - & - & - & 0 & 100\% & 0 & 100\% & 0 & 100\% & 0 & 100\% & 0 & 100\% & 0 & 100\% & - & - & - & - &  100\% &  100\%\\

14 & Memory Management Mismatch  & 0 & 100\% & 0 & 100\% & 0 & 100\% & 0 & 100\% & 0 & 100\% & 8 & 50\%  & 1 & 94\% & 4 & 81\%  & 0 & 100\% & 2 & 75\%  & - & - & - & - & - & - & - & - & 98\% & 76\% \\

15 & Local References Abuse  & 0 & 100\% & 0 & 100\% & 0 & 100\% & 1 & 80\%  & - & - & - & - & 0 & 100\% & 1 & 80\%  & - & - & - & - & - & - & - & - & 2 & 78\% & 2 & 78\% & 92\% & 79\% \\ \hline
\end{tabular}
}
\end{table*}
\end{landscape}

\subsection*{Acknowledgment}

This work has been supported by the Natural Sciences and Engineering Research Council of Canada.



\end{document}